\documentclass[a4paper, 12 pt, oneside, fleqn]{paper}
\title{The Multiphase Buoyant Plume Solution of the Dusty Gas Model.}
\author{Matteo Cerminara\thanks{matteo.cerminara@gmail.com}\\ \small{\em
Istituto Nazionale di Geofisica e Vulcanologia, Sezione di Pisa}}
\date{\today}

\usepackage{array}
\usepackage{bm}
\usepackage{bmpsize}
\usepackage{booktabs}
\usepackage[font = scriptsize, format=hang, labelfont = {sf, bf}]{caption}
\usepackage{chngpage}
\usepackage{enumitem}
\usepackage{eurosym}
\usepackage{fancyhdr}
\usepackage{float}
\usepackage{footmisc}
\usepackage{indentfirst}
\usepackage{layaureo}
\usepackage{listings}
\usepackage{longtable}
\usepackage{supertabular}
\usepackage{microtype}
\usepackage{mparhack}
\usepackage{showlabels}
\usepackage{siunitx}
\usepackage{subfig}
\usepackage{tabularx}
\usepackage{url}
\usepackage[english]{varioref}
\usepackage{xcolor}
\usepackage{wrapfig}
\usepackage{empheq} 
\usepackage{cancel} 
\usepackage[titletoc]{appendix}
\usepackage{natbib} 

\usepackage[latin1]{inputenc}
\usepackage{lmodern}
\usepackage[T1]{fontenc}
\usepackage[english]{babel}

\usepackage{graphicx}

\usepackage{textcomp}
\usepackage{verbatim}
\usepackage{latexsym}
\usepackage{amsmath, amssymb, amsthm, eucal, multicol}
\usepackage{esdiff}
\usepackage{amsfonts}

\usepackage[pagebackref=false]{hyperref}

\usepackage{multibib}
\newcites{web}{Web sites}

\newcommand*\widefbox[1]{\fbox{\hspace{0.5em}#1\hspace{0.5em}}} 

\newcommand{\de}{\mathrm{d}}

\numberwithin{equation}{section} \numberwithin{figure}{section}
\numberwithin{table}{section}

\renewcommand{\vec}{\bm}

\newenvironment{sistema}%
{\left\lbrace\begin{array}{@{}l@{}}}%
{\end{array}\right.}

\newcount\colveccount
\newcommand*\colvec[1]{
        \global\colveccount#1
        \begin{pmatrix}
        \colvecnext
}
\def\colvecnext#1{
        #1
        \global\advance\colveccount-1
        \ifnum\colveccount>0
                \\
                \expandafter\colvecnext
        \else
                \end{pmatrix}
        \fi
}

\newcommand{\footLabel}[2]{%
    \addtocounter{footnote}{1}%
    \footnotetext[\thefootnote]{%
        \addtocounter{footnote}{-1}%
        \refstepcounter{footnote}\label{#1}%
        #2%
    }%
}

\newcommand{\footRef}[1]{%
    $^{\ref{#1}}$%
}

\DeclareMathOperator{\sign}{sign}

\newcommand{\miss}{\textcolor{red}{\underline{\bf MISS}}}

\newcommand{\ddt}{\partial_t}
\newcommand{\Div}{\nabla\cdot}

\newcommand{\Oplm}{\Omega_\textup{plm}}

\newcommand{\Rey}{\textup{Re}}
\newcommand{\Pra}{\textup{Pr}}
\newcommand{\Prat}{\textup{Pr}_\textup{t}}
\newcommand{\St}{\textup{St}}

\newcommand{\Nu}{\textup{Nu}}
\newcommand{\Eu}{\textup{Eu}}
\newcommand{\Fr}{\textup{Fr}}
\newcommand{\Ri}{\textup{Ri}}
\newcommand{\Ec}{\textup{Ec}}
\newcommand{\Co}{\textup{Co}}
\newcommand{\Ma}{\textup{Ma}}
\newcommand{\BV}{\ddot{N}}
\newcommand{\strat}{\ddot{\ell}}

\newcommand{\Cv}{C_\textup{v}}
\newcommand{\Cp}{C_\textup{p}}
\newcommand{\Cs}{C_\textup{s}}

\newcommand{\gammac}{\gamma_\textup{c}}
\newcommand{\Cm}{C_\textup{m}}
\newcommand{\Ca}{C_\alpha}

\newcommand{\Rm}{R_\textup{m}}

\newcommand{\phic}{\phi_\textup{c}}

\newcommand{\rhom}{\rho_\textup{m}}

\newcommand{\rhombar}{\bar{\rho}_\textup{m}}

\newcommand{\ubb}{\bar{\vec{u}}_\textup{m}}

\newcommand{\Ta}{T_\alpha}

\newcommand{\kg}{k_\textup{g}}

\newcommand{\chis}{\chi_\textup{s}}
\newcommand{\chie}{\chi_\textup{e}}
\newcommand{\psie}{\psi_\textup{e}}

\newcommand{\taus}{\tau_\textup{s}}
\newcommand{\etaK}{\eta_\textup{K}}
\newcommand{\etaX}{\eta_\varkappa}
\newcommand{\tauK}{\tau_\eta}

\newcommand{\mut}{\mu_\textup{t}}
\newcommand{\ept}{\epsilon_\textup{t}}
\newcommand{\plume}{\delta_\textup{p}}
\newcommand{\jet}{\delta_\textup{j}}
\newcommand{\qpt}{a_\textup{p}}
\newcommand{\ellM}{\ell_\textup{M}}

\newcommand{\Tbb}{T_{\beta}}
\newcommand{\Cbb}{C_{\beta}}

\newcommand{\Ue}{U_\epsilon}
\newcommand{\Qe}{Q_\textup{e}}
\newcommand{\Qs}{Q_\textup{s}}
\newcommand{\Ye}{Y_\textup{e}}
\newcommand{\Ys}{Y_\textup{s}}
\newcommand{\Yeo}{Y_\textup{e,0}}
\newcommand{\Yso}{Y_\textup{s,0}}

\begin{document}
\maketitle
\begin{abstract}
Starting from the balance equations of mass, momentum and energy we formulate an integral 1D model
for a poly-disperse mixture injected in the atmosphere. We write all the equations, either in
their most general formulation or in the more simplified, taking particular care in considering all
the underlying hypothesis in order to make clear when it is possible and appropriate to use them.
Moreover, we put all the equations in a non-dimensional form, making explicit all the dimensionless
parameters that drive the dynamics of these phenomena. In particular, we find parameters to measure:
the goodness of the Boussinesq approximation, the injected mass flow, the column stability
and his eventual collapse, and the importance of the atmospheric stratification, the initial kinetic
energy and the gravitational potential energy. We show that setting to zero some of these
parameters, it is possible to recover some of the existing jet and plume models for single-phase
flows. Moreover, we write a simplified set of equations for which it is possible to find
analytical solutions that can be used to describe also the dynamics of multiphase ``weak-plumes''.

Starting from the paper~\cite{morton1956} the study on jets and plumes has been carried out by a lot
of different researcher involved in a variety of disciplines. Indeed, these kind of phenomena are
quite ubiquitous in nature...
\end{abstract}

\section{The main assumptions.}\label{sec:assumptions}
In order to use the Dusty Gas model we have to assume:
\begin{itemize}
 \item Local equilibrium.
 \item All the phases, either solid or gaseous, move with the same velocity field $\vec{u}(\vec{x},t)$. \cite{Marble1970} shows that this assumption it is valid even for the solid phase if the Stokes time $\taus \equiv \frac{\hat{\rho}_\textup{s}}{\hat{\rho}_\textup{g}} \frac{d_\textup{s}^2}{18\nu}$ is small compared to the smallest time scale of the evolution problem.
 \item All the phases, either solid or gaseous, have the same temperature field $T(\vec{x},t)$. \cite{Marble1970} shows that this assumption it is valid even for the solid phase if the thermal relaxation time $\tau_{T,\textup{s}} = \frac{\hat{\rho}_\textup{s} C_\textup{s}}{k_\textup{g}} \frac{d_\textup{s}^2}{12}$ is small compared to the smallest time scale of the evolution problem.
\end{itemize}
Here we are interested in the mean behavior of a turbulent buoyant plume. Writing that solution we will use the following assumptions (see \cite{morton1956, morton1959, Wilson1976, list1982, papanicolaou1988, woods1988, fannelop2003, kaminski2005, Ishimine2006, plourde2008}):
\begin{itemize}
 \item Reynold number is big enough and turbulence is fully developed, so that will be possible to disregard thermal conduction and shear dissipation.
 \item Pressure is constant in horizontal section.
 \item The profiles of mean vertical velocity and mean density in horizontal sections are of similar form at all heights.
 \item The mean velocity field outside and near the plume is horizontal. We will need to make additional assumption on the dependence of the rate of entrainment at the edge of the plume to some characteristic velocity at that height.
 \item Stationary flow.
 \item Radial symmetry around the source.
\end{itemize}

\section{The multiphase Dusty-Gas equations.}
Using the hypothesis given in the previous section, the Dusty-Gas model~\citep{Marble1970} simplifies:
\begin{subequations}\label{eq:dustyCancel}
\begin{align}
  &\textcolor{blue}{\cancel{\ddt\rho_i}} + \Div (\rho_i \vec{u}) = 0\,, \quad \quad i\in\mathcal{I}\\
    &\textcolor{blue}{\cancel{\ddt\rho_j}} + \Div (\rho_j \vec{u}) = 0\,, \quad \quad j\in\mathcal{J}\\
        &\textcolor{blue}{\cancel{\ddt\rhom}} + \Div (\rhom \vec{u}) = 0\,,\\
 &\textcolor{blue}{\cancel{\ddt \big(\rhom\, \vec{u}\big)}} + \Div \big(\rhom\, \vec{u} \otimes \vec{u} + p\mathbb{I} \big) = \textcolor{blue}{\cancel{\Div \mathbb{T}}} + \rhom\, \vec{g}\,,\\
 &\textcolor{blue}{\cancel{\ddt \big(\rhom E_\textup{m}\big)}} + \Div \big[\big(\rhom E_\textup{m} + p\big)\,\vec{u}\big] = \textcolor{blue}{\cancel{\Div(u\cdot \mathbb{T}) - \Div q}} + \rhom\, \vec{u}\cdot \vec{g}\,.
\end{align}
\end{subequations}
As suggested in \cite{woods1988}, it is convenient to use the specific enthalpy $h_\textup{m} = e_\textup{m} + \frac{p}{\rhom} = (\Cm + \Rm) T$ instead of the specific energy $e_\textup{m}$. We define the specific heat at constant pressure of the mixture consequently:
\begin{equation}
C_{\textup{p},\textup{m}} = \Cm+\Rm = \sum_{i\in\mathcal{I}} [y_i (C_i+R_i)] + \sum_{j\in\mathcal{J}} (y_j C_j)\,,
\end{equation}
so that $h_\textup{m} = C_{\textup{p},\textup{m}} T$.
In this way, Eqs. \eqref{eq:dustyCancel} reduces to:
\begin{subequations}\label{eq:Eqs}
\begin{align} 
  & \Div (\rho_i \vec{u}) = 0\,, \quad \quad i\in\mathcal{I} \vspace{2pt}\label{eq:1Dplume_rhoi}\\
  & \Div (\rho_j \vec{u}) = 0\,, \quad \quad j\in\mathcal{J} \vspace{2pt}\label{eq:1Dplume_rhoj}\\
  & \Div \big(\rhom\, \vec{u} \otimes \vec{u} + p\mathbb{I} \big) = \rhom\, \vec{g} \vspace{2pt}\label{eq:1Dplume_mom}\\
  & \Div \big[\rhom\big(\frac{|\vec{u}|^2}{2} + h_\textup{m}\big)\,\vec{u}\big] = \rhom\, \vec{u}\cdot \vec{g}\,.\label{eq:1Dplume_en}
\end{align}
\end{subequations}

\section{The Buoyant Plume Solution.}\label{sec:buoyantPlumeSolution}
Coherently with hypothesis of Section \ref{sec:assumptions}, we will look for a solution of Eqs. \eqref{eq:Eqs} in the following form:
\begin{align}
 y_k(r,z) & = \begin{cases}1\,, & \mbox{if } r \geq b(z) \; \mbox{and}\; k = 1\\
 Y_\alpha(z)\,, & \mbox{if } r < b(z) \; \mbox{and}\; k = 1\\
 0\,,& \mbox{if } r \geq b(z) \; \mbox{and}\; k \neq 1\\
 Y_k(z)\,,& \mbox{if } r < b(z) \; \mbox{and}\; k \neq 1\\\end{cases}\label{eq:1Dplume_yk}\\
 \rhom(r,z) & = \begin{cases} \beta(z)\,, & \mbox{if } 0\leq r < b(z)\\ \alpha(z)\,, & \mbox{if } r \geq b(z) \end{cases}\label{eq:rhoS}\\
 \vec{u}(r,z) & = 
 \begin{cases} 
     +U(z)\hat{z}\,, & \mbox{if } 0\leq r < b(z)\\
     -\Ue(z) \hat{r}\,, & \mbox{if } r = b(z)\\ 
     -u_\epsilon(r,z) \hat{r}\,, & \mbox{if } r > b(z)\\ 
     u_\epsilon = U_\epsilon & \mbox{if } r \to b(z)\\
     u_\epsilon \to 0 & \mbox{if } r \gg b(z) 
 \end{cases}\label{eq:uS}\\
 p(r,z) & = p(z)\label{eq:pS}\\
 T(r,z) & = \begin{cases} \Tbb(z)\,, & \mbox{if } 0\leq r < b(z)\\ \Ta(z)\,, & \mbox{if } r \geq b(z) \label{eq:TS}\end{cases}
\end{align}
where $k=i=1$ is the phase index corresponding to the atmospheric gas, while $k\neq 1$ is the generic index of a phase ejected by the plume vent. Here  we used the so called purely ``Top Hat'' auto-similar profile. In general -- as shown in~\cite{morton1959} -- it is possible to use better profiles. Experiments show (see e.g.~\cite{papanicolaou1988}) that the auto-similar Gaussian profile best fit data for a wide range of velocity measurements.  Moreover, experiments are better reproduced choosing two different plume radius (say $b(z)$ and $\lambda b(z)$) for the density and the velocity profile; the temperature profile should be determined by the equation of state of the fluid. Nevertheless, even if these modification could be done in Eqs. \eqref{eq:1Dplume_yk}--\eqref{eq:TS}, here we decided -- for simplicity -- to use the ``Top Hat'' profile. \miss ({\em add comments to introduce Eq. \eqref{eq:1DplumeSumY} and the comments on entrainment, aggregation and settling that are included in the paper. Moreover, add something pointing out that we are neglecting the presence of humidity in the atmosphere})

Here $\Ue$ is an entrainment velocity. We shall write it as
\begin{equation}\label{eq:Ue}
 \Ue = \varkappa\, U \etaX\left(\beta/\alpha\right)
\end{equation}
where $\varkappa$ is a dimensionless entrainment coefficient and $\etaX$ is an arbitrary function of the density ratio (see e.g.~\cite{fannelop2003}). When $\etaX = 1$ we have the model of \cite{morton1956}, if $\etaX(x) = \sqrt{x}$ we get the model \cite{ricou1961}.

It useful to notice that inside the plume, the dusty gas constant $\Rm$ and specific heat at constant volume $C_{\textup{m}}$ can be written:
\begin{align}
& R_\beta = Y_\alpha R_\alpha + \sum_{i=2}^I (Y_i R_i) + \sum_\mathcal{J}(Y_j R_j) = Y_\alpha R_\alpha + \sum_{i=2}^I (Y_i R_i)\,,\\
& C_{\textup{v},\beta} = Y_\alpha C_{v,\alpha} + \sum_{i=2}^I (Y_i C_i) + \sum_\mathcal{J}(Y_j C_j)\,,
\end{align}
where $R_\alpha$ and $C_{\textup{v},\alpha}$ are respectively the gas constant and the specific heat at constant volume for the atmosphere. We also define the specific heat at constant pressure of the atmosphere and of the plume:
\begin{align}
& C_\alpha = C_{v,\alpha} + R_\alpha\,,\\
&C_\beta = C_{v,\beta} + R_\beta = Y_\alpha C_\alpha + \sum_{i=2}^I \left(Y_i (C_i+ R_i)\right) + \sum_\mathcal{J}(Y_j C_j)\,.
\end{align}

\subsection{The mean conservation equations.}
For each altitude $z \in [0, L]$, we choose a control volume defined as the cylinder of fixed radius $B > b(z)$ centered above the source $\mathbb{C} = \{(r,z) \in [0,B]\times[z,z+\delta z]\}$. Using Eqs. \eqref{eq:1Dplume_rhoi}, \eqref{eq:1Dplume_rhoj}, \eqref{eq:rhoS} and \eqref{eq:uS}, and the Gauss theorem, we find:
\begin{multline*}
 0 = \displaystyle\int_\mathbb{C}\left( \sum_{i\in\mathcal{I}} \Div (\rho_i \vec{u}) + \sum_{j\in\mathcal{J}} \Div (\rho_j \vec{u}) \right)=  \int_\mathbb{C} \Div (\rhom \vec{u}) = \\
 = \beta U \pi b^2 |_{z + \delta_z} - \beta U \pi b^2 |_z - \alpha u_\epsilon(B,z)2\pi B \delta z\,.
\end{multline*}
Now, dividing for $\delta z$, sending it to $0$ and then $B \to b(z)$, we get total mass flux conservation:
\begin{equation}\label{eq:Q1}
 \de_z (Q) \equiv \de_z (\beta U b^2) = 2 \alpha b \Ue\,.
\end{equation}
In the general case, the source eject solid phases that are not in the atmosphere and some gaseous phase that is not included in the ambient composition. Identifying such a phases, respectively, with the index $i \in [2; I]$ and $j \in \mathcal{J}=[I+1; I+J]$, and using again Eqs. \eqref{eq:1Dplume_rhoi}, \eqref{eq:1Dplume_rhoj}, \eqref{eq:rhoS} and \eqref{eq:uS}, we find that the following mass fluxes are conserved (we are neglecting particle aggregation and fallout):
\begin{align}\label{eq:Qj1}
& \de_z (Q_i) \equiv \de_z (Y_i \beta U b^2) = 0\,, \quad \forall i \in [2;I]\,,\\
&  \de_z (Q_j) \equiv \de_z (Y_j \beta U b^2) = 0\,, \quad \forall j \in \mathcal{J}\,,
\end{align}
while for the atmospheric phase $i=1=\alpha$:
\begin{equation}\label{eq:Qalpha1}
\de_z (Q_\alpha) \equiv \de_z (Y_\alpha\beta U b^2) = 2 \alpha b \Ue\,.
\end{equation}
Since the mass flow rate of the erupted gases and particles are conserved, it is useful to define their mass flow rate and mass fraction (respectively $Q_{\textup{e},\textup{s}}$ and $Y_{\textup{e},\textup{s}}$):
\begin{align}
& \Qe \equiv \sum_{i=2}^I Q_{i,0} = \sum_{i=2}^I Q_i = \sum_{i=2}^I Y_i \beta U b^2 = Q\,\sum_{i=2}^I Y_i \equiv Q\,\Ye\,,
\label{eq:Qe}\\
& \Qs \equiv \sum_\mathcal{J} Q_{j,0} = \sum_\mathcal{J} Q_j = \sum_\mathcal{J} Y_j \beta U b^2 = Q\,\sum_\mathcal{J} Y_j \equiv Q\,\Ys\,,
\label{eq:Qs}
\end{align}

Putting together Eqs.~\eqref{eq:Q1},~\eqref{eq:Qj1},~\eqref{eq:Qalpha1} and
\begin{equation}\label{eq:1DplumeSumY}
Y_\alpha + \sum_{i=2}^I Y_i + \sum_{j=I+1}^{I+J} Y_j = Y_\alpha + \Ye + \Ys =1\,,
\end{equation}
we obtain a relationship giving the mass flow rate $Q_\alpha(z)$ as a function of only vent conditions ($Q_i(0) \equiv Q_{i,0},\quad Q_j(0)\equiv Q_{j,0}$) and $Q(z)$:
\begin{equation}\label{eq:Qalpha}
Q_\alpha(z) = Q(z) - \left(\sum_{i=2}^I Q_i(z) + \sum_\mathcal{J} Q_j(z)\right) = Q(z) - \left(\Qe - \Qs\right)\,.
\end{equation}
By dividing Eq.~\eqref{eq:Qe},~\eqref{eq:Qs} and \eqref{eq:Qalpha} by $Q$ we obtain a relationship giving us all the mass fraction as a function of only vent conditions ($Q_{\textup{e},\textup{s}}$) and the total mass flow rate:
\begin{subequations}\label{eq:1DplumeYk}
\begin{align}
& \Ye(z) = \frac{\Qe}{Q(z)}\,,\\
& \Ys(z) = \frac{\Qs}{Q(z)}\,,\\
& Y_\alpha(z) = 1 - \frac{\Qe + \Qs}{Q(z)}\,.
\end{align}
\end{subequations}

Dealing with the momentum, the vertical component of Eq. \eqref{eq:1Dplume_mom} and Eqs. \eqref{eq:rhoS} \eqref{eq:uS} \eqref{eq:pS} yields:
\begin{multline}
\frac{1}{\delta z}\int_\mathbb{C} -\beta g = -\pi\beta g b^2 - \pi\alpha g (B^2 - b^2) = \frac{1}{\delta z} \int_\mathbb{C} \Div \big(\beta\, u_z \vec{u} + p\hat{z} \big) = \\
\frac{\pi}{\delta z}\big[(\beta U^2 b^2 + p B^2)_{z + \delta z} - (\beta U^2 b^2 + p B^2)_{z}\big] \xrightarrow{\delta z \to 0} \de_z (\pi\beta U^2 b^2) + \pi B^2 \de_z p\,.
\end{multline}
Again, we take the limit $B \to b(z)$, obtaining
\begin{equation}\label{eq:M1}
 \de_z (\beta U^2 b^2) = (\alpha - \beta)g b^2.
\end{equation}
Here we used $\de_z p = -\alpha g$, stated by Eq. \eqref{eq:1Dplume_mom} together with $p(r,z) = p(z)$ and $u \to 0$ when $r \gg b(z)$.

Turning to the energy balance \eqref{eq:1Dplume_en} and using the same techniques, we find:
\begin{equation}\label{eq:F1}
 \de_z \left[ b^2 \beta U\left(\frac{U^2}{2} + h_{\beta}\right) \right] = 2 \alpha b \Ue \left(\textcolor{blue}{\cancel{\frac{\Ue^2}{2}}} + h_\alpha \right) - g\beta U b^2\,,
\end{equation}
where $h_\beta \equiv C_\beta \Tbb$ and $h_\alpha = \Ca \Ta$. We neglect the term proportional to $\Ue^2$, to be compared to that proportional to $U^2$, because the entrainment velocity $\Ue$ is typically one order of magnitude smaller than $U$.

Eq. \eqref{eq:F1} could be written in different ways using \eqref{eq:Q1} and \eqref{eq:M1}:
\begin{equation}\label{eq:F2}
 \de_z \left( \beta U b^2\,\Cbb\Tbb \right) = \left(\Ca \Ta\right) \de_z(\beta U b^2) + \frac{U^2}{2} \de_z(\beta U b^2) - g\alpha U b^2\,,
\end{equation}
that is equivalent to Eq. (8) in \cite{woods1988}, or
\begin{equation}\label{eq:F3}
 \de_z \left( \beta U b^2\,(\Cbb\Tbb - \Ca\Ta) \right) = -\beta U b^2\,\de_z\left(\Ca \Ta\right) + \frac{U^2}{2} \de_z(\beta U b^2) - g\alpha U b^2\,,
\end{equation}
where the dependence on the buoyancy flux and ambient stratification is highlighted.

Finally, we have that Eqs. \eqref{eq:1Dplume_yk}--\eqref{eq:TS} are one mean solution of \eqref{eq:Eqs} if
\begin{equation}\label{eq:Woods}
 \begin{sistema}
  \de_z (\Qe) = 0\,,\\
   \de_z (\Qs) = 0\,,\\
  \de_z (\beta U b^2) = 2 \alpha b \Ue \vspace{2pt}\\
  \de_z (\beta U^2 b^2) = (\alpha - \beta)g b^2 \vspace{2pt}\\
  \de_z \left( \beta U b^2\,(\Cbb\Tbb - \Ca\Ta) \right) = -\beta U b^2\,\de_z\left(\Ca \Ta\right) + \frac{U^2}{2} \de_z(\beta U b^2) - g\alpha U b^2\,.
 \end{sistema}
\end{equation}
By noting again that $\Qe$ and $\Qs$ are conserved and that Eqs.~\eqref{eq:1DplumeYk} hold, here the unknowns are $\beta(z)$, $U(z)$, $b(z)$ and $\Tbb(z)$, provided the knowledge of the ambient density $\alpha$, the ambient temperature $\Ta$ and the dependence of $\Ue$ on the other unknowns (the entrainment model). We are still lacking in one condition. The equation of state of the various phases together with the full expanded plume hypothesis -- $p(r,z) = p(z)$ -- will give us that last needed condition.

\section{The Gas-Particle Plume model.}\label{sec:GasParticlePlume}
In order to close the latter system of equations, we can use solution \eqref{eq:1Dplume_yk}--\eqref{eq:TS} with the constitutive law for the dusty gas pressure. Since in Eq.~\eqref{eq:pS} we have assumed $p(r,z) = p(z)$, we have that -- at a given height -- the pressure inside the plume is the same of that outside the plume:
\begin{equation}
p = \beta R_\beta \Tbb = \alpha R_\alpha \Ta\,.
\end{equation}
Thus, we can rewrite the plume internal-external enthalpy differential as follows:
\begin{equation}\label{eq:buoyancy_1}
\beta(\Cbb\Tbb - \Ca\Ta) = \alpha \Ca\Ta \frac{R_\alpha \Cbb}{R_\beta \Ca} - \beta \Ca\Ta\,.
\end{equation}
We define the thermodynamic properties of the ejected gas and of the particles as follows
\begin{align}
& R_\textup{e} = \frac{1}{\Ye}\,\sum_{i=2}^I Y_i\, R_i\,,\\
& C_\textup{e} = \frac{1}{\Ye}\,\sum_{i=2}^I Y_i\, (C_i + R_i)\,,\\
& C_\textup{s} = \frac{1}{\Ys}\sum_\mathcal{J} Y_j\, C_j\,,
\end{align}
noticing that all these quantities are -- coherently -- conserved along $z$\footnote{It is sufficient to multiply both numerator and denominator of the right hand sides by $Q$, and notice that $Y_k Q = Q_k = Q_{k,0}$.}. In this way thermodynamic properties of the mixture can be written in terms of the thermodynamic properties of the three components, for example:
\begin{equation}
\Cbb = Y_\alpha \Ca + \Ye C_\textup{e} + \Ys \Cs\,.
\end{equation}
Using these definitions plus $\chis = \frac{\Cs}{\Ca}$, $\chie = \frac{C_\textup{e}}{\Ca}$, $\psie = \frac{R_\textup{e}}{R_\alpha}$, and Eqs.~\eqref{eq:1DplumeSumY}, \eqref{eq:Qe}, \eqref{eq:Qs}, we can write in a convenient form Eq.\eqref{eq:buoyancy_1}:
\begin{equation}\label{eq:enthalpy}
 \beta (\Cbb\Tbb - \Ca \Ta) = \Ca \Ta \left[(\alpha - \beta) + \alpha \frac{\chis \Qs + (\chie - \psie)\Qe}{(Q - \Qs) + (\psie - 1)\Qe}\right]\,.
\end{equation}
Now, defining the relative flux of enthalpy
\begin{equation}\label{eq:enthalpyFlux}
 F = \left[(\alpha - \beta) + \alpha \frac{\chis \Qs + (\chie - \psie)\Qe}{(Q - \Qs) + (\psie - 1)\Qe}\right]U b^2
\end{equation}
equation \eqref{eq:F3} can be rearranged
\begin{equation}
 F' = -(F + Q) \frac{\de_z (\Ca\Ta)}{\Ca\Ta} + \frac{U^2 Q'}{2\Ca \Ta} - \frac{\alpha g U b^2}{\Ca\Ta}\,.
\end{equation}
It is useful to define
\begin{align}
& Q_\psi = -\Qs + (\psie - 1)\Qe\,, \\
& Q_\chi = (\chis - 1) \Qs + (\chie - 1)\Qe\,,
\end{align}
which are constants along $z$, so that
\begin{equation}\label{eq:enthalpyFlux2}
 F = \left[(\alpha - \beta) + \alpha \frac{Q_\chi - Q_\psi}{Q + Q_\psi}\right]U b^2\,.
\end{equation}
This expression for $F$ represents a modification of the buoyancy flux for a dusty-gas plume in the general non-Boussinesq case (cf.~\cite{Cerminara2015ir}). It takes the classic form $(\alpha - \beta)U b^2$ (\cite{fannelop2003}, \cite{kaminski2005}) for a single-component gas plume (in such a case $Q_\chi = 0$ and $Q_\psi = 0$). For this reason we will refer to the relative flux of enthalpy $F$ as the {\em dusty gas buoyancy flux}, a generalization for the multiphase case of the standard buoyancy flux.

This new quantity $F$, together with the mass flux $Q = \beta U b^2$ and the momentum flux $M = \beta U^2 b^2$ allow us to close problem~\eqref{eq:Woods} in their terms:
\begin{subequations}
\begin{empheq}[box=\widefbox]{align}
& Q' = 2\Ue(\alpha, Q, M, F) \sqrt{\dfrac{\alpha Q (F + Q)(Q + Q_\psi)}{M [Q + Q_\chi]}} \vspace{8pt}\\
& M' = \dfrac{g F Q}{M}\left[ 1 - \dfrac{(F + Q)(Q_\chi - Q_\psi)}{F[Q + Q_\chi]} \right]
\vspace{8pt}\\
& F' = -(F + Q) \dfrac{(\Ca\Ta)'}{\Ca\Ta} + \dfrac{M^2 Q'}{2\Ca \Ta Q^2} -
\dfrac{g(F+Q)(Q+Q_\psi)}{\Ca\Ta(Q + Q_\chi)}\,,
\end{empheq}
\end{subequations}
where $U = \frac{M}{Q}$, $b= \sqrt{\frac{Q(F+Q)(Q+Q_\psi)}{\alpha M (Q + Q_\chi)}}$ and $\beta =
\alpha \frac{Q [Q + Q_\chi]}{(F + Q)(Q + Q_\psi)}$.

\section{Non-dimensionalization.}
It is useful to transform the latter problem in dimensionless form. We choose $Q(z) = Q_0 q(\zeta)$,
$M(z) = M_0 m(\zeta)$, $F(z) = F_0 f(\zeta)$ and $z = \ell_0 \zeta$ ($\ell_0 = \frac{Q_0}{\sqrt{\alpha_0 M_0}}$), where
$(\cdot)_0$ refers to the vent height. In this way, we have $q(0) = m(0) = f(0) = 1$. It is worth noting that $\zeta=0$ can correspond to the actual vent elevation as to any height above the vent (cf. \cite{Cerminara2015ir}). 
The model in non-dimensional form then is
\begin{subequations}\label{eq:fullModel}
 \begin{align}
 & q' = v_q \etaX \sqrt{a(\zeta)\,\dfrac{m(\phi f + q) (q + q_\psi)}{q(q + q_\chi)}}\\
 & m' = v_m \dfrac{q}{m}\left( f - \gammac\dfrac{(\phi f + q)}{(q +
q_\chi)}\right)\\
 & f' = \frac{v_f}{t_\alpha(\zeta)}\,\left[(\phi f + q)\left(\theta_f(\zeta) - \frac{q+q_\psi}{q+q_\chi }\right) + \frac{\phi}{2 v_m} \frac{m^2 q'}{q^2}\right]\,,
 \label{eq:fullModel_f}
 \end{align}
\end{subequations}
where $\etaX$ -- defined in Eq. \eqref{eq:Ue} -- is the entrainment function, potentially depending on the other variables and parameters; $a(\zeta) \equiv \alpha(\ell_0\zeta)/\alpha_0$, $t_\alpha(\zeta) = T_\alpha(\ell_0\zeta)/T_{\alpha,0}$, $\phi \equiv
F_0/Q_0$, $q_\psi \equiv Q_\psi/Q_0$, $q_\chi \equiv Q_\chi/Q_0$, $\gammac \equiv \frac{Q_\chi - Q_\psi}{F_0}$, $\theta_f(\zeta) \equiv
-\frac{1}{v_f\phi}t_\alpha'(\zeta)$ and
\begin{align}
 &v_q = 2\varkappa 
 \label{eq:1Dplume_vq}\\
 & v_m = \frac{g F_0 Q_0 \ell_0}{M_0^2} = \frac{\phi g \ell_0}{U_0^2} = \Ri
 \label{eq:1Dplume_vm}\\
 & v_f = \frac{g Q_0 \ell_0}{F_0\, \Ca T_{\alpha,0}}=\frac{g \ell_0}{\phi \Ca T_{\alpha,0}} = \frac{g \ell_0}{C_{\beta,0} T_{\beta,0} - \Ca T_{\alpha,0}} = \frac{g \ell_0}{\Delta h_0}  = \frac{\Ec}{\Fr^2}\,.
 \label{eq:1Dplume_vf}
\end{align}
We call these last three parameters the rate of variation respectively of $q,\,m,\,f$. In Eq.~\eqref{eq:1Dplume_vm}, we have given a modified definition of the Richardson number $\Ri = \phi g \ell_0/U_0^2$, because $\phi g = g'$ in the monophase case ($g'$ being the reduced gravity). In Eq.~\eqref{eq:1Dplume_vf} we used the definition of the Froude number $\Fr = U_0^2/g \ell_0$ and of the Eckert number $U_0^2/\Delta h_0$, where $\Delta h_0 = C_{\beta,0} T_{\beta,0} - \Ca T_{\alpha,0}$ is the enthalpy anomaly at the vent. Moreover, we have used Eqs.~\eqref{eq:enthalpy}, \eqref{eq:enthalpyFlux} implying \mbox{$\phi \Ca T_{\alpha,0} = C_{\beta,0} T_{\beta,0} - \Ca T_{\alpha,0}\,.$} It is also useful to rewrite the physical variables as a function of these new parameters:
\begin{subequations}\label{eq:fullModel_physicalParameters}
\begin{align}
 & U = \frac{M_0}{Q_0} \frac{m}{q}\\
 & b = \ell_0 \sqrt{\frac{q(\phi f + q)(q+q_\psi)}{a\, m (q+q_\chi)}}\\
 & \beta = \alpha\,\frac{q(q+q_\chi)}{(\phi f + q)(q+q_\psi)}\\
 & T_\beta = T_{\alpha} \frac{\phi f + q}{q+q_\chi}\\
 & Y_{\textup{e}\,(\textup{s})} = \frac{Y_{\textup{e},0\,(\textup{s},0)}}{q} \,.
\end{align}
\end{subequations}

It is worth noting that $q_\chi,\,q_\psi > -1$ because the specific heats and gas constants are positive ($\chi_\cdot,\, \psi_\cdot > 0$) and the sum of the initial mass fraction is smaller than 1 (cf. definition of $q_\chi,\,q_\psi$ in Tab. \ref{tab:plume_parameters}). Moreover, $\phi > -1$ because $C_{\beta,0} T_{\beta,0} > 0$. Even if these are the general conditions for such parameters, in Tab.~\ref{tab:plume_parameters} there are summarized the possible ranges for volcanic eruptions.

Using $\de_z p = -\alpha g$ and the ideal gas law it is possible to obtain the density stratification as a function of the temperature:
\begin{equation}\label{eq:stratifiedDensity}
a(\zeta) = t_\alpha^{-1}(\zeta)\,\exp\left(-\frac{g \ell_0}{R_\alpha T_{\alpha,0}}\int_0^\zeta t_\alpha^{-1}(\zeta')\, \de \zeta'\right)\,.
\end{equation}
For example, if the non-dimensional atmospheric thermal gradient $\theta = \theta_\alpha \ell_0/T_{\alpha,0}$ is constant, we have $t_\alpha(\zeta) = 1 - \theta \zeta$ and:
\begin{equation}
a(\zeta) = (1 - \theta \zeta)^{\frac{g}{R_\alpha \theta_\alpha} - 1}\,,
\end{equation}
and $\theta_f(\zeta) = \theta_f = \theta/v_f\phi$.

It is also useful to define the Brunt-V\"ais\"all\"a frequency $\BV$. Recalling that the potential temperature is
\begin{equation}
t_{\textup{p},\alpha}(\zeta) = t_\alpha(\zeta)\, (a(\zeta) t_\alpha(\zeta))^{-\frac{R_\alpha}{\Ca}}\,,
\end{equation}
we obtain
\begin{equation}\label{eq:brunt-vaisalla}
\BV^2 = \frac{g}{b_0}\,\ln(t_{\textup{p},\alpha})'(\zeta) = \frac{g^2}{\Ca T_{\alpha,0}}\frac{1 - \theta_{f}(\zeta)}{t_\alpha(\zeta)} \,.
\end{equation}
This frequency depends on the height $z$, but it can be approximately be considered as a constant because it vary slowly in our atmosphere: $\approx 10$ \% of variation in the troposphere. In what follows we call $\BV_0$ its constant approximation. Using standard average conditions for the troposphere, we find \mbox{$\BV_0\simeq 1.13*10^{-2}$ Hz}.
Studying plumes in a stratified atmosphere (cf. Sec.~\ref{sec:plumeStratified}), it is useful to define
\begin{equation}\label{eq:1Dplume_vf0}
v_f \frac{1 - \theta_{f}}{t_\alpha} = \frac{\ell_0\BV^2}{\phi\, g} \simeq \frac{\ell_0\BV_0^2}{\phi\, g} = \frac{1}{\phi \strat} \equiv v_{f,0} \,,
\end{equation}
showing that the new parameter $v_{f,0}$ can be recovered by knowing the enthalpy anomaly $\phi$ and the non-dimensional stratification length scale $\strat \equiv g/\BV_0^2 \ell_0$. In other words, the more $v_{f,0}$ increases the more the vent dimensions corrected with the enthalpy anomaly are comparable with the stratification length scale.

\begin{table}[t]
\centering
\begin{tabular}{cccr}
\toprule
parameter & explicit form & range of variability & description\\
\midrule
$\phi$ & $\dfrac{\Cbb T_{\beta,0} - \Ca T_{\alpha,0}}{\Ca T_{\alpha,0}}$ & $0.3\div 5$ & 
\begin{tabular}[l]{@{}c@{}}enthalpy anomaly\\ (non-Boussinesqness)\end{tabular}\vspace{5pt}\\
$q_\psi$ & $-\Yso + (\psie-1)\Yeo$ & $-1\div 1$ & \begin{tabular}[l]{@{}c@{}}mass flux anomaly\\ due to gas constants\end{tabular}\vspace{5pt}\\
$q_\chi$ & $(\chis - 1) \Yso + (\chie - 1)\Yeo$ & $-1\div 1$ & \begin{tabular}[l]{@{}c@{}}mass flux anomaly\\ due to specific heats\end{tabular}\vspace{5pt}\\
$v_q/2$ & $\varkappa$ & $0.05\div 0.3$ & \begin{tabular}[l]{@{}c@{}}entrainment\\ coefficient\end{tabular}\vspace{5pt}\\
$v_m$ & $\dfrac{\phi\,g \ell_0}{U_0^2}$ & $10^{-4}\div 10$ & \begin{tabular}[l]{@{}c@{}}modified\\ Richardson number\end{tabular}\vspace{5pt}\\
$\strat$ & $\dfrac{g}{\BV_0^2 \ell_0}$ & $10^2\div 10^5$ & \begin{tabular}[l]{@{}c@{}}stratification\\ length-scale\end{tabular}\vspace{5pt}\\
\bottomrule
\end{tabular}
\caption{Independent parameters for a multiphase plume in a stratified atmosphere.}\label{tab:plume_parameters}
\end{table}

All these non-dimensional parameters characterize the multiphase plume and give us the possibility to classify through them all the possible regimes. We summarize in Tab.~\ref{tab:plume_parameters} six of them, which are {\bf the independent non-dimensional parameters sufficient to characterize a multiphase plume}. In order to fix ideas, we show there the range of variability of those independent parameters for Strombolian to Plinian volcanic eruptions.

Indeed, the knowledge of these parameters and of the thermodynamic properties of the atmosphere allows us to retrieve the physical dimensional parameters. We report here all the  inversion relationships needed:
\begin{subequations}
\begin{align}
& \ell_0 = \frac{g}{\BV_0^2\ddot{\ell}} \qquad\mbox{see footnote\footRef{note1}}\\
& b_0 = \ell_0 \sqrt{\frac{(1 + \phi)(1+q_\psi)}{1+q_\chi}}\label{eq:plumeInvertRadius}\\
& \beta_0 = \alpha_0\,\frac{1+q_\chi}{(1 + \phi)(1+q_\psi)}\\
& U_0 = \sqrt{\frac{g\phi\ell_0}{v_m}}\\
& T_{\beta,0} = T_{\alpha,0}\,\frac{1 + \phi}{1+q_\chi}\\
& Q_0 = \beta_0 U_0 b_0^2\\
& M_0 = \beta_0 U_0^2 b_0^2\\
& F_0 = \phi Q_0\\
& \gammac = \frac{q_\chi - q_\psi}{\phi}\\
& v_f = \frac{g \ell_0}{\phi\,\Ca T_{\alpha,0}}\\
& v_{f,0} = \frac{1}{\phi\strat}\\
& \Yeo = \frac{q_\chi + (\chis - 1)q_\psi}{(\chie - 1) + (\chis - 1)(\psie - 1)}\qquad \mbox{see footnote\footRef{note2}}\\
& \Yso = (\psie - 1)\Yeo - q_\psi\\
& Y_{\alpha,0} = 1 - \Yso - \Yeo\,.
\end{align}
\end{subequations}
\footLabel{note1}{When stratification is disregarded, no reference length scales are present in the non-dimensional system, thus $b_0$ must be given and $\ell_0$ can be recovered from Eq.~\eqref{eq:plumeInvertRadius}.}
\footLabel{note2}{In order to have the mass fraction of ejected gas and solids, their thermodynamic properties must be known: namely their specific heat and the gas constant of the ejected gas.}
In \citet{Cerminara2015ir}, we have used these inversion relationships to obtain the vent condition of a real volcanic eruption occurred at Santiaguito (Santa Maria Volcano, Guatemala).

In this thesis, we will study only two of all the possible entrainment models introduced in the literature:
\begin{itemize}
\item \citet{morton1956}, where $\etaX = 1$
\item \citet{ricou1961}, where $\etaX = \etaX (\beta/\alpha) = \displaystyle\left(\dfrac{q(q+q_\chi)}{(q+\phi f)(q+q_\psi)}\right)^{\frac{1}{2}}$
\end{itemize}
More complex models have been studied in volcanology and fluid dynamics. One example can be found in \cite{Carazzo2008} where $\etaX$ depends on the local Richardson number.

It is worth noting that the mass flux $q(\zeta)$ is a strictly increasing function as long as $\etaK$ is positive, while the sign of $m'(\zeta)$ depends on the buoyancy sign:
\begin{equation}
\mathrm{sign}(\textup{buoyancy}) = \mathrm{sign}\left( f - \gammac\dfrac{(\phi f + q)}{(q +
q_\chi)}\right)\,,
\end{equation}
because $v_m,\,q,\,m$ are strictly positive. For an analysis on the plume buoyancy behavior see Sec.~\ref{sec:buoyancy}. In Sec.~\ref{sec:plumeStratified} we will study in detail the evolution of the plume variables under the Boussinesq approximation. However, something can be noted even at this point of the analysis by looking at the full system~\eqref{eq:fullModel}: 1) the mass flow $q(z)$ is a strictly increasing function because the entrainment models we are using are positive functions; 2) the momentum flux $m(z)$ has derivative equal to zero when the buoyancy become zero. It can be due to two causes, buoyancy reversal or neutral buoyancy level. We denote $\zeta_\textup{nbl}$ the neutral buoyancy level; 3) when $m(z) = 0$ system~\eqref{eq:fullModel} encounters a singularity. In that point the plume reaches its maximum height $\zeta_\textup{max}$; 3) the enthalpy flux is a strictly decreasing function, because usually in applications the term containing $\left(\theta_f(\zeta) - \frac{q+q_\psi}{q+q_\chi }\right)$ is dominant and negative.

In the next sections we discuss some of the approximations applicable to problem \eqref{eq:fullModel}. In particular we find that $\gammac$ is the parameter related to the column instability -- if $\gammac > 1$ then the volcanic column will collapse -- and that $\phi$ is the parameter measuring the non-Boussinesqness of the mixture -- if $\phi \ll 1$ then the Boussinesq approximation holds. Moreover, $q_\psi$ and $q_\chi$ are the parameters measuring the multiphaseness of the mixture -- if $|q_\psi| \simeq |q_\chi|  \ll 1 $ the plume can be considered as a single phase one.

In this thesis we will study three different volcanic eruption and one experimental plume that we denote, from the weaker to the stronger: \textsf{[forcedPlume]}, \textsf{[Santiaguito], [weakPlume], [strongPlume]}. We report in Tab.~\ref{tab:volcanic_plumes} all the parameters for these volcanic eruptions, respectively: 1) the physical parameters at the vent -- radius, density, temperature, velocity and mass fractions; 2) the mass, momentum and enthalpy flows; the non-dimensionalization length scale and the multiphase Morton length scale (see below); 3) the six independent non-dimensional parameters; 4) the non-dimensional dependent parameters; 5) the non-dimensional plume maximum and neutral buoyancy level height, as obtained from system~\eqref{eq:fullModel} with~\citet{ricou1961} entrainment model
\footnote{While for \textsf{[forcedPlume]}, \textsf{[Santiaguito], [weakPlume]} we have used a constant atmospheric thermal gradient, for \textsf{[strongPlume]} the atmospheric temperature profile is a little bit more complex, because we have included in it the presence of the tropopause~\citep[cf.][]{costa_etal_2015}.}.
\begin{table}[ph!]
\centering
\begin{tabular}{ccccc}
\toprule
parameter & \textsf{[forcedPlume]} & \textsf{[Santiaguito]} & \textsf{[weakPlume]} & \textsf{[strongPlume]}\\
\midrule
$b_0$ [m] & 0.03175 & 22.9 & 26.9 & 703 \vspace{5pt}\\
$\beta_0$ [kg/m$^3$] & 0.622 & 1.05 & 4.87 & 3.51 \vspace{5pt}\\
$\alpha_0$ [kg/m$^3$] & 1.177 & 0.972 & 1.100 & 1.011 \vspace{5pt}\\
$T_{\beta,0}$ [K] & 568 & 375 & 1273 & 1053 \vspace{5pt}\\
$T_{\alpha,0}$ [K] & 300 & 288 & 270.92 & 294.66 \vspace{5pt}\\
$U_0$ [m/s] & 0.881 & 7.29 & 135 & 275 \vspace{5pt}\\
$R_\alpha$ [m$^2$/s$^2$K] & 287 & 287 & 287 & 287 \vspace{5pt}\\
$C_\alpha$ [m$^2$/s$^2$K] & 1004.5 & 998 & 1004 & 1004 \vspace{5pt}\\
$\psie$ & -- & 1.61 & 1.61 & 1.61 \vspace{5pt}\\
$\chie$ & -- & 1.866 & 1.803 & 1.803 \vspace{5pt}\\
$\chis$ & -- & 1.102 & 1.096 & 1.096 \vspace{5pt}\\
$\Yeo$ & 0 &  0.196 & 0.03 & 0.05 \vspace{5pt}\\
$\Yso$ & 0 & 0.410 & 0.97 & 0.95 \vspace{5pt}\\
$Y_{\alpha,0}$ & 1 & 0.394 & 0 & 0 \vspace{5pt}\\
$\BV_0$ [Hz]& $1.14*10^{-2}$ & $1.43*10^{-2}$ & $1.40*10^{-2}$ & $2.14*10^{-2}$ \vspace{5pt}\\
\midrule
$\pi Q_0$ [kg/s] & $1.74*10^{-3}$ & $1.26*10^4$ & $1.5*10^6$ & $1.5*10^9$ \vspace{5pt}\\
$\pi M_0$ [kg m/s$^2$] & $1.53*10^{-3}$ & $9.19*10^4$ & $2.02*10^8$ & $4.12*10^{11}$ \vspace{5pt}\\
$\pi F_0$ [kg/s] & $1.55*10^{-3}$ & $7.28*10^3$ & $6.35*10^6$ & $4.56*10^9$\vspace{5pt}\\
$\ell_0$ [m] & 0.02308 & 23.8 & 56.6 & 1310\vspace{5pt}\\
$L_\textup{M}$ [m] & 0.0854 & 18.4 & 352 & 4070\vspace{5pt}\\
\midrule
$\phi$ & 0.893 & 0.58 & 4.25 & 3.04 \vspace{5pt}\\
$q_\psi$ & 0 & -0.290 & -0.952 & -0.920 \vspace{5pt}\\
$q_\chi$ & 0 & 0.212 & 0.117 & 0.131 \vspace{5pt}\\
$v_q$ & 0.28 & 0.659 & 0.2 & 0.2 \vspace{5pt}\\
$v_m$ & 0.261 & 2.54 & 0.129 & 0.517 \vspace{5pt}\\
$\strat$ & $3.29*10^{6}$ & 2020 & 886 & 16.4 \vspace{5pt}\\
\midrule
$\gammac$ & 0 & 0.869 & 0.252 & 0.345 \vspace{5pt}\\
$v_f$ & $8.41*10^{-7}$ & $1.41*10^{-3}$ & $4.81*10^{-4}$ & $1.43*10^{-2}$ \vspace{5pt}\\
$v_{f,0}$ & $3.40*10^{-7}$  & $8.58*10^{-4}$ & $2.66*10^{-4}$ & $2.01*10^{-2}$ \vspace{5pt}\\
\midrule
$\zeta_\textup{max}$ & 1665 & 23.98 & 160.6 & 20.68 \vspace{5pt}\\
$\zeta_\textup{max}/\zeta_\textup{nbl}$ & 1.318 & 1.306 & 1.354 & 1.523 \vspace{5pt}\\
\bottomrule
\end{tabular}
\caption{Relevant parameters of the plumes studied in this thesis.}\label{tab:volcanic_plumes}
\end{table}

\subsection{Monophase plume.}
If the thermodynamic properties of the ejected fluid are similar to them of the ambient fluid then $|q_\psi| \simeq |q_\chi|  \ll 1$. In this case, model \eqref{eq:fullModel} becomes:
\begin{subequations}\label{eq:1Dplume_monophase}
\begin{align}
& q' = v_q \etaX \sqrt{a(z)\frac{m(\phi f + q)}{q}}\\
& m' = v_m \frac{q f}{m}\\
& f' = \frac{v_f}{t_\alpha(z)}\left[(\phi f + q)(\theta_f(z) - 1) + \frac{\phi}{2 v_m}\frac{m^2 q'}{q^2}\right]\,,
\label{eq:1Dplume_f_monophase}
\end{align}
\end{subequations}
where
\begin{align*}
& \etaX = 1 & &\mbox{(\cite{morton1956})}\\
& \etaX = \sqrt{\frac{q}{\phi f+q}} & &\mbox{(\cite{ricou1961})}\,.
\end{align*}

It is worth noting that in the single phase case $\Cbb = \Ca$ and $R_\beta = R_\alpha$. Thus, the initial enthalpy anomaly reduces to the initial thermal anomaly or equivalently to the density anomaly:
\begin{equation}
\phi = \frac{T_{\beta,0} - T_{\alpha,0}}{T_{\alpha,0}} \equiv \frac{\Delta T_0}{T_{\alpha,0}} = \frac{\alpha_0 -\beta_0}{\beta_0}\,.
\end{equation}
Consequently the reduced gravity becomes $g' = \phi g$.

\subsection{Jet regime}
In the jet regime -- defined as the one where $m = f = 1$ --~\citet{woods1988} pointed out that the~\citet{ricou1961} model can be used. In this case, Eqs.~\eqref{eq:fullModel} simplify a lot,
becoming:
\begin{align}\label{eq:fullModel_jet}
 q' & = v_q & m' & = 0 & f' = 0\,,
\end{align}
with the easy solution $q(\zeta) = v_q\,\zeta + 1$.

Substituting this solution in Eqs.~\eqref{eq:fullModel} and proceeding with the dimensional
analysis, it is possible to find $\ell_M$, the dimensionless transition length scale between the jet
and the plume regime. It is the length scale for which the momentum variation becomes important.
From the momentum equation we find:
\begin{align}\label{eq:mortonL}
 \frac{1}{\ellM} & \simeq v_m (v_q \ellM + 1) \simeq v_m v_q \ellM \quad \Rightarrow \quad \ellM = (v_q v_m)^{-\frac{1}{2}}\,,
\end{align}
from which, back to dimensional units:
\begin{equation}
 L_M = \left( \frac{U_0^2 \ell_0}{2\varkappa \phi g}\right)^\frac{1}{2}\,. \label{eq:1Dplume_LM_monophase}
\end{equation}
This quantity became equivalent to that defined in \citet{morton1959} when
$q_\psi = q_\chi = 0$ and $\beta \simeq \alpha$.

The typical length scale of stratification $\ell_\textup{S}$ for a jet can be found by using a similar dimensional analysis for Eq.~\eqref{eq:fullModel_f}
\begin{equation}
\frac{1}{\ell_\textup{S}} = v_{f,0}(\phi + 1 + v_q\ell_\textup{S}) \simeq v_{f,0}v_q \ell_\textup{S} \quad \Rightarrow \quad \ell_S = \left(v_q v_{f,0}\right)^{-\frac{1}{2}}\,,
\end{equation}
or
\begin{equation}\label{eq:1Dplume_jet}
\frac{\ell_\textup{S}}{\ellM} = \left(\frac{v_m}{v_{f,0}}\right)^{\frac{1}{2}} = \frac{\phi g}{U_0 \BV} \equiv \jet\,.
\end{equation}
This parameter is comparing the rate of variation of $m$ and $f$. We have that if $\jet < 1$ than stratification have a role in the jet-like part of the plume, on the contrary, if $\jet > 1$ stratification is important just in the plume-like part of the plume. We will comment better this length scale in the section below dedicated to the plume height.

Usually in jets, atmospheric stratification is not important because of their limited height ($\jet > 1$). We want to explore now when the kinetic correction term could be important. Contrarily to the last two terms, the second term in square brackets in Eq.~\eqref{eq:fullModel_f} become less important as $\zeta$ grows. In particular it decreases with $q'/q^2\propto \zeta^{-2}$. Defining the typical length scale for this term $\ell_K$, we have:
\begin{equation}
 \frac{1}{\ell_K} \simeq \frac{\phi v_f v_q}{2 v_m(1 + v_q \ell_K)^2} \quad \Rightarrow \quad \ell_K = \frac{1}{v_q}\left[\left(\frac{\phi v_f}{4 v_m}-1\right)\pm \sqrt{\left(\frac{\phi v_f}{4 v_m}-1\right)^2-1}\right]\,,
\end{equation}
admitting a positive solution if and only if
\begin{equation}
\frac{\phi v_f}{v_m} = \frac{U_0^2}{\Delta h_0} > 8 \quad \Rightarrow \quad \ell_K \simeq
\begin{cases}
\dfrac{4 v_m}{v_q \phi v_f} = \dfrac{2 \Delta h_0}{\varkappa U_0^2}\lesssim 1
\vspace{5pt}\\
\dfrac{\phi v_f}{2 v_q v_m} = \dfrac{U_0^2}{4\varkappa \Delta h_0}\gg 1\,.
\end{cases}
\end{equation}
 Thus, the kinetic correction can be important just near the vent or very far from it and  only when $\Delta h_0 \ll U_0^2$ ($\Ec \gg 1$). In other words, this correction can be important for ``cold and fast'' jets and far from the jet center. Generally, in volcanic plumes the $\Ec$ number is small, thus the kinetic correction can be disregarded.

\subsection{Non stratified plume regime}
If stratification and the last term in square brackets of Eq.~\eqref{eq:fullModel_f} can be disregarded, $f = 1$ and model~\eqref{eq:fullModel} becomes
\begin{subequations}\label{eq:fullModel_nonStratified}
 \begin{align}
 & q' = v_q \etaX \sqrt{\dfrac{m(\phi + q) (q + q_\psi)}{q(q + q_\chi)}}
 \label{eq:fullModel_nonStratified_q}\\
 & m' = v_m \dfrac{q}{m}\left( 1 - \gammac\dfrac{(\phi + q)}{(q +
q_\chi)}\right)\\
 & f' = 0\,.
 \end{align}
\end{subequations}
This ordinary differential equation has a first integral of motion\footnote{A first integral of motion is a quantity remaining constant along the motion described by the differential equation. It is also called {\em constant of motion}.} $\mathcal{U}$ in both the considered cases for $\etaX$. We found respectively for the entrainment models of \citet{morton1956} and \citet{ricou1961}:
\begin{subequations}
\begin{align}
& \mathcal{U}_\textup{MTT} = 2\displaystyle\int \left( 1 - \gammac\dfrac{(\phi + q)}{(q +
q_\chi)}\right)\sqrt{\dfrac{q(q + q_\chi)}{(\phi + q) (q + q_\psi)}}\,\de q - \frac{4v_q}{5v_m}m^{5/2}\\
& \mathcal{U}_\textup{RS} = q^{2}(1-\gammac) -2\gammac(\phi-q_\chi)\left[q - q_\chi\ln(|q+q_\chi|)\right] - \frac{4v_q}{5v_m}m^{5/2}\,.
\label{eq:fullModel_nonStratified_U}
\end{align}
\end{subequations}
Using this first integral of motion in Eq.~\eqref{eq:fullModel_nonStratified_q} it is possible to find an implicit solution for the height of the form $\zeta = \zeta(q)$. For the Ricou entrainment model, defining
\begin{equation}\label{eq:firstIntegral_l}
l(q) = q^{2}(1-\gammac) -2\gammac(\phi-q_\chi)\left[q - q_\chi\ln(|q+q_\chi|)\right]\,,
\end{equation}
 and substituting the corresponding first integral of motion found in Eq.~\eqref{eq:fullModel_nonStratified_U}
\begin{equation}\label{eq:firstIntegral_nonStratified_lq}
\mathcal{U}_\textup{RS}(q,m) =  l(q) - \frac{4v_q}{5v_m}m^{5/2} = \mathcal{U}_\textup{RS}(1,1) = l(1) - \frac{4v_q}{5v_m}\,,
\end{equation}
into Eq.~\eqref{eq:fullModel_nonStratified_q}, we found the following implicit solution:
\begin{equation}\label{eq:zetaNonStratified}
\zeta = \zeta(q) = \int_1^q \de x\, \frac{1}{v_q}\left[\frac{5v_m}{4v_q}\left(l(x) - \mathcal{U}_\textup{RS}\right)\right]^{-\frac{1}{5}}\,.
\end{equation}
Using this solution it is possible to find the height at which the Boussinesq approximation starts to hold: $\zeta = \zeta_\textup{Bou}$. We choose the value $q = q_\textup{Bou} = 10\, \textup{max}(|\phi |,\,|q_\chi |,\, |q_\psi |)$. In Tab.~\ref{tab:plume_stability} are reported the value we obtain for the examples considered in this thesis. By comparing those values with $\zeta_\textup{max}$ reported in Tab.~\ref{tab:volcanic_plumes} it is possible to have an idea of the part of the plume where the Boussinesq regime holds.

Under the same hypothesis of this section, the monophase case~\eqref{eq:1Dplume_monophase} becomes equivalent to the model studied in~\citet{fannelop2003}:
\begin{align}
& q' = v_q \etaX \sqrt{\frac{m(\phi + q)}{q}}\\
& m' = v_m \frac{q}{m}\\
& f' = 0\,.
\end{align}
For the entrainment models of \citet{morton1956} and \citet{ricou1961} the first integral of motion are respectively:
\begin{align}
& \mathcal{U}_\textup{MTT} = \left(q-\frac{3}{2}\phi\right)\sqrt{q(q+\phi)} + \frac{3}{2}\phi^2\ln\left(\sqrt{q} + \sqrt{q + \phi}\right) - \frac{4v_q}{5v_m}m^{5/2}\\
& \mathcal{U}_\textup{RS} = q^{2} - \frac{4v_q}{5v_m}m^{5/2}\,.
\end{align}

\subsection{Buoyancy reversal and plume stability}\label{sec:buoyancy}
In this section, we consider the plume model behavior near the vent, where it is not possible to use the approximation $q \gg |\phi |,\,|q_\chi |,\,|q_\psi |$ (see next section) but $f \simeq 1$ as done in the previous section. Here we will use the Richou entrainment model because we are near the vent, however the present analysis is independent from the entrainment model used since the sign of the buoyancy does not depend on $\etaK$. In model \eqref{eq:fullModel_nonStratified}, the sign of the buoyancy force is determined by:
\begin{equation}
\mathrm{sign}(\textup{buoyancy}) = \mathrm{sign}\left(1 - \gammac\dfrac{(\phi + q)}{(q +
q_\chi)}\right) = \mathrm{sign}(l'(q))\,.
\end{equation}
Here, $l(q)$ is the first integral function defined in Eq.~\eqref{eq:firstIntegral_l}. When $l'(q) < 0$, the plume is negatively buoyant and $m$ decreases. If we arrive to the condition $l(q) = \mathcal{U}_\textup{RS}$ then $m\to 0$ because the first integral $\mathcal{U}_\textup{RS}$ must be constant. Thus the plume stops (or collapses) and it is not able to reverse its buoyancy. 

We can better understand the behavior of the non-stratified multiphase plume by  analyzing all the possible configurations. For this purpose, it is useful to define
\begin{align}
\gamma^{*} & \equiv \frac{1+q_\chi}{1+\phi} = \frac{T_{0,\alpha}}{T_{0,\beta}}\,, & q_\textup{min} = \frac{\gammac\phi - q_\chi}{1-\gammac}\,,
\end{align}
where $l'(q_\textup{min}) = 0$.
We enumerate the following situations for $q \geq 1$ (recall that $q(\zeta) \geq 1$ because it is a strictly increasing function and $q_\chi,\,\phi > -1$) by denoting ``C'' the cases when the plume collapses and by ``B'' the cases when the plume can reach and sustain the condition of positive buoyancy:
\begin{enumerate}
\item[1B)] {\em positive buoyant}. If $\gammac \leq 1 \, \wedge\, \gammac < \gamma^*$

then $l'(q) > 0 \quad\forall \, q\geq 1$ and the plume rises indefinitely.

\item[2B)] {\em zero, then immediately positive buoyant}. If $\gammac = \gamma^* < 1\quad (\Rightarrow\,\phi > q_\chi)$

then $l'(q) > 0 \quad \forall \, q > 1,\quad l'(q) = 0\quad \mbox{if}\quad q=1$

\item[3B)] {\em jet with zero buoyancy}. If $\gammac = \gamma^* = 1 \quad (\Rightarrow\,\phi = q_\chi)$

then $l'(q) = 0$ and the plume behaves as a jet.

\item[4BC)] {\em from negative to positive buoyancy}. If $\gamma^* < \gammac < 1 \quad  (\Rightarrow\,\phi > q_\chi)$

then $l'(q) < 0\quad \mbox{when} \quad q < q_\textup{min}$, the minimum of $l(q)$ is reached in $q = q_\textup{min}$ and $l'(q) > 0 \quad \mbox{when} \quad q > q_\textup{min}$. In this case inversion of the buoyancy sign can be possible if the minimum value of $l(q)$ is above the first integral: $l(q_\textup{min}) - \mathcal{U}_\textup{RS} >0$. In the opposite situation $l(q_\textup{min}) - \mathcal{U}_\textup{RS} < 0$ the plume is not able to invert its buoyancy and it collapses when $m = 0$, thus when $l(q) = \mathcal{U}_\textup{RS}$.

\item[5C)] {\em from positive to negative buoyancy}. If $1 < \gammac < \gamma^* \quad (\Rightarrow \phi < q_\chi) $

then $l'(q) > 0 \quad \mbox{when} \quad q < q_\textup{min}$, the maximum of $l(q)$ is reached in $q = q_\textup{min}$ and $l'(q) < 0 \quad \mbox{when} \quad q > q_\textup{min}$. In this case the plume always collapses going from positive to negative buoyancy.

\item[6C)] {\em zero, then immediately negative buoyant}. If $\gammac = \gamma^* > 1 \quad (\Rightarrow\, \phi < q_\chi)$

then $l'(q) < 0 \quad \forall \, q > 1,\quad l'(q) = 0\quad \mbox{if}\quad q=1$

\item[7C)] {\em negative buoyant}. If $\gammac \geq 1 \, \wedge \, \gammac > \gamma^*$

then $l'(q) < 0 \quad\forall \, q\geq 1$ and the plume collapses being always negative buoyant.

\end{enumerate}
Thus, we can summarize that: 1) if $\gammac > 1$ the plume starts or becomes negative buoyant and collapses; 2) $\gamma^*$ must be compared with $\gammac$ to know the initial buoyancy of the plume: if $\gammac < \gamma^* (>)$ then the plume is initially positive (negative) buoyant; 3) if $\gammac < 1$ then the plume is or can become positive buoyant, buoyancy reversal occurs if $l(q_\textup{min}) - \mathcal{U}_\textup{RS} > 0$. In Tab.~\ref{tab:plume_stability} we report all of these parameters for the plumes studied in this thesis. While \textsf{[forcedPlume]} is positive buoyant, the other three plumes are initially negative buoyant. For all of them, buoyancy reversal occurs.
\begin{table}[t]
\centering
\begin{tabular}{ccccc}
\toprule
parameter & \textsf{[forcedPlume]} & \textsf{[Santiaguito]} & \textsf{[weakPlume]} & \textsf{[strongPlume]}\\ 
\midrule
$\zeta_\textup{Bou}$ & 15.77 & 7.07 & 82.2 & 53.9 \vspace{5pt}\\
$\gammac$ & 0 & 0.869 & 0.252 & 0.345 \vspace{5pt}\\
$\gamma^*$ & 0.528 & 0.768 & 0.213 & 0.280 \vspace{5pt}\\
$q_\textup{min}$ & -- & 2.22 & 1.27 & 1.40 \vspace{5pt}\\
$l(q_\textup{min})-\mathcal{U}_\textup{RS}$ & -- & 0.0388 & 1.19 & 0.214 \vspace{5pt}\\
$a_q$ & 0.860 & 1.59 & 1.65 & 0.473 \vspace{5pt}\\
\bottomrule
\end{tabular}
\caption{Column stability parameters for the plumes studied in this thesis.}
\label{tab:plume_stability}
\end{table}

\subsection{Non stratified Boussinesq regime}
\begin{figure}
\centering
\includegraphics[width=0.8\columnwidth]{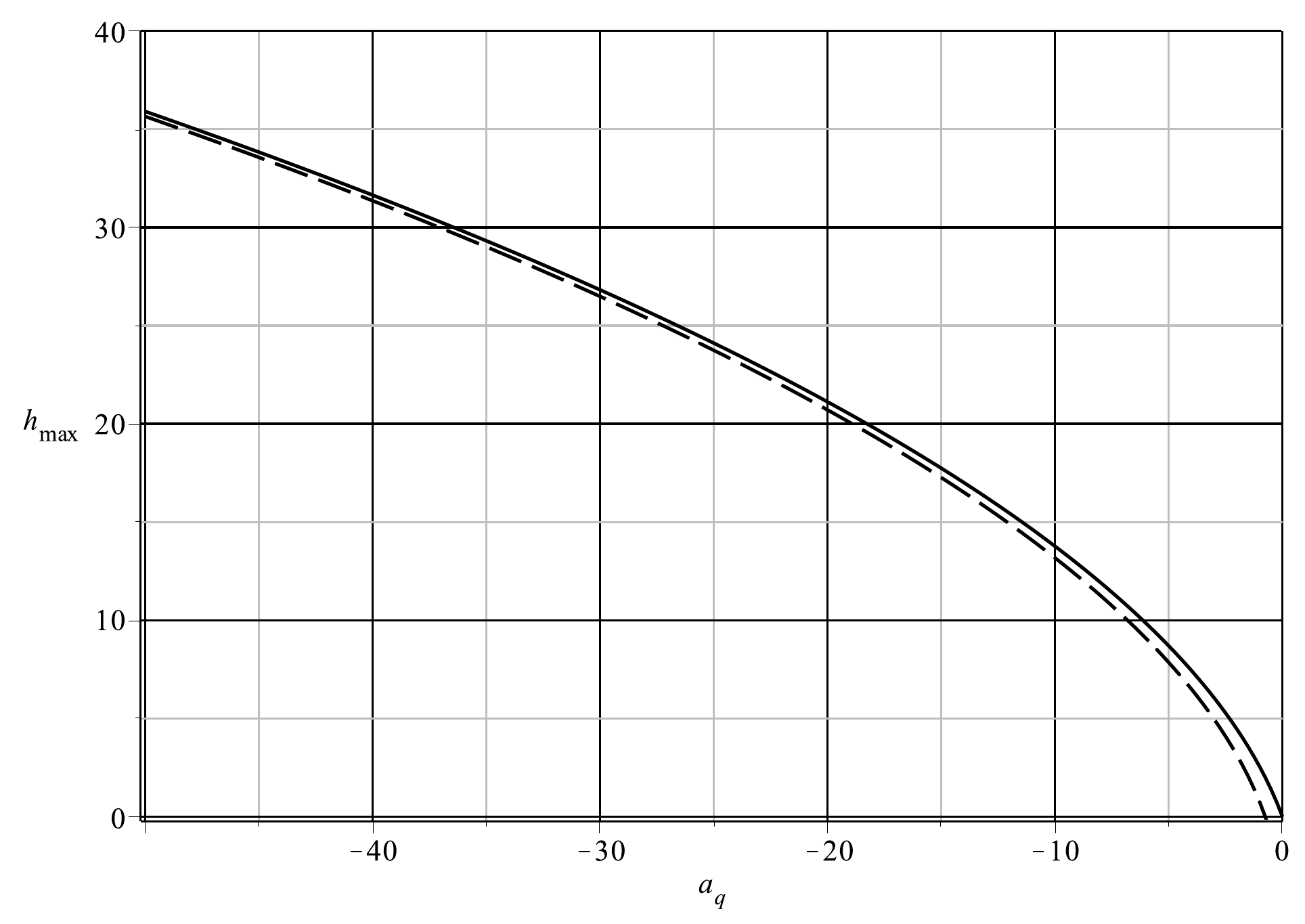}
\caption{The height of collapse of a multiphase plume in a non-stratified stable atmosphere as a function of the parameter $a_q$ defined in Eq.~\eqref{eq:firstIntegral_boussinesq_aq}. Here we compare the exact formula Eq.~\eqref{eq:fullModel_boussinesq_plumeHeight} with its asymptotic expansion Eq.~\eqref{eq:fullModel_boussinesq_plumeHeight_2}, in the case $v_q = 0.2$.}
\label{fig:plumeCollapse}
\end{figure}

In the Boussinesq limit, we have that $q \gg |\phi |\,,|q_\chi |\,,|q_\psi |$. 
It is worth noting that under this  approximation the reduced gravity $g'$ can be written via $\phi$:
\begin{equation}
\phi g \simeq \frac{\alpha_0 - \beta_0}{\alpha_0} g = g'\,.
\end{equation}
Moreover, the two entrainment models we are considering become equivalent and Eqs.~\eqref{eq:fullModel} reduces to:
\begin{subequations}\label{eq:fullModel_boussinesq}
 \begin{align}
 & q' = v_q \sqrt{m}\label{eq:fullModel_boussinesq_q}\\
 & m' = v_m\left( 1 - \gammac\right) \dfrac{q}{m}\label{eq:fullModel_boussinesq_m}\\
 & f' = 0\,.
 \end{align}
\end{subequations}
which is the multiphase version of the celebrated model introduced by \citet{morton1956}:
\begin{subequations}
\begin{align}
& q' = v_q \sqrt{m}\\
& m' = v_m \frac{q}{m}\\
& f' = 0\,.
\end{align}
\end{subequations}
Thus, we have found that the equations for a multiphase plume in a calm environment under the Boussinesq approximation are equivalent to the monophase \citet{morton1956} model with the following modification:
\begin{equation}
v_m \to v_m(1-\gammac)\,.
\end{equation}
Model~\eqref{eq:fullModel_boussinesq} has the following first integral:
\begin{align}
& \mathcal{U}_\textup{MTT} = \mathcal{U}_\textup{RS} = \mathcal{U} = q^{2} - \frac{4v_q}{5v_m(1-\gammac)}m^{5/2}
\label{eq:firstIntegral_boussinesq}\\
& \mathcal{U} = 1 - a_q\\
& a_q \equiv \frac{4v_q}{5v_m(1-\gammac)}\,,
\label{eq:firstIntegral_boussinesq_aq}
 \end{align}
The values of $a_q$ for the plume examples studied in this thesis are reported in Tab.~\ref{tab:plume_stability}.
From this expression and Eq.\eqref{eq:fullModel_boussinesq_q}, we found the implicit solution:
\begin{equation}\label{eq:zetaNonStratifiedBou}
\zeta = \zeta(q) = \frac{|a_q|^\frac{1}{5}}{v_q}\int_1^q \de x\, \left|x^2 - 1 + a_q\right|^{-\frac{1}{5}}\,.
\end{equation}
This solution has two branches, depending on the sign of $(1-\gammac)$, thus on the sign of $a_q$. If $a_q < 0$, the column is unstable with implicit solution (cf. App.~\ref{app:hypergeometric} for the definition of the Gaussian hypergeometric functions $\mathfrak{F}_b$ and $\mathfrak{G}_b$):
\begin{equation}
\zeta = \frac{(-a_q)^\frac{1}{5}}{v_q (1-a_q)^\frac{1}{5}}\left[q\,\mathfrak{F}_{-\frac{1}{5}}\left(\frac{q^2}{1-a_q}\right)-\mathfrak{F}_{-\frac{1}{5}}\left(\frac{1}{1-a_q}\right)\right]\,.
\end{equation}
The maximum height is reached when $q_\textup{max} = \sqrt{1-a_q}$:
\begin{equation}\label{eq:fullModel_boussinesq_plumeHeight}
H_\textup{max}/\ell_0 = \frac{(-a_q)^\frac{1}{5}}{v_q (1-a_q)^\frac{1}{5}}\left[\sqrt{1-a_q}\,\mathfrak{F}_{-\frac{1}{5}}\left(1\right)-\mathfrak{F}_{-\frac{1}{5}}\left(\frac{1}{1-a_q}\right)\right]\,.
\end{equation}
In Fig.~\ref{fig:plumeCollapse} we show the behavior of $H_\textup{max}/\ell_0$ for $v_q=0.2$ and we compare it with the following asymptotic expansion ($\mathfrak{F}_{-1/5}(1) \simeq 1.150$):
\begin{equation}\label{eq:fullModel_boussinesq_plumeHeight_2}
H_\textup{max}/\ell_0 = \frac{1}{v_q}\left(\mathfrak{F}_{-\frac{1}{5}}\left(1\right)\,\sqrt{-a_q} - 1\right) + O\left((-a_q)^{-\frac{1}{2}}\right)\,.
\end{equation}
Thus, the maximum height of a collapsing multiphase plume in Boussinesq regime behaves approximately as $\sqrt{-a_q}$.

\begin{figure}
\centering
\subfloat[][Virtual radius $b_\textup{v}$]{\includegraphics[width=0.5\columnwidth]{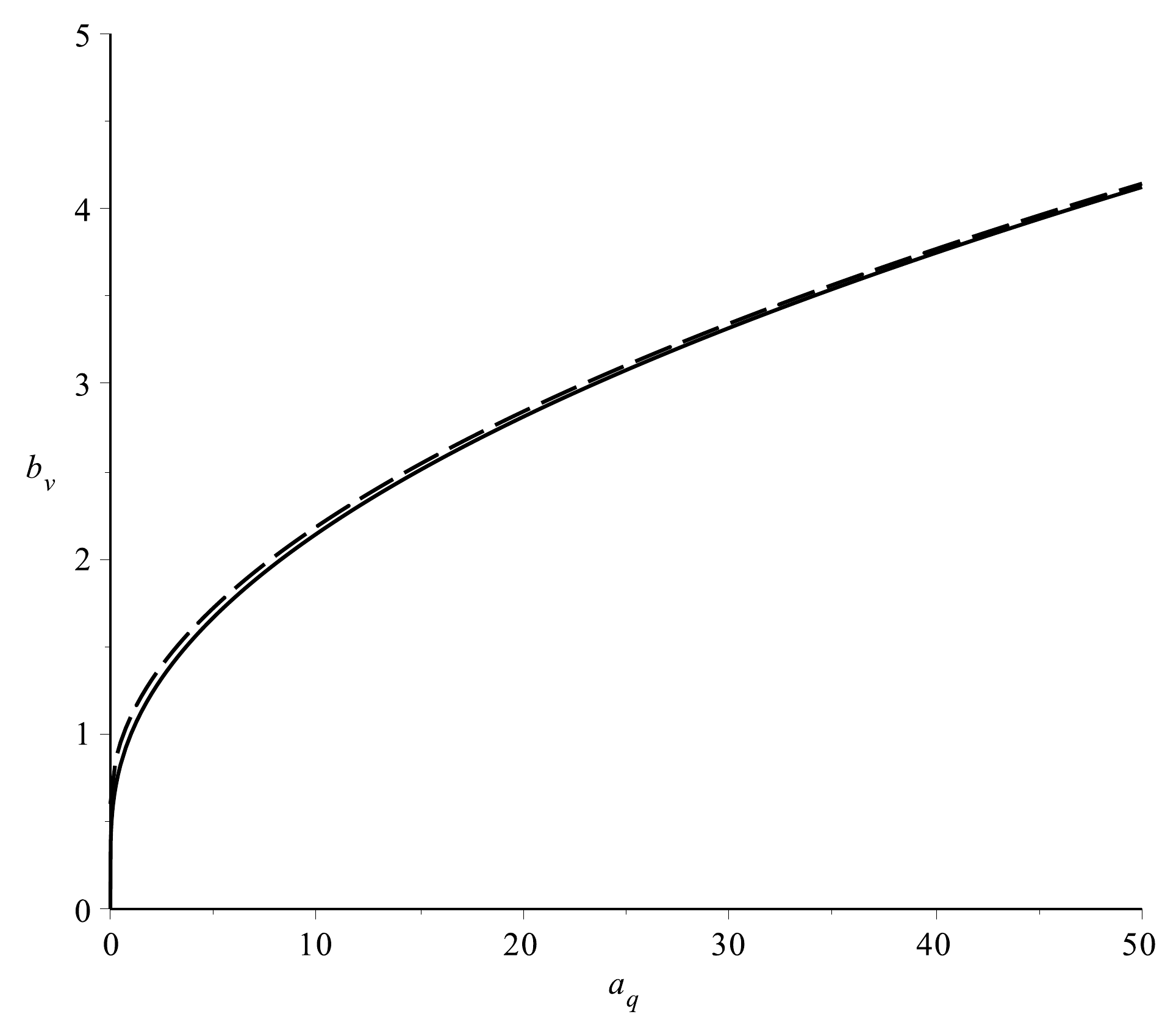}\label{fig:virtualRadius}}
\subfloat[][Necking height $\zeta_\textup{neck}$]{\includegraphics[width=0.5\columnwidth]{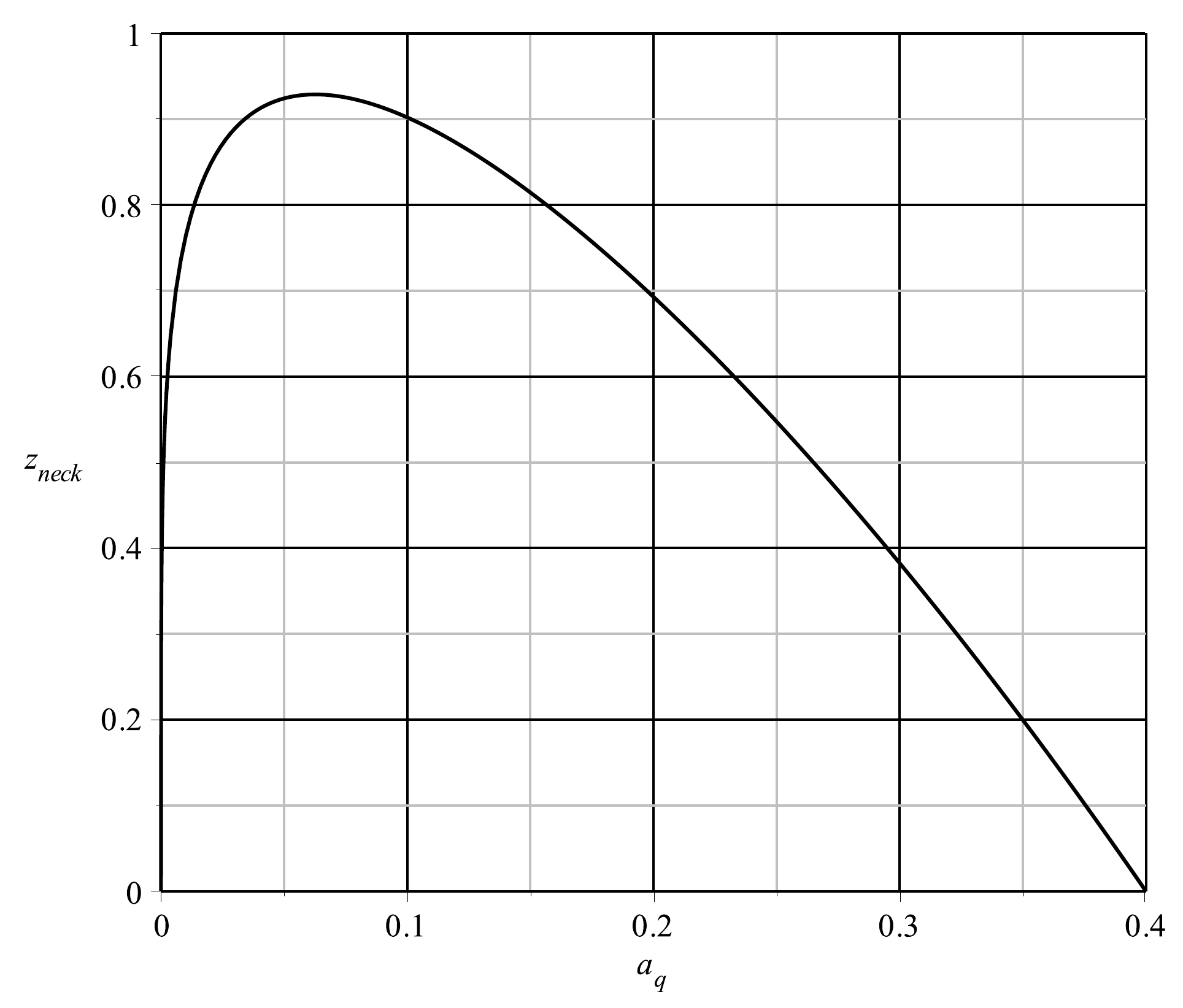}\label{fig:plumeNeck}}
\caption{a) The virtual radius $b_\textup{v}$ as a function of $a_q$. The virtual radius tends to zero when $a_q \to 0$ and increases with a square root law as $a_q$ increases (cf. Eq.~\eqref{eq:virtualRadius}). b) Height of the plume radius necking $\zeta_\textup{neck}$ as predicted by Eq.~\eqref{eq:plumeNeck}.}
\end{figure}

On the other hand, if $a_q >0$, the column is stable, rising indefinitely with this law (see App.~\ref{app:hypergeometric}):
\begin{equation}
\zeta = \frac{5}{3v_q} a_q^\frac{1}{5}\left[q^\frac{3}{5}\mathfrak{G}_{-\frac{1}{5}}\left(\frac{1-a_q}{q^2}\right)-\mathfrak{G}_{-\frac{1}{5}}\left(1-a_q\right)\right]\,.
\end{equation}
The asymptotic expansion $\mathfrak{G}(x) = 1 + O(x)$ allows us to find the self-similar solution:
\begin{subequations}\label{eq:asymptoticQbased}
\begin{align}
& q(\zeta) = \left(\frac{3 v_q}{5 a_q^\frac{1}{5}}\zeta + \mathfrak{G}_{-\frac{1}{5}}(1-a_q)\right)^\frac{5}{3} \propto \zeta^\frac{5}{3}\\
& m(\zeta) = \left[\frac{1}{a_q}\left(q^2(\zeta) - 1\right) + 1\right]^\frac{2}{5}\propto \zeta^\frac{4}{3}\,.
\end{align}
\end{subequations}
From here it is possible to extract the asymptotic plume radius evolution:
\begin{equation}
 b(\zeta) = \frac{q(\zeta)}{\sqrt{m(\zeta)}} = \frac{3}{5}v_q\,\zeta + a_q^\frac{1}{5}\,\mathfrak{G}_{-\frac{1}{5}}(1-a_q)\,.
\end{equation}
In this formula, we can recognize the famous result of~\citet{morton1956}: the plume spread $b'(\zeta)$ is
asymptotically constant and equal to $\frac{3}{5}v_q = \frac{6}{5}\varkappa$. Moreover we found the initial
virtual radius of the asymptotic plume and its asymptotic approximation, 
\begin{equation}\label{eq:virtualRadius}
b_\textup{v} = a_q^{1/5}\,\mathfrak{G}_{-\frac{1}{5}}(1-a_q) \simeq 0.5012 \sqrt{a_q} + 0.6\,. 
\end{equation}
The virtual plume radius is the intercept between $z=0$ and the radius of the equivalent plume spreading from a point source at $z = z_\textup{v} = -\frac{5 a_q^{1/5}}{3 v_q}\, \mathfrak{G}_{-\frac{1}{5}}(1-a_q)$. In Fig.~\ref{fig:virtualRadius} we show the behavior of $b_\textup{v}(a_q)$ and of its asymptotic approximation.
Finally, it is worth noting that the derivative of the plume radius has a simple expression thanks to the first integral~\eqref{eq:firstIntegral_boussinesq}
\begin{equation}
b'(\zeta) = v_q \left[\frac{3}{5}-\frac{2(1-a_q)}{5 a_q m^{5/2}}\right]\,,
\end{equation}
from which
\begin{equation}
b'(0) = v_q \left[\frac{3}{5}-\frac{2(1-a_q)}{5a_q}\right]\,,
\end{equation}
is the plume radius slope at $\zeta=0$. Another important property is the necking height $\zeta=\zeta_\textup{neck}$, where $b'(\zeta_\textup{neck}) = 0$. It exists only when $0< a_q < 2/5$:
\begin{equation}\label{eq:plumeNeck}
\zeta_\textup{neck} = \frac{5}{3v_q} a_q^\frac{1}{5}\left[\left(\frac{5}{3}(1-a_q)\right)^\frac{3}{10}\mathfrak{G}_{-\frac{1}{5}}\left(\frac{3}{5}\right)-\mathfrak{G}_{-\frac{1}{5}}\left(1-a_q\right)\right]\,.
\end{equation}
As shown in Fig.~\ref{fig:plumeNeck}, the necking height never exceeds $\zeta=1$.

\begin{figure}[t]
\centering
\includegraphics[width=0.8\columnwidth]{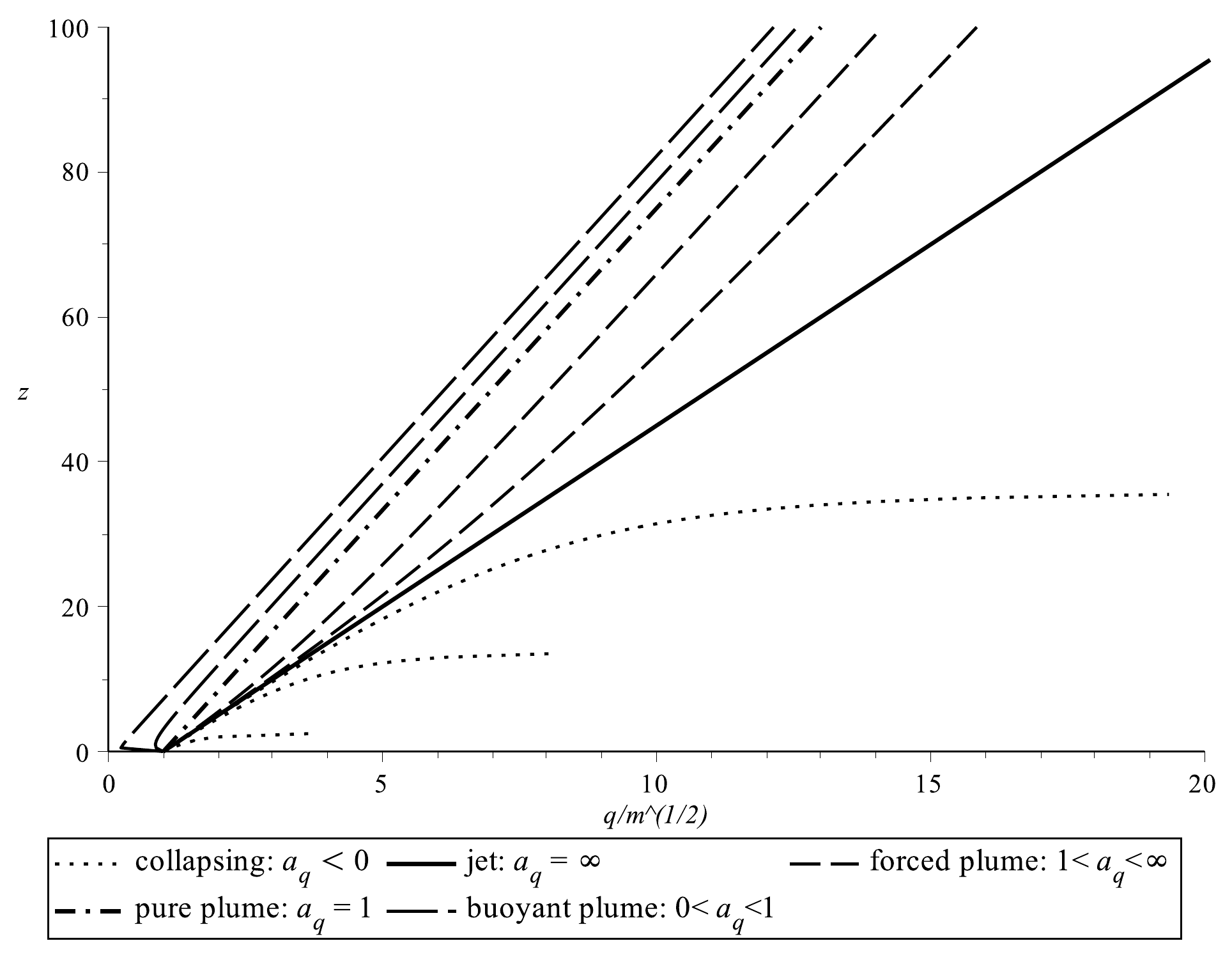}
\caption{Evolution of the plume radius $b(z) = q/\sqrt{m}$ in all the admissible regimes of model~\eqref{eq:fullModel_boussinesq} with $v_q = 0.2$. Starting from the lower graph, we choose: $a_q = -1,\, -10,\, -50,\, \infty,\, 50,\, 10,\, 1,\, 0.1,\, 0.0001\,.$}
\label{fig:nonStratifiedBoussinesq}
\end{figure}

We summarize in Fig.~\ref{fig:nonStratifiedBoussinesq}
all the possible regimes of model~\eqref{eq:fullModel_boussinesq}. Ranging from $a_q = 0^-$ to $a_q = 0^+$ passing through $a_q = \infty$, we have shown that: 1) ({\em collapsing regime}) when $a_q < 0$ the plume is collapsing, $b'(0) > v_q$, and its height increases as $a_q$ decreases (cf. Fig.~\ref{fig:plumeCollapse}); 2) ({\em jet regime}) when $a_q \to \infty$ then model Eq.~\eqref{eq:fullModel_boussinesq} reduces to the jet model~\eqref{eq:fullModel_jet} with $b(z) = v_q z +1$; 3) ({\em forced plume regime}) when $a_q >1$ the initial slope is $\frac{3v_q}{5} < b'(0) < v_q$, and the plume starts behaving as a jet until $z < \ell_\textup{M}$ (cf. ~\eqref{eq:mortonL} and \citet{morton1959}), then it moves to the plume-like behavior. As shown in Figs.~\ref{fig:nonStratifiedBoussinesq}, \ref{fig:virtualRadius}, $\ell_\textup{M}$ and $b_\textup{v}$ increase with $a_q$; 4) ({\em pure plume regime}) when $a_q = 1$ the solution of model~\eqref{eq:fullModel_boussinesq} highly simplifies and asymptotic expansions coincide with the exact solution. In particular, we have $b(z) = \frac{3 v_q}{5}z + 1$. There is not a jet-like interval in this regime; 5) ({\em buoyant plume regime}) when $0 < a_q < 1$ we have $b'(0) < \frac{3v_q}{5}$, and the plume radius reach its asymptotic slope $\frac{3v_q}{5}$ rapidly, after a small necking interval. In particular, if $0 < a_q < 2/5$ there exist $\zeta_\textup{neck} > 0$ where $b'(\zeta_\textup{neck}) = 0$.

\subsection{Boussinesq plume regime in a stratified environment}\label{sec:plumeStratified}
The Boussinesq approximation, with atmospheric stratification reduces~\eqref{eq:fullModel} to:
\begin{align}
& q' = v_q \sqrt{a(\zeta)\,m}
\label{eq:1Dplume_q_stratified}\\
& m' = v_m \frac{q}{m}(f-\gammac)\\
& f' = - v_f\frac{1 - \theta_f(\zeta)}{t_\alpha(\zeta)}\,q\,.
\end{align}
If we consider the atmospheric stratification only at the first order, we can apply the following approximation to the latter system (cf. Eqs.~\eqref{eq:brunt-vaisalla} and~\eqref{eq:1Dplume_vf0}):
\begin{align}
& a(\zeta) \simeq 1& & v_f\frac{1 - \theta_f(\zeta)}{t_\alpha(\zeta)} \simeq v_{f,0}\,,
\end{align}
allowing us to write the multiphase plume model in a stratified calm atmosphere:
\begin{subequations}\label{eq:1Dplume_stratified}
\begin{empheq}[box=\widefbox]{align}
& q' = v_q \sqrt{m}\\
& m' = v_m \frac{q}{m}(f-\gammac)\\
& f' = - v_{f,0}\,q\,.
\label{eq:1Dplume_stratified_f}
\end{empheq}
\end{subequations}
This model reduces to the same model introduced by~\citet{morton1959} in the monophase case:
\begin{subequations}\label{eq:1Dplume_stratified_monophase}
\begin{align}
& q' = v_q \sqrt{m}\\
& m' = v_m \frac{qf}{m}\\
& f' = - v_{f,0}\,q\,,
\end{align}
\end{subequations}
where $v_{f,0}$ is proportional to the Brunt-V\"ais\"all\"a frequency $\BV^2$ (cf.~\citet{Woods2010} and Eq~\eqref{eq:1Dplume_vf0}).

In order to find the first integrals of motion, we write system~\eqref{eq:1Dplume_stratified} in this form:
\begin{equation}\label{eq:1Dplume_first_integral}
\frac{\de q}{v_q \sqrt{m}} = \frac{m\,\de m}{v_m q (f-\gammac)} = -\frac{\de f}{v_{f,0} q}\,.
\end{equation}
By using the last equation multiplied by $q (f-\gammac)$, we obtain the first conserved quantity (recall that $f_0 = m_0 = 1$):
\begin{equation}\label{eq:1Dplume_Um}
\mathcal{U}_m = \frac{v_{f,0}}{v_m}m^2 + (f - \gammac)^2 = (1 - \gammac)^2 + \frac{v_{f,0}}{v_m}\,.
\end{equation}
$\mathcal{U}_m$ is a very interesting quantity, because it holds whatever the entrainment model is. Indeed, we have found it just by using the conservation of mass and enthalpy in system~\eqref{eq:1Dplume_stratified}, which are independent from the entrainment model. Moreover, this conserved quantity is telling us that $m$ reaches its maximum value
\begin{equation}\label{eq:m_max}
m_\textup{max} = \sqrt{1+ \frac{v_m}{v_{f,0}}(1-\gammac)^2}\,,
\end{equation}
when $f=\gammac$. In other words, the flux of momentum is maximum when the flux of buoyancy $(f-\gammac)$ is zero: {\em neutral buoyancy level}.
\\
Additionally, this first integral of motion tells us the value of the enthalpy flux when the plume reaches its maximum height. We define the maximum height of the plume as the point $\zeta=\zeta_\textup{max}$ where $m=0$, thus the minimum value of the enthalpy flux should be
\begin{equation}\label{eq:f_max}
f(\zeta_\textup{max}) \equiv f_\textup{min} = \gammac - \sqrt{\mathcal{U}_m}\,,
\end{equation}
because $f$ is a strictly decreasing function of $\zeta$ (cf. Eq.~\eqref{eq:1Dplume_stratified_f}). Thus, increasing the height $\zeta$ from 0 to $\zeta_\textup{max}$ let $f$ decrease from 1 to $f_\textup{min}$; while $m$ increases from 1 ($f=1$) to $m_\textup{max}$ ($f=\gammac$), then it decreases to 0 when $f=f_\textup{min}$. These observations, will be very useful in the next sections of this chapter.

Moving back to Eq.~\eqref{eq:1Dplume_first_integral}, it is easy to show that:
\begin{equation}
q\,\de q = -\frac{v_q}{v_{f,0}} \sqrt{m}\,\de f = -\frac{v_q v_m^{1/4}}{v_{f,0}^{5/4}} \left(\mathcal{U}_m - (f - \gammac)^2\right)^{1/4}\,\de f\,,
\end{equation}
from which we obtain another first integral of motion:
\begin{equation}\label{eq:1Dplume_Uq_monophase}
\mathcal{U}_q = q^2 + \frac{2v_q v_m^{1/4}}{v_{f,0}^{5/4}}\mathcal{U}_m^{1/4}(f-\gammac)\, \mathfrak{F}_{\frac{1}{4}}\left(\frac{(f-\gammac)^2}{\mathcal{U}_m}\right)\,,
\end{equation}
where $\mathfrak{F}_{\frac{1}{4}}(x) =\, _2F_1\left(-\frac{1}{4},\frac{1}{2};\frac{3}{2};x\right)$ is the hypergeometric function defined when $x < 1$ in App.~\ref{app:hypergeometric} and $\mathfrak{F}_{\frac{1}{4}}(1) = \pi^{3/2}\sqrt{2}/(6\,\Gamma^2(3/4)) \simeq 0.8740$\footnote{Here $\Gamma(x)$ is the Gamma function.}.
Noting that $x\mathfrak{F}_\frac{1}{4}(x^2)$ is a strictly increasing function bounded in $[-1,1]$, we have that, as $f$ decrease from 1 to $\gammac-\sqrt{\mathcal{U}_m}$, $q$ must increase from $1$ to
\begin{equation}
q_\textup{max}^2 = 1 + \frac{2v_q v_m^{1/4}}{v_{f,0}^{5/4}}\mathcal{U}_m^{1/4}\left[(1-\gamma_c)\mathfrak{F}_{\frac{1}{4}}\left(\frac{(1-\gammac)^2}{\mathcal{U}_m}\right) + \sqrt{\mathcal{U}_m}\,\mathfrak{F}_\frac{1}{4}(1)\right]\,.
\end{equation}

\begin{table}[t]
\centering
\begin{tabular}{ccccc}
\toprule
parameter & \textsf{[forcedPlume]} & \textsf{[Santiaguito]} & \textsf{[weakPlume]} & \textsf{[strongPlume]}\\ 
\midrule
$1-x_0$ & $6.521*10^{-7}$ & $9.163*10^{-3}$ & $1.832*10^{-3}$ & $4.245*10^{-2}$ \vspace{5pt}\\
$\tilde{q}_0$ & $2.054*10^{-8}$ & $1.753*10^{-3}$ & $2.178*10^{-4}$ & $3.940*10^{-2}$ \vspace{5pt}\\
$\plume$ & $1.114*10^{-3}$ & 0.1363 & $6.062*10^{-2}$ & 0.3010 \vspace{5pt}\\
$\qpt$ & 0.9321 & 0.5183 & 0.4828 & 1.691 \vspace{5pt}\\
\midrule
$\zeta_\textup{max}$ & 1526 & 21.35 & 141.9 & 17.89 \vspace{5pt}\\
$\zeta_\textup{max}^{(1)}$ & 1524 & 20.54 & 139.3 & 15.89 \vspace{5pt}\\
$\zeta_\textup{max}^{(0)}$ & 1532 & 24.16 & 151.5 & 24.33 \vspace{5pt}\\
$\zeta_\textup{max}/\zeta_\textup{nbl}$ & 1.318 & 1.375 & 1.345 & 1.488 \vspace{5pt}\\
$\zeta_\textup{max}^{(1)}/\zeta_\textup{nbl}^{(1)}$ & 1.318 & 1.394 & 1.354 & 1.582 \vspace{5pt}\\
\bottomrule
\end{tabular}
\caption{The main parameters defined in this section for the four plume examples of this thesis.}
\label{tab:plume_height}
\end{table}

By using again Eq.~\eqref{eq:1Dplume_stratified_f} with ~\eqref{eq:1Dplume_Uq_monophase}, we have found the implicit solution of problem~\eqref{eq:1Dplume_stratified}:
\begin{equation}\label{eq:1Dplume_stratified_solution}
\zeta = \frac{1}{v_{f,0}}{\displaystyle \int\limits_f^{1}\,\de f'\left[\mathcal{U}_q - \frac{2v_q v_m^{1/4}}{v_{f,0}^{5/4}}\mathcal{U}_m^{1/4}\,(f'-\gammac)\, \mathfrak{F}_{\frac{1}{4}}\left(\frac{(f'-\gammac)^2}{\mathcal{U}_m}\right)\right]^{-\frac{1}{2}}}\,.
\end{equation}

In order to better understand the behavior of the solution in different regimes, it is useful to define (see also Eq.~\eqref{eq:1Dplume_jet}):
\begin{align}\label{eq:1Dplume_epsilon}
& \plume\equiv \left(\frac{v_{f,0}}{(1-\gammac)^2 v_m}\right)^{\frac{1}{2}} = \frac{1}{|1-\gammac|}\frac{U_0 \BV}{\phi g} & & \mbox{plume limit parameter}\\
& \jet \equiv (|1-\gammac|\plume)^{-1} = \frac{\phi g}{U_0 \BV} & & \mbox{jet limit parameter}
\end{align}
which are comparing $U_\phi = U_0/\phi$ with $U_g = g/\BV\simeq 925$ m/s and $\gammac$ with 1. As we will show in the next section, when $\plume$ is small ($U_\phi \ll U_g$ and $\gammac < 1$) the solution has mainly a plume-like behavior, on the contrary, when $\jet \ll 1$, the solution behaves manly as a jet.

When we are in the plume limit regime ($\plume \ll 1$), any power of $\mathcal{U}_m$ can be simplified to (see Eq.~\eqref{eq:1Dplume_Um}):
\begin{equation}
\mathcal{U}_m^\gamma = |1-\gammac|^{2\gamma}\left(1 + \plume^2\right)^\gamma = |1-\gammac|^{2\gamma}\left(1 + \gamma \plume^2 + O(\plume^4)\right)\,.
\end{equation}
This approximation, leads to the limit
\begin{subequations}
\begin{align}
& q_\textup{max} \simeq
\begin{cases}
 2\sqrt{\dfrac{v_q}{v_m(1-\gammac)}\mathfrak{F}_{\frac{1}{4}}(1)}\, \plume^{-5/4} & \mbox{if} \quad \gammac < 1\vspace{5pt}\\
 1 & \mbox{if} \quad \gammac > 1
 \end{cases}\\
& m_\textup{max} \simeq \plume^{-1}\\
& f_\textup{min} \simeq 
\begin{cases}
2\gammac -1 & \mbox{if} \quad \gammac < 1\vspace{5pt}\\
1 & \mbox{if} \quad \gammac > 1\,.
\end{cases}
\end{align}
\end{subequations}
Thus, in this regime we recognize two distinct behaviors: when $\gammac > 1$ the multiphase plume is too heavy and slow to reach its height of positive buoyancy and it collapses. On the contrary, when $\gammac < 1$, the plume is able to reach its buoyancy reversal height and it can rise into the atmosphere. During its ascent, $f$ varies approximately in $[2\gammac-1,1]$, while $q$ and $m$ reach a much larger value the more $\plume$ is small.

On the other hand, in the jet limit regime ($\jet \ll 1$) we have:
\begin{subequations}
\begin{align}
& \mathcal{U}_m^\gamma = \left((1-\gammac)^2 + \jet^{-2}\right)^\gamma \simeq \jet^{-2\gamma}\\
& q_\textup{max} \simeq 1 + \frac{v_q}{v_m}\mathfrak{F}_{\frac{1}{4}}(1)\,\jet\\
& m_\textup{max} \simeq 1 + \frac{1}{2}(1-\gammac)^2 \jet^2\\
& f_\textup{min} \simeq - \jet^{-1}\,.
\end{align}
\end{subequations}
In this case $q$ and $m$ reach maximum values near $1$, while $f$ decreases the more the more $\jet$ is small.

\subsubsection{Plume height}
Eq.~\eqref{eq:1Dplume_stratified_solution} gives us the opportunity to write an analytic expression for the maximum height reached by a plume described by Eqs.~\eqref{eq:1Dplume_stratified}. Indeed, the maximum plume height (m=0) is reached when when $f = f_\textup{min}$ (cf. Eq.~\eqref{eq:f_max}). Thus, by substituting $f = f_\textup{min}$ in the integral lower limit, and  performing a change of variable in the integral with $x = (f-\gammac)/\sqrt{\mathcal{U}_m}$, we obtain (see definition for $\mathcal{U}_m$ in Eq.~\eqref{eq:1Dplume_Um}):
\begin{subequations}\label{eq:1Dplume_height_stratrified}
\begin{empheq}[box=\widefbox]{align}
& \zeta_\textup{max} = \frac{1}{v_q^{\frac{1}{2}}(v_m v_{f,0})^{\frac{1}{4}}}\left(\frac{v_m(1-\gammac)^2 + v_{f,0}}{v_{f,0}}\right)^{\frac{1}{8}}\,\mathfrak{h}(x_0,\tilde{q}_0)\\
& \mathfrak{h}(x_0,\tilde{q}_0) = \frac{1}{\sqrt{2}}{\displaystyle \int\limits_{-1}^{x_0}\,\de x\left[\tilde{q}_0 + x_0\, \mathfrak{F}_q(x_0^2) - x \mathfrak{F}_q(x^2)\right]^{-\frac{1}{2}}}\,,\\
& x_0 = (1-\gammac)\left(\frac{v_m}{v_m(1-\gammac)^2 + v_{f,0}}\right)^{\frac{1}{2}}
\label{eq:1Dplume_x0}\\
& \tilde{q}_0 = \frac{(v_m v_{f,0})^{\frac{1}{2}}}{2v_q} \left(\frac{v_{f,0}}{v_m(1-\gammac)^2 + v_{f,0}}\right)^{\frac{3}{4}}\,.
\label{eq:1Dplume_q0t}
\end{empheq}
\end{subequations}
where $\mathfrak{h}(x_0,\tilde{q}_0)$ is a function defined in $[-1,1]\times[0,\infty)$.
It is worth noting that with this substitution the neutral buoyancy level height can be easily obtained by substituting the lower bound of the integral $x=-1$ with $x = 0$ (cf. Eqs.~\eqref{eq:m_max} and~\eqref{eq:f_max}).

\begin{figure}[t]
\centering
\includegraphics[width=0.8\columnwidth]{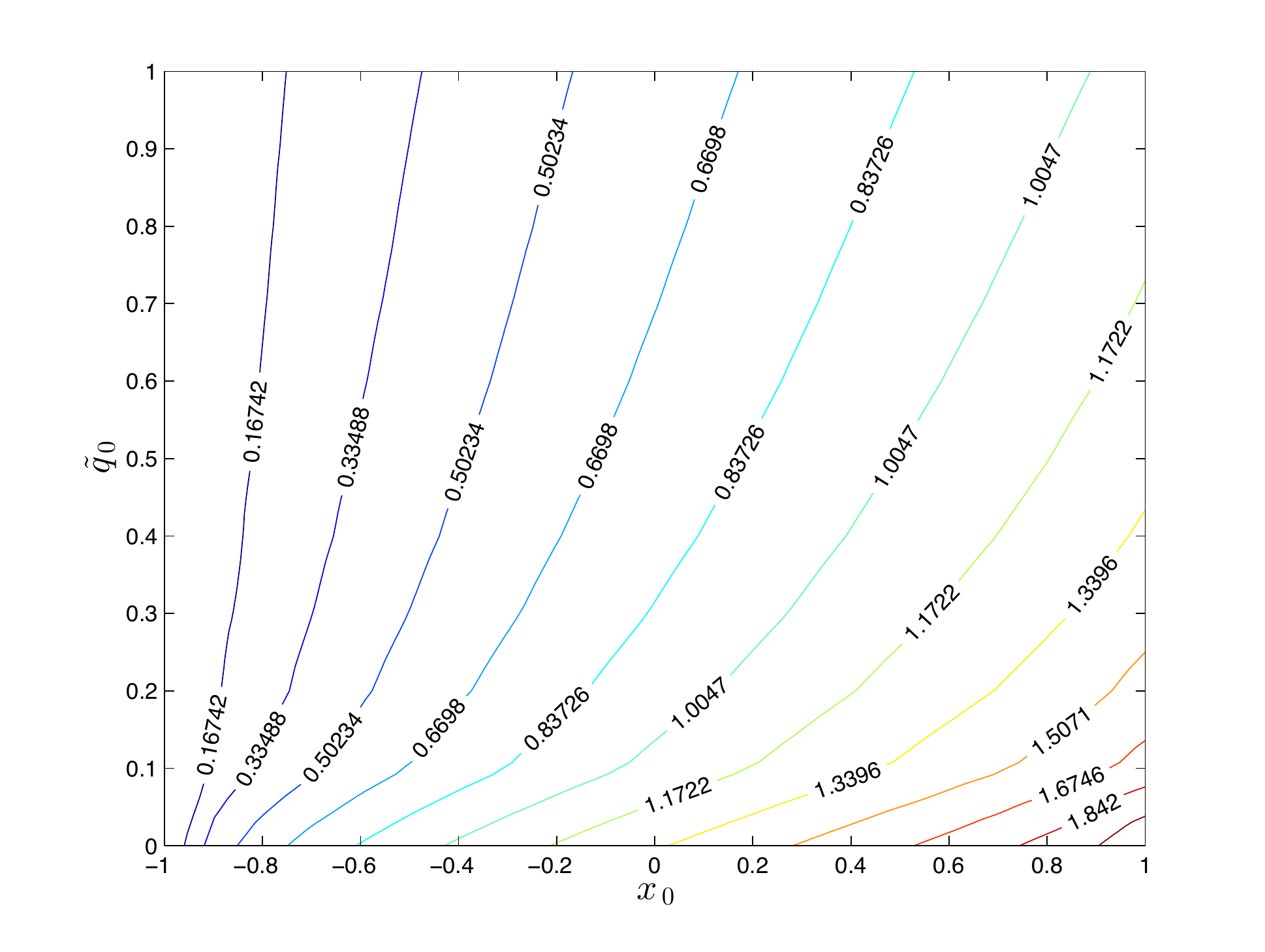}
\caption{Contour plot of the plume height function $\mathfrak{h}(x_0,\tilde{q_0})$ defined in Eq.~\eqref{eq:1Dplume_height_stratrified}. This function assumes its maximum in $\mathfrak{h}(1,0) = \Gamma_1 \simeq 2.572$, and it is a strictly decreasing function of $\tilde{q}_0$. When $x_0\to 1^{-}$ we are in the {\em plume regime}; $x_0 \to 0$}\label{fig:height_monophase}
\end{figure}

In Fig.~\ref{fig:height_monophase} we represent the values assumed by $\mathfrak{h}(x_0,\tilde{q}_0)$ in $(x_0,\tilde{q}_0) \in (-1,1)\times (0,1)$. We notice that this function has a maximum in $\mathfrak{h}(1,0) = \Gamma_1 \simeq 2.572$. Approaching this point, the function increases suddenly.
This figure must be read keeping in mind four main regimes: 1) $x_0 \to 1^{-}$ when $\gammac < 1$ and $\plume \ll 1$. In this case we are in the {\em plume regime} near the singular point $(x_0,\tilde{q}_0 = (1,0)$, thus the column initially has enough momentum to reach its buoyancy reversal height and enough enthalpy to rise until its maximum; 2) when $\gammac > 1$ and $\plume \ll 1$, we are in the {\em collapsing plume regime} near the point $(x_0,\tilde{q}_0) = (-1,0)$; 3) when $\jet \ll 1$ we are in the {\em jet regime}, near the line $x_0 = 0$. In general, $\gammac$ is the parameter controlling the column stability: when $\gammac < 1$ then $0 < x_0 < 1$, the column is not collapsing and when $x_0 \to 1$ the column behaves as a plume, while $x_0 \to 0^{+}$, the column behaves as a jet.

The expression for the plume height we have found is the multiphase version of to that found in~\citet{morton1959}. The behavior of $\mathfrak{h}$ near $(x_0,\tilde{q}_0) = (1,0)$ is the more interesting from a volcanological point of view, and it can be studied by using asymptotic expansion techniques for $\plume \ll 1$ (plume regime). In this case, Eqs.~\eqref{eq:1Dplume_height_stratrified} can be highly simplified. Indeed by using Eq.~\eqref{eq:1Dplume_epsilon}, we have:
\begin{align}
& x_0 = \sign(1-\gammac)\left(1 - \frac{1}{2}\plume^2 + O(\plume^4)\right) \simeq 1 -\frac{1}{2}\plume^2 \\
& \tilde{q}_0 = |1-\gammac|\,\frac{v_m}{2 v_q} \plume^{5/2} + O(\plume^{9/2}) \simeq (1-\gammac)\,\frac{v_m}{2 v_q} \plume^{5/2} = \frac{1}{2}\qpt \plume^{5/2}\\
& \qpt \equiv (1-\gammac)\,\frac{v_m}{v_q}\,,\qquad \mbox{see footnote\footnotemark}
\end{align}
because $\gammac < 1$ near $x_0 = 1$
\footnotetext{Recall that $\qpt = \frac{4}{5 a_q}$, see Eq.~\eqref{eq:firstIntegral_boussinesq_aq}.}.
Moreover, if $x\simeq 1$, the hypergeometric function can be approximated as follows:
\begin{equation}
x\mathfrak{F}(x^2) = \int (1 - x^2)^{\frac{1}{4}}\,\de x \simeq 2^{1/4}\int (1-x)^{\frac{1}{4}} = - \frac{2^{9/4}}{5}(1-x)^{\frac{5}{4}} + \mathfrak{F}(1)\,.
\end{equation}
With these information and $\qpt$ small enough, say
\begin{equation}
\qpt  < 2^{3/4} + \frac{4}{5} \simeq 2.5\,, 
\end{equation}
it is possible to show that:
\begin{equation}
\frac{1}{\sqrt{2}}{\displaystyle \int\limits_{-1}^{x_0}\,\de x\left[\tilde{q}_0 + x_0\, \mathfrak{F}_q(x_0^2) - x \mathfrak{F}_q(x^2)\right]^{-\frac{1}{2}}} \simeq \Gamma_1\left[ 1 - \Gamma_2\left(1 + \qpt^{\,5/12}\right)\,\plume^{3/4}\right]\,,
\end{equation}
where
\begin{align*}
& \Gamma_1 = \frac{1}{\sqrt{2}}{\displaystyle \int\limits_{-1}^{1}\,\de x\left[\mathfrak{F}_q(1) - x \mathfrak{F}_q(x^2)\right]^{-\frac{1}{2}}} \simeq 2.572\\
& \Gamma_2  \simeq 0.3802\,.
\end{align*}
In this ``plume regime'', the analytic formulation for the plume height given in~\eqref{eq:1Dplume_height_stratrified} simplifies to the first order approximation:
\begin{equation}\label{eq:1Dplume_height_stratified_2}
\zeta_\textup{max}^{(1)} = H^{(1)}_\textup{max}/\ell_0 = \frac{\Gamma_1}{v_q\, \qpt^{\,\frac{1}{2}}\,\plume^{\frac{3}{4}}}\left[ 1 - \Gamma_2\left(1 + \qpt^{\,\frac{5}{12}}\right)\plume^{\frac{3}{4}}\right]\,,
\end{equation}
while the zeroth order approximation is:
\begin{equation}
\zeta_\textup{max}^{(0)} = H^{(0)}_\textup{max}/\ell_0 = \frac{\Gamma_1}{v_q\, \qpt^{\,\frac{1}{2}}\,\plume^{\frac{3}{4}}}\,.
\end{equation}
This last approximation holds in the limit $\plume\to 0$, which is equivalent to the pure plume solution with initial mass and momentum equal to zero and finite initial flux of buoyancy.

\begin{figure}
\centering
\includegraphics[width=\columnwidth]{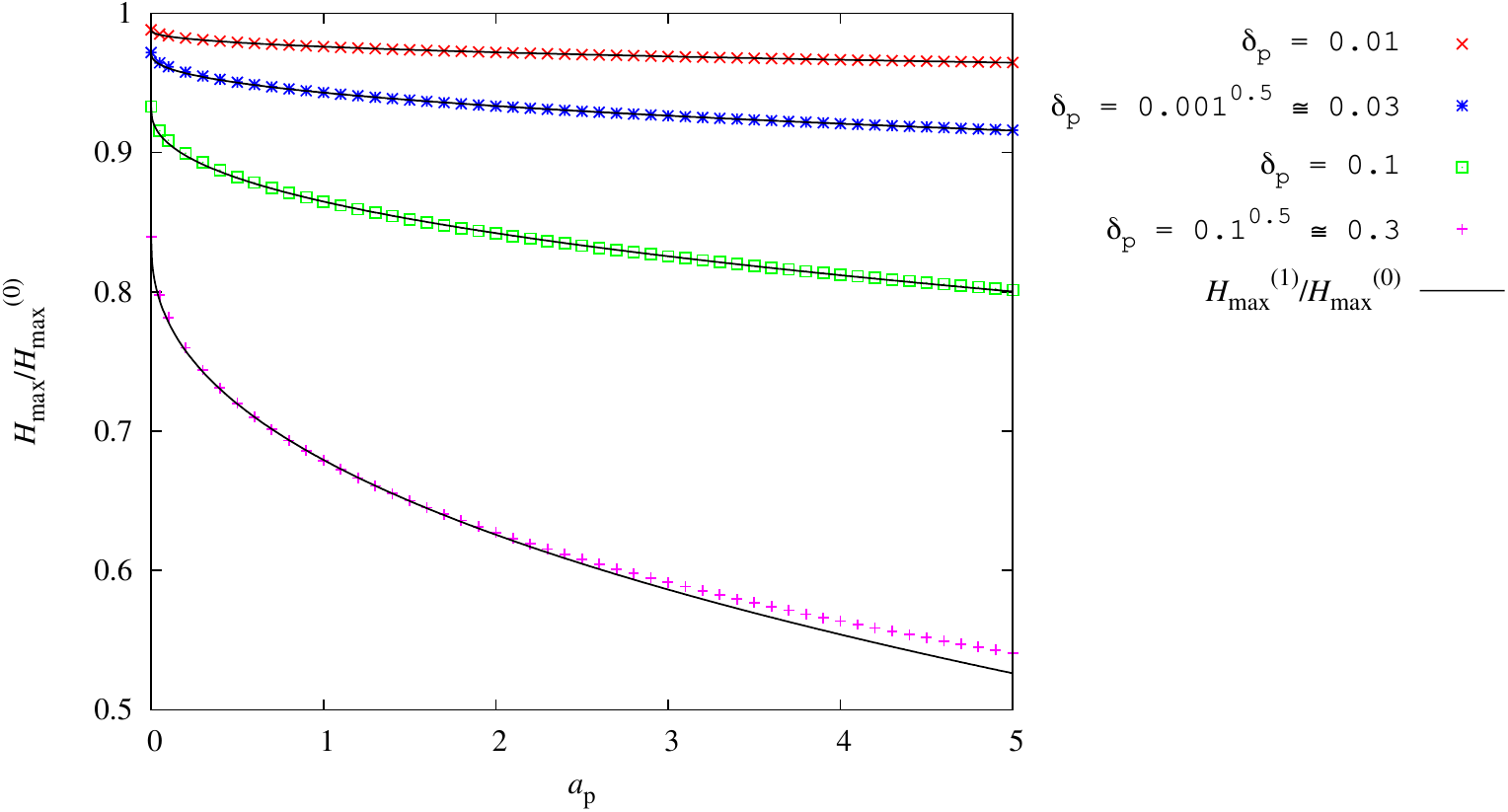}
\caption{Comparison of the exact formula Eq.~\eqref{eq:1Dplume_height_stratrified} for the plume height of model~\eqref{eq:1Dplume_stratified} with the first order approximation Eq.~\eqref{eq:1Dplume_height_stratified_2} over the zeroth order approximation Eq.~\eqref{eq:1Dplume_height_stratified_0}. }\label{fig:firstOrder}
\end{figure}

In Fig.~\ref{fig:firstOrder} we show the good behavior of Eq.~\eqref{eq:1Dplume_height_stratified_2} when $\plume < 0.3$ and $\qpt < 5$. It i worth noting from Tab.~\ref{tab:volcanic_plumes} that this parameter range is the most interesting from the point of view of volcanic plumes. Fig.~\ref{fig:firstOrder} compares the first order, the zeroth order and the exact solution~\eqref{eq:1Dplume_height_stratrified}. It shows that the first order approximation behaves very well in the selected parameter range. On the other hand, we point out that considering the first order approximation instead of the zeroth order allows to avoid an error up to $100\,\%$ when $\plume \simeq 0.3$ and $\qpt = 5$ ($H_\textup{max} \simeq H^{(1)}_\textup{max} \simeq 0.5 H^{(0)}_\textup{max}$). We observe also that Fig.~\ref{fig:firstOrder} is a zoom on the singularity at the bottom right of Fig.~\ref{fig:height_monophase}, since $\tilde{q}_0 \propto \plume^{5/2}$.

 In the literature, the problem of obtaining the maximum plume height starting from the monophase ($\gammac = 0$) formulation of the plume model in a stratified environment, Eq~\eqref{eq:1Dplume_stratified_monophase} has been studied in~\citet{morton1956}. He found \mbox{$\zeta_{\textup{max},\textup{M}} \simeq 2.805$} in his non-dimensionalization. We can recover the same result in the zero order approximation, by noting that the conversion factor from our non-dimensionalization to that used by~\citet{morton1956} is
\begin{equation}
\zeta_\textup{M} = 2^{\frac{1}{8}} v_{f,0}^{\frac{3}{8}}v_q^{\frac{1}{2}}v_m^{\frac{1}{8}}\,\zeta = 2^{\frac{1}{8}} v_q \qpt^{\,\frac{1}{2}} \plume^\frac{3}{4} \,\zeta\,,
\end{equation}
from which $\zeta_{\textup{max},\textup{M}} = 2^{\frac{1}{8}}\,\Gamma_1 \simeq 2.805$.
Turning to dimensional variables, at the zeroth order we have recovered the famous relationship:
\begin{equation}\label{eq:1Dplume_height_stratified_0}
H^{(0)}_\textup{max} =  \frac{\Gamma_1}{\sqrt{2\varkappa}} \left(\frac{ \phi g U_0 \ell_0^2}{\BV_0^3}\right)^{\frac{1}{4}} = \frac{\Gamma_1}{\sqrt{2\varkappa}} \left(\frac{\phi g Q_0}{\alpha_0\BV_0^3}\right)^{\frac{1}{4}}\,,
\end{equation}
telling that the maximum plume height to the power four is proportional to the mass flow rate times the enthalpy anomaly and inversely proportional to the cube of the Brunt-V\"ais\"all\"a frequency. In the monophase case, when the~\citet{ricou1961} entrainment model can be considered a good approximation for the dynamics of the first part of the plume, this result is valid even if the Boussinesq approximation is not valid (see Eq.~\eqref{eq:1Dplume_q_stratified}). 

In volcanological applications the zero order formula is widely used. We have found a correction to that formula, for the multiphase case in both the zeroth and first order formulation. In dimensional variables, the multiphase first order formulation of the plume height reads:
\begin{align}
& H_\textup{max} = \frac{\Gamma_1}{\sqrt{2\varkappa}} \left(\frac{\phi^* g Q_0}{\alpha_0\BV_0^3}\right)^{\frac{1}{4}}\,\left\lbrace 1 - \Gamma_2\left[1 + \left(\frac{\phi^* g \ell_0}{2\varkappa\,U_0^2}\right)^\frac{5}{12}\right]\left(\frac{U_0 \BV_0}{\phi^* g}\right)^{\frac{3}{4}}\right\rbrace\\
& \phi^* \equiv (1-\gammac)\phi = \phi - \left[\chis\Yso + (\chie-\psie)\Yeo\right]\,.
\end{align}
which strongly increase the accuracy of the plume height, keeping a simple analytic formulation. The only difference between the monophase and the multiphase formulation is in the factor $(1-\gammac)$, through the substitution $\phi \to \phi^*$.

We remind that this Taylor series approximation holds when $\plume \ll 1$ which is equivalent to \mbox{$U_0/\phi < g/\BV_0\simeq 925\,\textup{m/s}$}. This last condition give us a lower limit for $\phi$ and than to the vent temperature:
\begin{equation}
\phi > \frac{U_0 \BV_0}{g} \quad \Rightarrow \quad \frac{\Delta T_0}{T_{\alpha,0}} > \frac{U_0 \BV_0}{g}\,.
\end{equation}
If the vent temperature is much smaller than this lower bound, than the plume behaves more likely to a jet, and integral~\eqref{eq:1Dplume_height_stratrified} must be evaluated without the approximation $\plume \ll 1$. 

When we are in the opposite condition $\jet = \plume^{-1} \to 0$ (jet limit), we have $x_0 \to \jet \ll 1$. In this regime, the function $\mathfrak{h}(x_0,\tilde{q}_0)$ does not have a strong singularity as in the case $x_0 \to 1$ (cf. Fig.~\ref{fig:height_monophase}) and Eq.~\eqref{eq:1Dplume_height_stratrified} can be safely approximated at the zeroth order as (use the fact that $x\mathfrak{F}(x^2)\simeq x$ in $x \in [-1,0]$):
\begin{align}
H_\textup{max}&  \simeq \ell_0 \frac{1}{v_q\left(\sqrt{\tilde{q_0}+\tilde{q}_0^2} + \tilde{q}_0\right)}\,, & \tilde{q}_0  & = \frac{(v_m v_{f,0})^\frac{1}{2}}{2v_q}=\frac{\ell_0 \BV_0}{4\varkappa U_0}\,.
\end{align}
If also $\tilde{q}_0 \ll 1$ this expression further simplifies giving the following expression for the maximum jet height:
\begin{equation}
H_\textup{max} \simeq \left(\frac{U_0 \ell_0}{\varkappa\BV_0}\right)^{\frac{1}{2}}\,.
\end{equation}
As a first order approximation one can use $\ell_0\simeq b_0$ and invert this expression to find the inlet velocity from the jet height.

\subsubsection{Neutral buoyancy level and plume height inversion}
By recalling that the neutral buoyancy level (nbl) is reached when $f=0$, it is easy to modify Eqs.~\eqref{eq:1Dplume_height_stratrified} and~\eqref{eq:1Dplume_height_stratified_2} to find $H_\textup{nbl}$:
\begin{align}
& H_\textup{nbl}/\ell_0 = \frac{1}{v_q^{\frac{1}{2}}(v_m v_{f,0})^{\frac{1}{4}}}\left(\frac{v_m(1-\gammac)^2 + v_{f,0}}{v_{f,0}}\right)^{\frac{1}{8}}\,\mathfrak{h}_\textup{nbl}(x_0,\tilde{q}_0)\\
& \mathfrak{h}_\textup{nbl}(x_0,\tilde{q}_0) = \frac{1}{\sqrt{2}}{\displaystyle \int\limits_{0}^{x_0}\,\de x\left[\tilde{q}_0 + x_0\, \mathfrak{F}_q(x_0^2) - x \mathfrak{F}_q(x^2)\right]^{-\frac{1}{2}}}\,,\\
& H^{(1)}_\textup{nbl}/\ell_0 = \frac{\Gamma_1}{v_q\, \qpt^{\,\frac{1}{2}}\,\plume^{\frac{3}{4}}}\left[ \Gamma_\textup{nbl} - \Gamma_2\left(1 + \qpt^{\,\frac{5}{12}}\right)\plume^{\frac{3}{4}}\right]\,,
\label{eq:height_monophase_nbl}\\
& \Gamma_\textup{nbl} = 1 - \frac{1}{\sqrt{2}\,\Gamma_1}\,{\displaystyle \int\limits_{-1}^{0}\,\de x\left[\mathfrak{F}_q(1) - x \mathfrak{F}_q(x^2)\right]^{-\frac{1}{2}}} \simeq 0.7596\,.
\end{align}
Thus we have found a first-order modification of the result of~\citet{Turner1979}:
\begin{equation}\label{eq:nblRatio}
\frac{H^{(1)}_\textup{max}}{H^{(1)}_\textup{nbl}} = \frac{1}{\Gamma_\textup{nbl}} + \frac{\Gamma_2(1 - \Gamma_\textup{nbl})}{\Gamma_\textup{nbl}^2}\left(1 + \qpt^{\,\frac{5}{12}}\right)\plume^{\frac{3}{4}}\,.
\end{equation}
At the zeroth order we find $H^{(0)}_\textup{max}/H^{(0)}_\textup{nbl} = 1/\Gamma_\textup{nbl} \simeq 1.316$ in agreement with $H_\textup{max}/H_\textup{nbl} = 1.3$ obtained by~\citet{Turner1979}.

This result is telling us that the ratio between the maximum plume height and its neutral buoyancy level is a constant $\Gamma_\textup{nbl}^{-1} \simeq 1.3$ when $\plume$ is small enough, and it grows with $\plume^{3/4}$.

\begin{figure}[t]
\centering
\includegraphics[width=0.7\columnwidth]{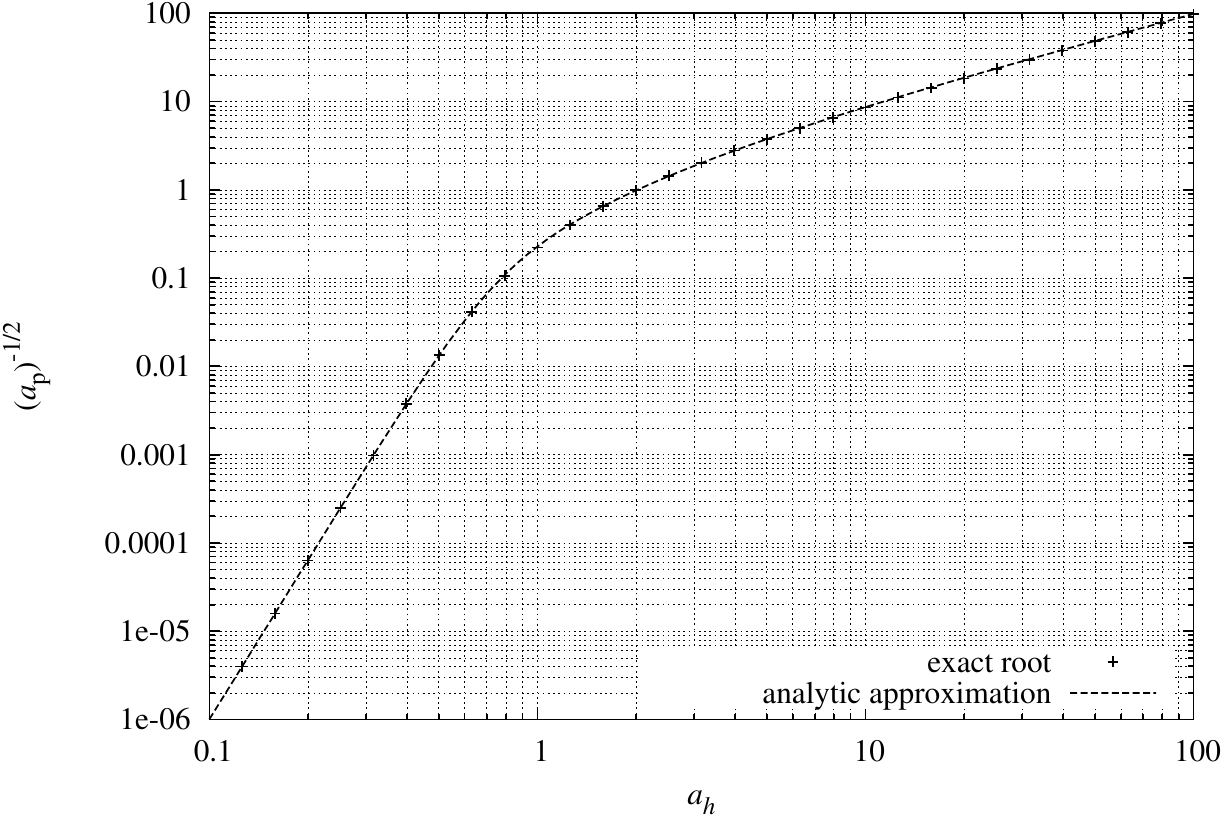}
\caption{Root of Eq.~\eqref{eq:root_height_monophase} as a function of $a_h$ and its analytic approximation, Eq.~\eqref{eq:height_monophase_root_approx}.}\label{fig:root_heigh_monophase}
\end{figure}

The neutral buoyancy level of a plume can be observed by measuring the height where the plume umbrella begins to spread up. If we know $H_\textup{nbl}$, $H_\textup{max}$, $\ell_0\simeq b_0$ and the entrainment $v_q= 2\varkappa$, it is possible to invert Eqs.~\eqref{eq:1Dplume_height_stratified_2} and~\eqref{eq:height_monophase_nbl} in order to find $\plume$ and $\qpt$ or equivalently $U_0$, $\phi$ and $\beta_0$. Defining $h_\textup{nbl} = H_\textup{max}/H_\textup{nbl}$ and $h_\textup{max} = H_\textup{max}/\ell_0$, we find
\begin{subequations}\label{eq:height_monophase_inversion}
\begin{align}
& \left(\qpt\right)^{-\frac{1}{2}} + \left(\qpt\right)^{-\frac{1}{12}} = a_h
\label{eq:root_height_monophase}\\
& a_h = \frac{v_q h_\textup{max}(h_\textup{nbl}\Gamma_\textup{nbl} - 1)}{\Gamma_1\Gamma_2 h_\textup{nbl}(1-\Gamma_\textup{nbl})}\\
& \left(\qpt\right)^{-\frac{1}{2}} \simeq \frac{a_h^6}{1-0.41a_h^2+1.4a_h^3+1.39a_h^4+a_h^5}
\label{eq:height_monophase_root_approx}\\
& \plume^\frac{3}{4} = \frac{\Gamma_1 h_\textup{nbl}(1-\Gamma_\textup{nbl})}{v_q h_\textup{max}(h_\textup{nbl} - 1)} \left(\qpt\right)^{-\frac{1}{2}}\\
& U_0 = \frac{\ell_0\BV_0}{v_q \qpt \plume}\\
& \phi^* = (1-\gammac)\phi = \frac{\BV^2 \ell_0}{v_q g\, \qpt \plume^2}\,,
\end{align}
\end{subequations}
a well posed problem when $h_\textup{nbl}>\Gamma_\textup{nbl}^{-1} \simeq 1.316$. The first equation can be solved looking for the unique positive root with respect $x=(\qpt)^{-1/2}$ (cf. Fig.~\ref{fig:root_heigh_monophase}). In Eq.~\eqref{eq:height_monophase_root_approx} we give an approximate analytic solution which has a good behavior both in the asymptotic ($a_h\to0$ and $a_h \to \infty$) and intermediate regime ($0.5<a_h<5$). In conclusion, the first order approximation for the plume height gives an additional information allowing to find both $U_0$ and $\phi^*$ in contrast with the zero order approximation which needs an additional hypothesis on $\phi^*$ to give the mass flux.

In order to fix ideas, we give an example. Suppose to have a monophase air plume with $U_0 = 30$ m/s, $T_0 = 373$ K ($\beta_0 = 0.947$ kg/m$^3$), $b_0 = 0.1$ m ejected in an atmosphere with $T_{\alpha,0} = 300$ K, $p_0 = 101325$ Pa and $\BV_0 = 1.015*10^{-2}$ Hz. Solving Eqs.~\eqref{eq:1Dplume_monophase} with the~\citet{ricou1961} model ($\varkappa = 0.14$), we obtain $H_\textup{max}/\ell_0\simeq 1387.2$ and $h_\textup{nbl} = 1.3461$, slightly bigger than $\Gamma_\textup{nbl}^{-1} \simeq 1.316$. Now, substituting $H_\textup{max}/\ell_0$, $h_\textup{nbl}$, $v_q = 0.28$ and $\ell_0 = 0.1$ m in Eqs.~\eqref{eq:height_monophase_inversion}, we can invert the problem recovering the initial velocity and density. With our first order approximation, we obtain:
\begin{align}
& U_{0,\textup{inverted}} \simeq 28\,\mbox{m/s}\\
& \beta_{0,\textup{inverted}} \simeq 0.887\,\mbox{kg/m}^3\,,
\end{align}
with less than $10\,\%$ of error with respect to the ``real'' values.

\subsection{Analytic solution for a non-Boussinesq plume in a stratified environment}
In this section we want to find an analytic solution approximating the behavior of model~\eqref{eq:fullModel} in its complete form, from the vent elevation up to the neutral buoyancy level. The strategy that we will follow here will bring to an update of the results we have presented in~\citet{Cerminara2015ir}.

Both Eqs.~\eqref{eq:zetaNonStratified} and~\eqref{eq:zetaNonStratifiedBou} admit the same asymptotic solution fulfilling the initial condition $q(0) = 1$
\footnote{In Eqs.~\eqref{eq:asymptoticQbased} are the asymptotic solution of system~\eqref{eq:fullModel_boussinesq}, written in a form such that it is possible to find the virtual radius $b_\textup{v}$. However, that solution does not fulfill initial conditions for $q$ and $m$. To write an asymptotic solution respecting the initial condition it is more convenient to use $q(\zeta)$ in the form given in this section.}:
\begin{equation}\label{eq:asymptoticQ}
q(\zeta) = \left(\frac{3 v_q}{5 a_q^{1/5}}\, \zeta + 1\right)^\frac{5}{3}\,,\qquad \mbox{where}\quad a_q = \frac{4 v_q}{5 v_m (1-\gammac)}\,.
\end{equation}
Thus this solution approximate the plume model~\eqref{eq:fullModel} in both the Boussinesq and non-Boussinesq regime. The difference between these two regimes appears in the asymptotic solution when we choose which first integral of motion to use, either $\mathcal{U}$ (Eq.~\eqref{eq:firstIntegral_boussinesq}) or $\mathcal{U}_\textup{RS}$ (Eq.~\eqref{eq:firstIntegral_nonStratified_lq}), thus in the form of $m$:
\begin{align}
& m(\zeta) = \left[\frac{1}{a_q} \left(q^2(\zeta) -1\right) + 1\right]^\frac{2}{5}\,, & & \mbox{or}
\label{eq:asymptoticM_Q}\\
& m(\zeta) = \left\lbrace\frac{1}{a_q} \left[(l_\textup{c}(q(\zeta)) - l_\textup{c}(1)\right] + 1\right\rbrace^\frac{2}{5}\,, & & \mbox{with}
\label{eq:asymptoticM}\\
& l_\textup{c}(q) = q^2 -\frac{2\gammac (\phi -q_\chi)}{1-\gammac}\left[q - q_\chi\ln (|q+q_\chi|)\right]\,.
\end{align}
These asymptotic expansions are equivalent to Eqs.~\eqref{eq:asymptoticQbased}, with correct initial conditions $m(0) = 1$ and $q(0)=1$. In what follows, we will use the latter Eq.~\eqref{eq:asymptoticM} as asymptotic expansion for the momentum flux, because it works better than the former equation in the non-Boussinesq regime. Indeed, even if this solution has been found by applying the approximation $q \gg 1$ to Eqs.~\eqref{eq:fullModel}, we want to extend its applicability to plumes in non-Boussinesq regime. We will describe a strategy to hold this task, after having introduced atmospheric stratification.

The only difference between Eqs.~\eqref{eq:fullModel_boussinesq} --~from where we have extracted the latter asymptotic solution~-- and the Eqs.~\eqref{eq:1Dplume_stratified} --~for a stratified atmosphere~-- is the variability of $f(\zeta)$. In the former system $f$ is considered as constant and equal~to~1, while in the latter one it is considered as a function $f = f(\zeta)$. However, we have seen in the previous section that $f(z)$ is a slowly varying function, because $v_{f,0}$ is usually very small with respect to the rate of variation of the other equations involved, namely $v_q$ and $v_m$. Thus, one strategy to look for an analytic solution of the problem in a stratified atmosphere could be to consider the asymptotic solution~\eqref{eq:asymptoticQ} valid also for problem~\eqref{eq:1Dplume_stratified}, and use it for finding $f(\zeta)$. In particular, substituting $q(\zeta)$ in \eqref{eq:1Dplume_stratified_f}, we obtain:
\begin{equation}\label{eq:asymptoticF}
f(\zeta) = 1 - \frac{v_{f,0}}{2(1-\gammac)v_m}(m(\zeta)^2 - 1)\,,
\end{equation}
with $m(\zeta)$ defined in Eqs.~\eqref{eq:asymptoticM_Q}.
Now, we recall the first integral of motion found in Eq.~\eqref{eq:1Dplume_Um}
\begin{equation}
\mathcal{U}_m = (1-\gammac)^2 + \frac{v_{f,0}}{v_m} = (f-\gammac)^2 + \frac{v_{f,0}}{v_m} m^2\,,
\end{equation}
and we try to substitute Eq.~\eqref{eq:asymptoticF} in it. We find:
\begin{equation}\label{eq:asymptotic_Um}
(f-\gammac)^2 = (1-\gammac)^2 + \frac{v_{f,0}}{v_m}(1 - m^2) + \frac{v_{f,0^2}}{4(1-\gammac)^2 v_m^2} (1 - m^2)^2\,.
\end{equation}
This result differs from Eq.~\eqref{eq:1Dplume_Um} just because of the term
\begin{equation}
\frac{v_{f,0}^2}{4(1-\gammac)^2 v_m^2} (1 - m^2)^2 = \frac{1}{4}(1-\gammac)^2 \plume^2 \,(1-m^2)^2\,,
\end{equation}
where we have used the definition of $\plume = v_{f,0}/(1-\gammac)^2 v_m$. The latter term is $O(\plume^2)$, thus it can be disregarded in the plume regime ($\plume \ll 1$) with respect the other two terms in the right-hand-side of Eq.~\eqref{eq:asymptotic_Um}, which are respectively $O(1)$ and $O(\plume)$. By noting that $\mathcal{U}_m$ is approximatively conserved by the asymptotic solution found in this section, we have corroborated the fact that this solution is approximating the complete solution in the plume regime. 

Having the enthalpy flux evolution $f(\zeta)$, it is possible to calculate the maximum plume height and neutral buoyancy level by using $m_\textup{max}$ and $f_\textup{min}$ given respectively in Eqs.~\eqref{eq:m_max} and~\eqref{eq:f_max}. In Tab.~\ref{tab:asymptotic_height} we recall the maximum plume height and neutral buoyancy level as obtained from model~\eqref{eq:fullModel}, comparing it with the asymptotic results $\zeta_\textup{max}^{(\textup{asy})}$, $\zeta_\textup{nbl}^{(\textup{asy})}$.
\begin{table}[t]
\centering
\begin{tabular}{ccccc}
\toprule
parameter & \textsf{[forcedPlume]} & \textsf{[Santiaguito]} & \textsf{[weakPlume]} & \textsf{[strongPlume]}\\ 
\midrule
$\zeta_\textup{max}$ & 1665 & 23.98 & 160.6 & 20.68 \vspace{5pt}\\
$\zeta_\textup{max}^{(\textup{asy})}$ & 1487 & 21.79 & 139.8 & 19.65 \vspace{5pt}\\
$\zeta_\textup{nbl}$ & 1264 & 18.36 & 118.6 & 13.58 \vspace{5pt}\\
$\zeta_\textup{nbl}^{(\textup{asy})}$ & 1145 & 16.53 & 106.1 & 14.55 \vspace{5pt}\\
\bottomrule
\end{tabular}
\caption{The main parameters defined in this section for the four plume examples of this thesis.}
\label{tab:asymptotic_height}
\end{table}

\begin{figure}[p]
\centering
\subfloat[][mass, momentum and enthalpy fluxes]{\includegraphics[width=0.45\columnwidth]{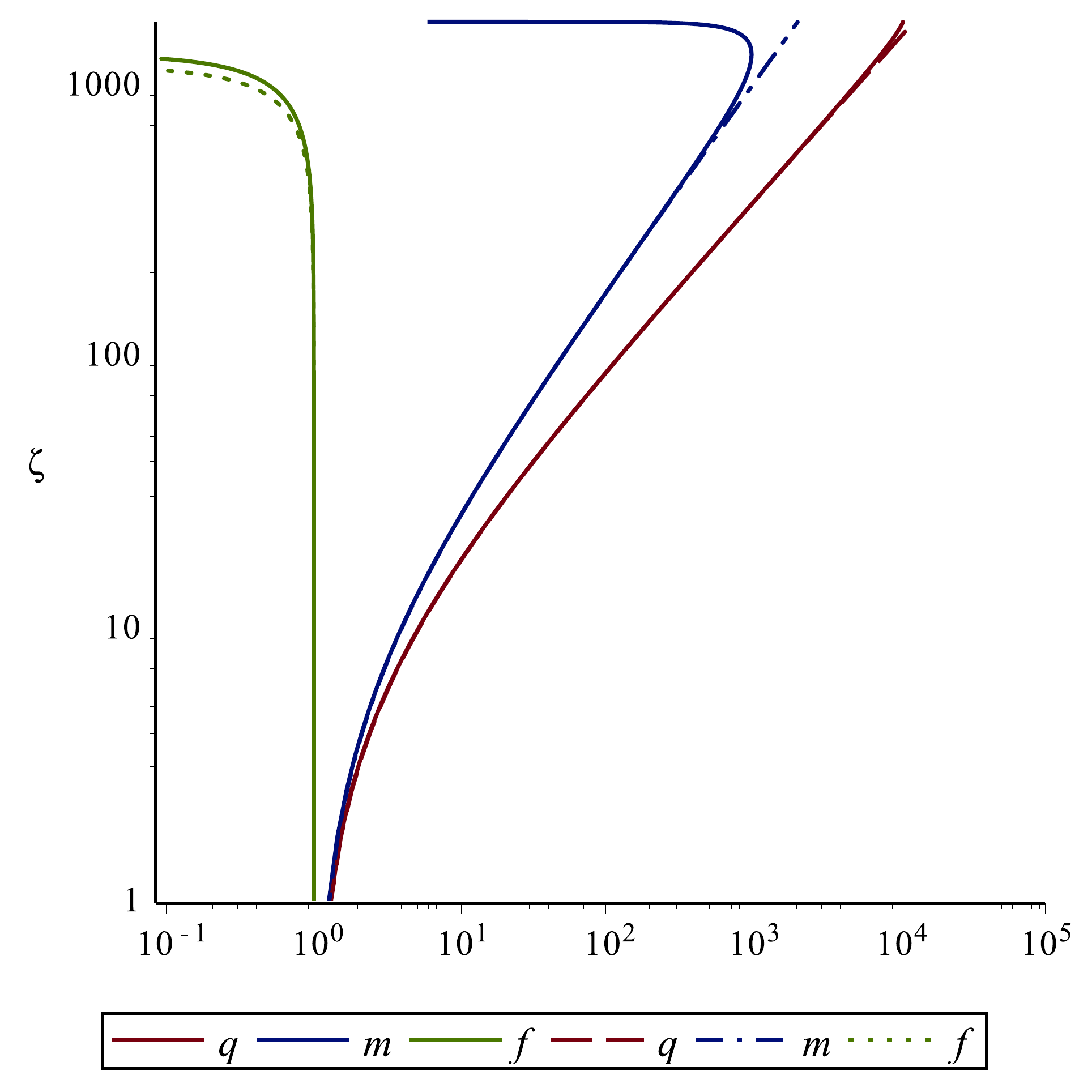}}\quad
\subfloat[][axial velocity]{\includegraphics[width=0.45\columnwidth]{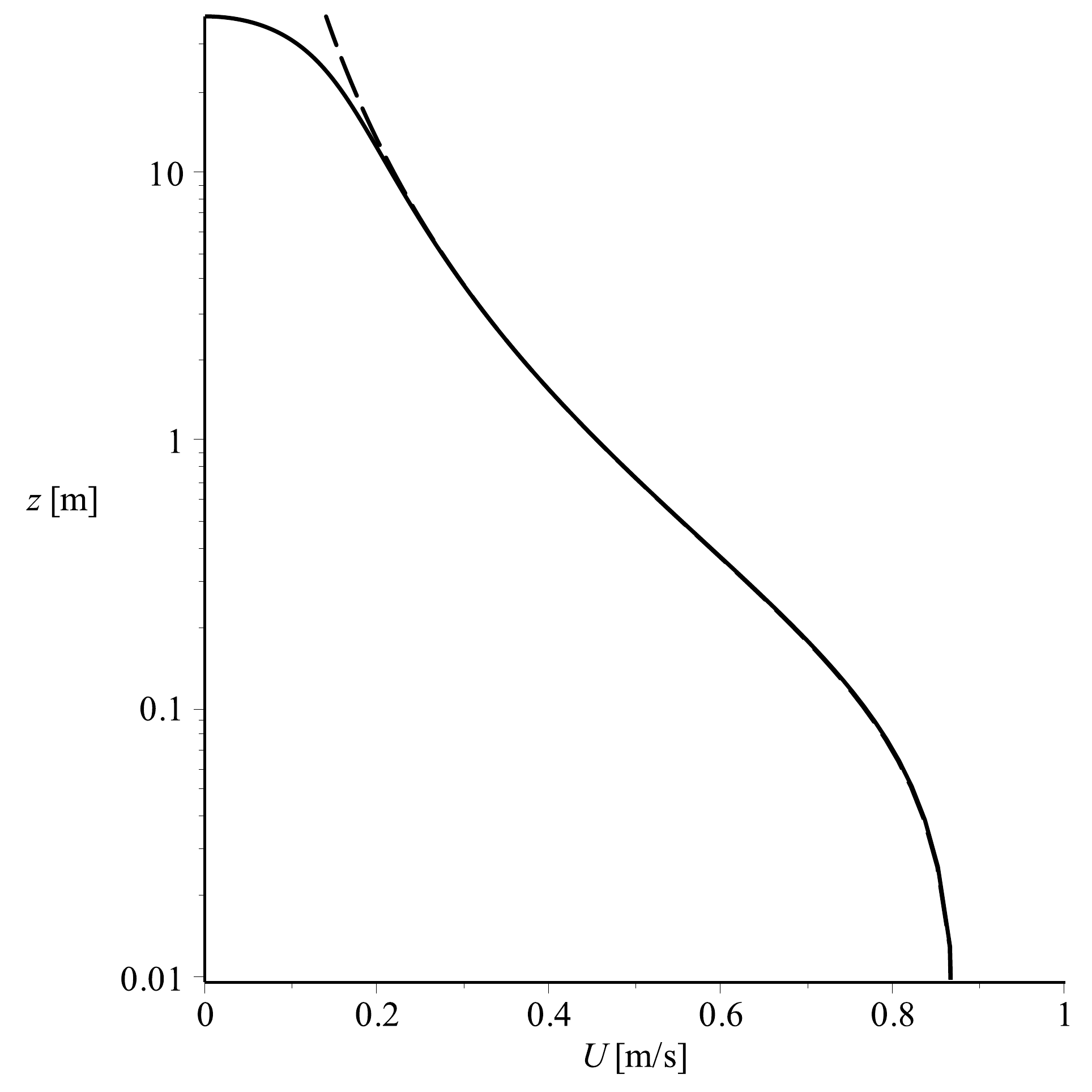}}\\
\subfloat[][plume radius]{\includegraphics[width=0.45\columnwidth]{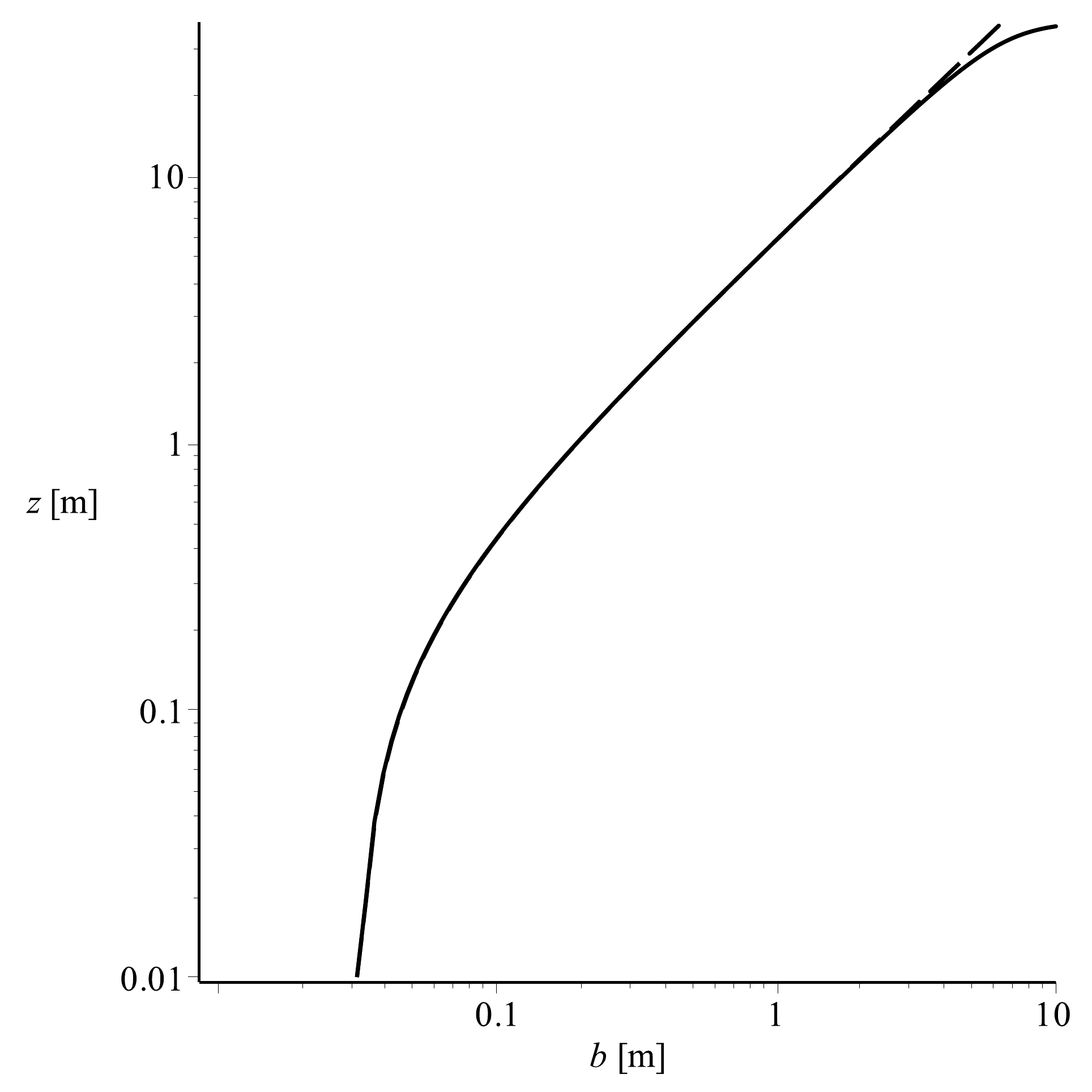}}\quad
\subfloat[][plume density]{\includegraphics[width=0.45\columnwidth]{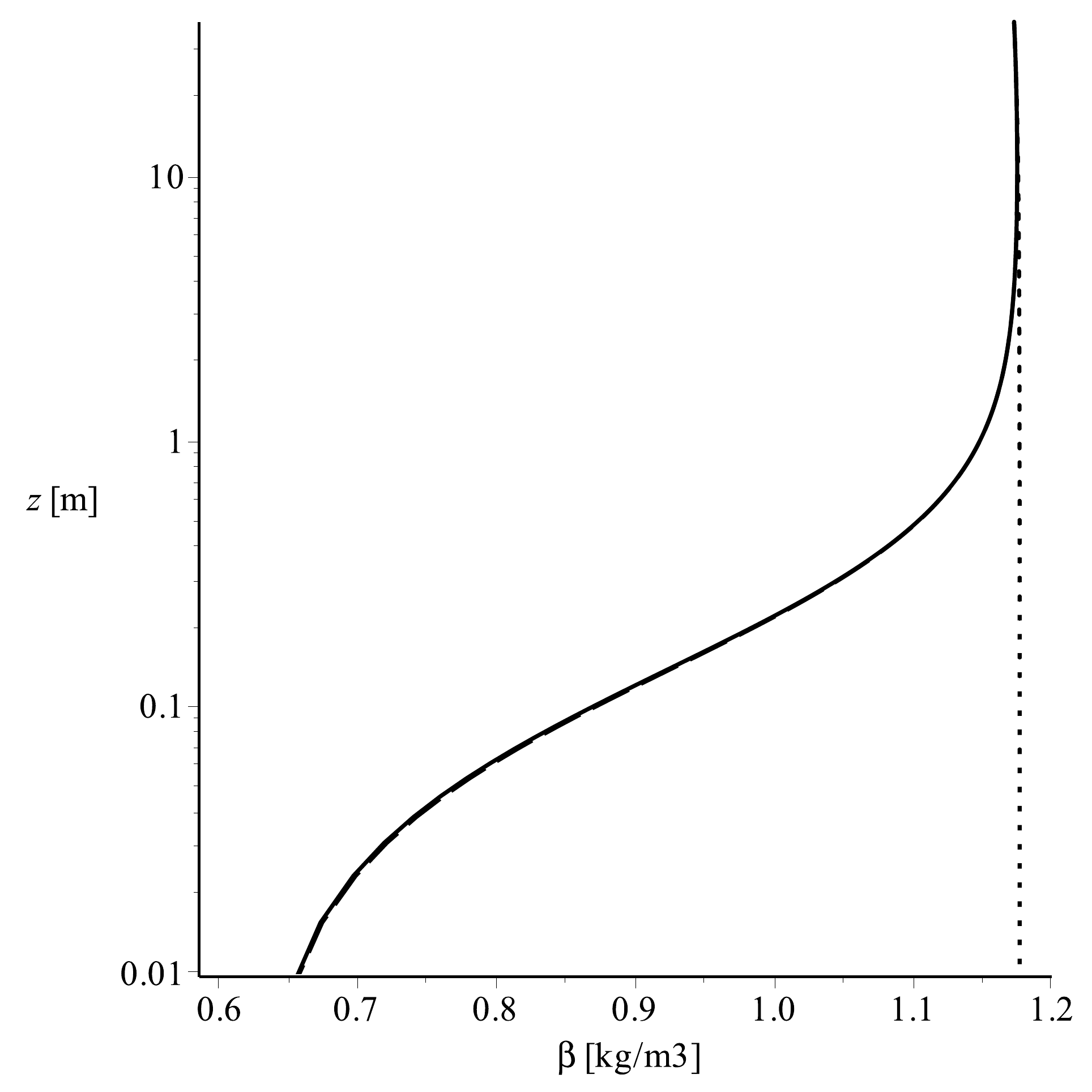}}\\
\subfloat[][plume temperature]{\includegraphics[width=0.45\columnwidth]{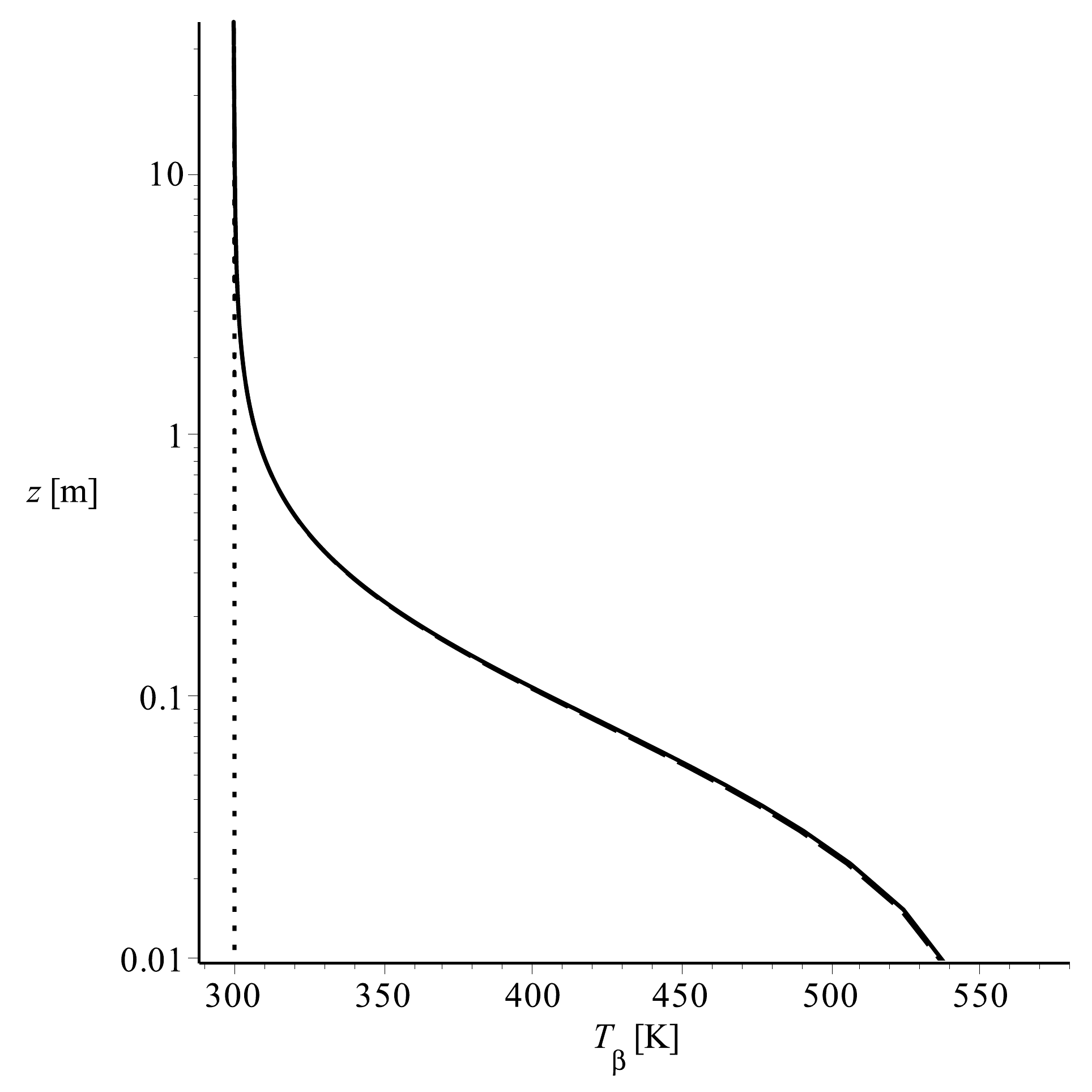}}\quad
\subfloat[][tracer mass fraction]{\includegraphics[width=0.45\columnwidth]{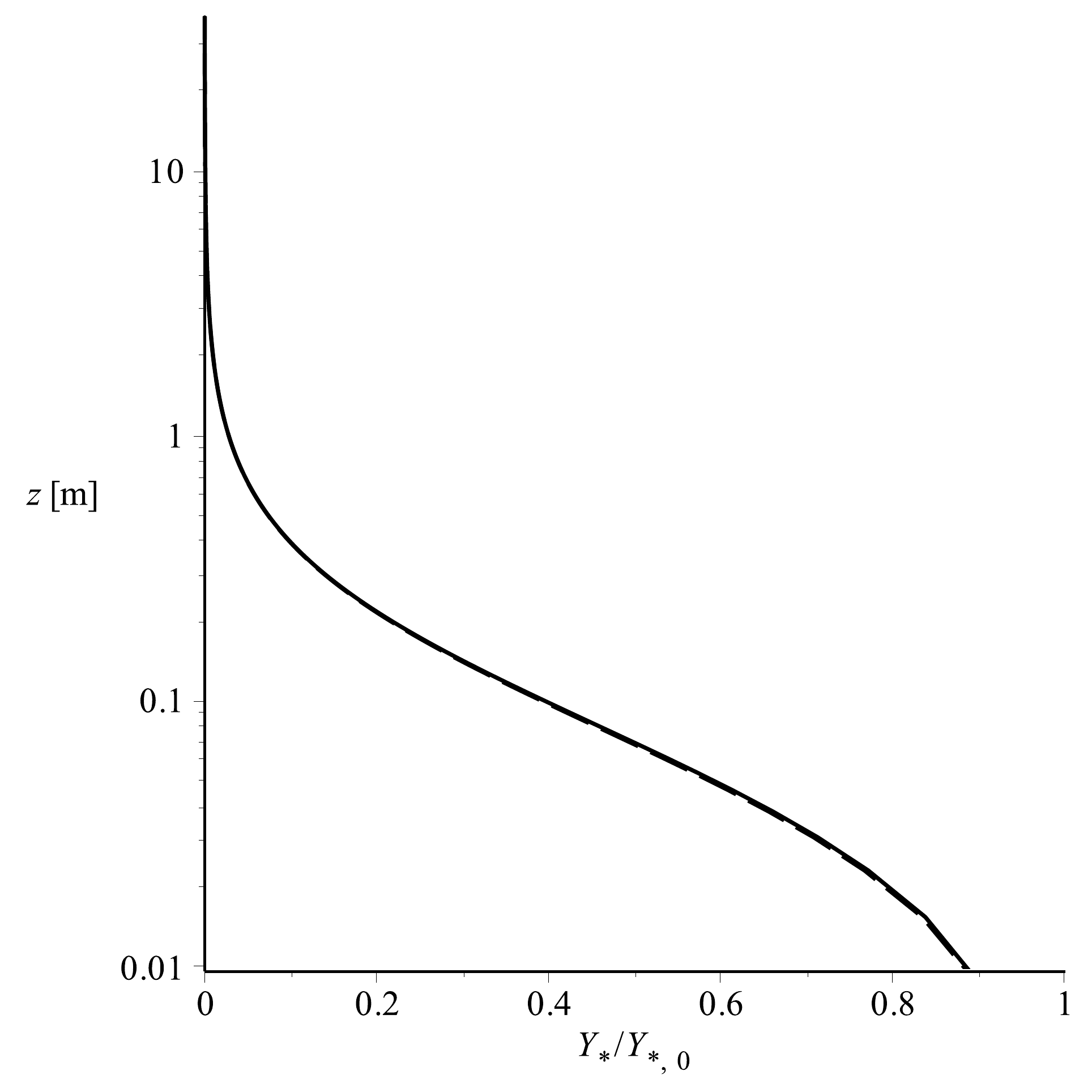}}
\caption{\textsf[{forcedPlume}]: Vertical evolution of the non-dimensional fluxes $q,\,m,\,f$ (log-log scale), of the plume radius $b$ (log-log scale) and of the dimensional physical parameters $U,\,\beta,\,T_\beta,\,Y_{\textup{e}\,(\textup{s})}$, in (linear-log) scale. Solid lines correspond to the numerical solution of model~\eqref{eq:fullModel}, while dashed lines are evaluated by using the analytic asymptotic solution Eqs.~\eqref{eq:asymptoticQ},~\eqref{eq:asymptoticM},~\eqref{eq:asymptoticF}.}
\label{fig:forcedplume_integral}
\end{figure}

\begin{figure}[p]
\centering
\subfloat[][mass, momentum and enthalpy fluxes]{\includegraphics[width=0.45\columnwidth]{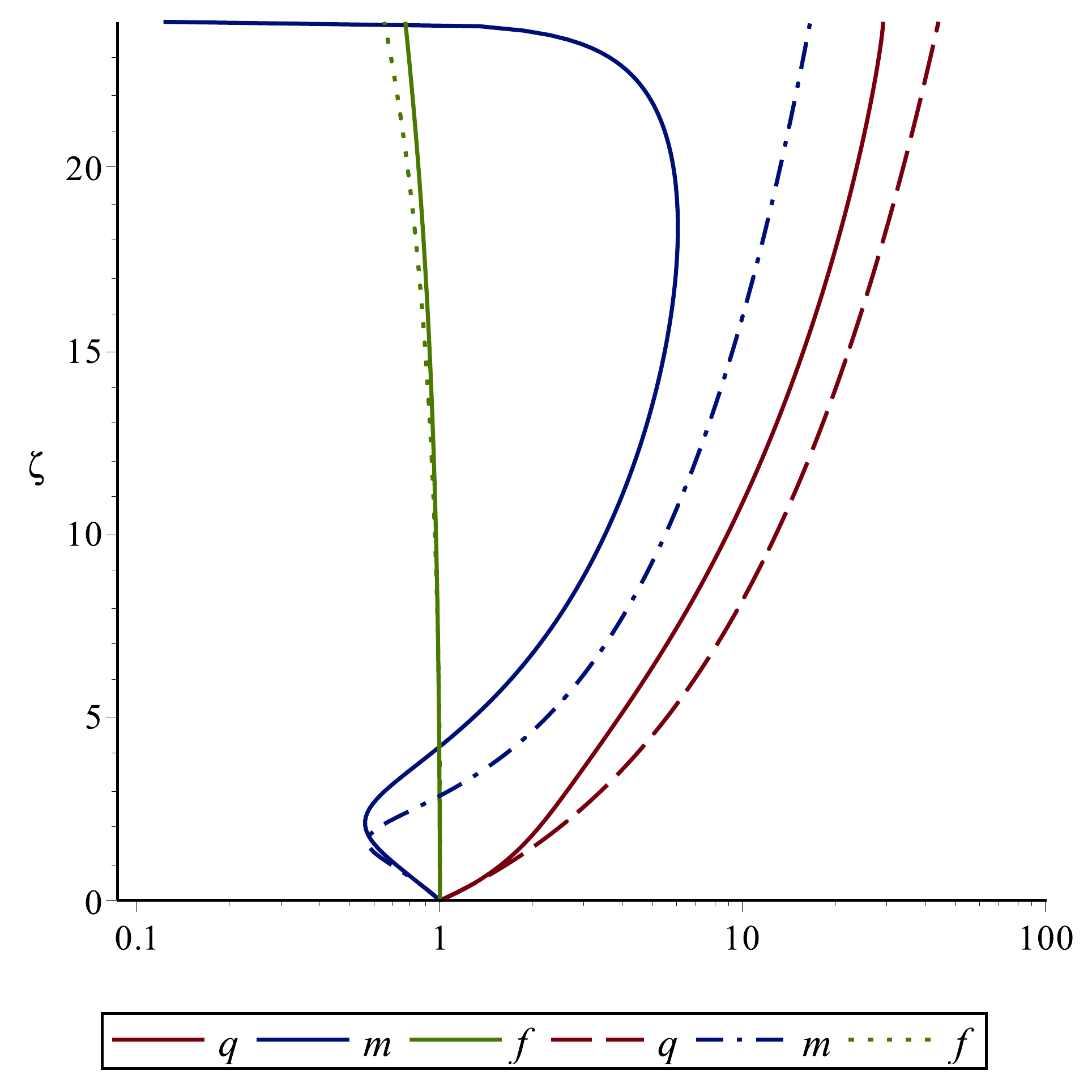}}\quad
\subfloat[][axial velocity]{\includegraphics[width=0.45\columnwidth]{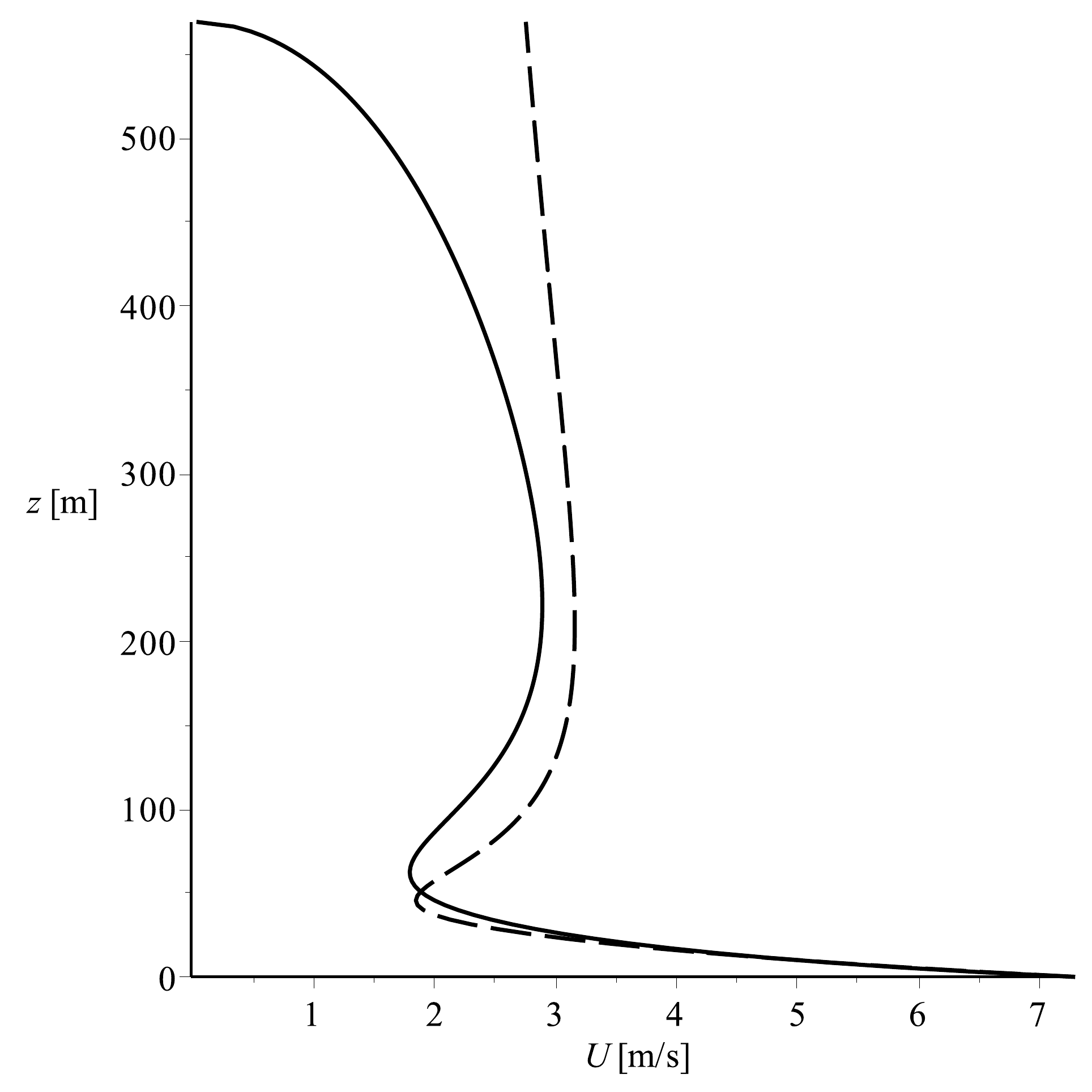}}\\
\subfloat[][plume radius]{\includegraphics[width=0.45\columnwidth]{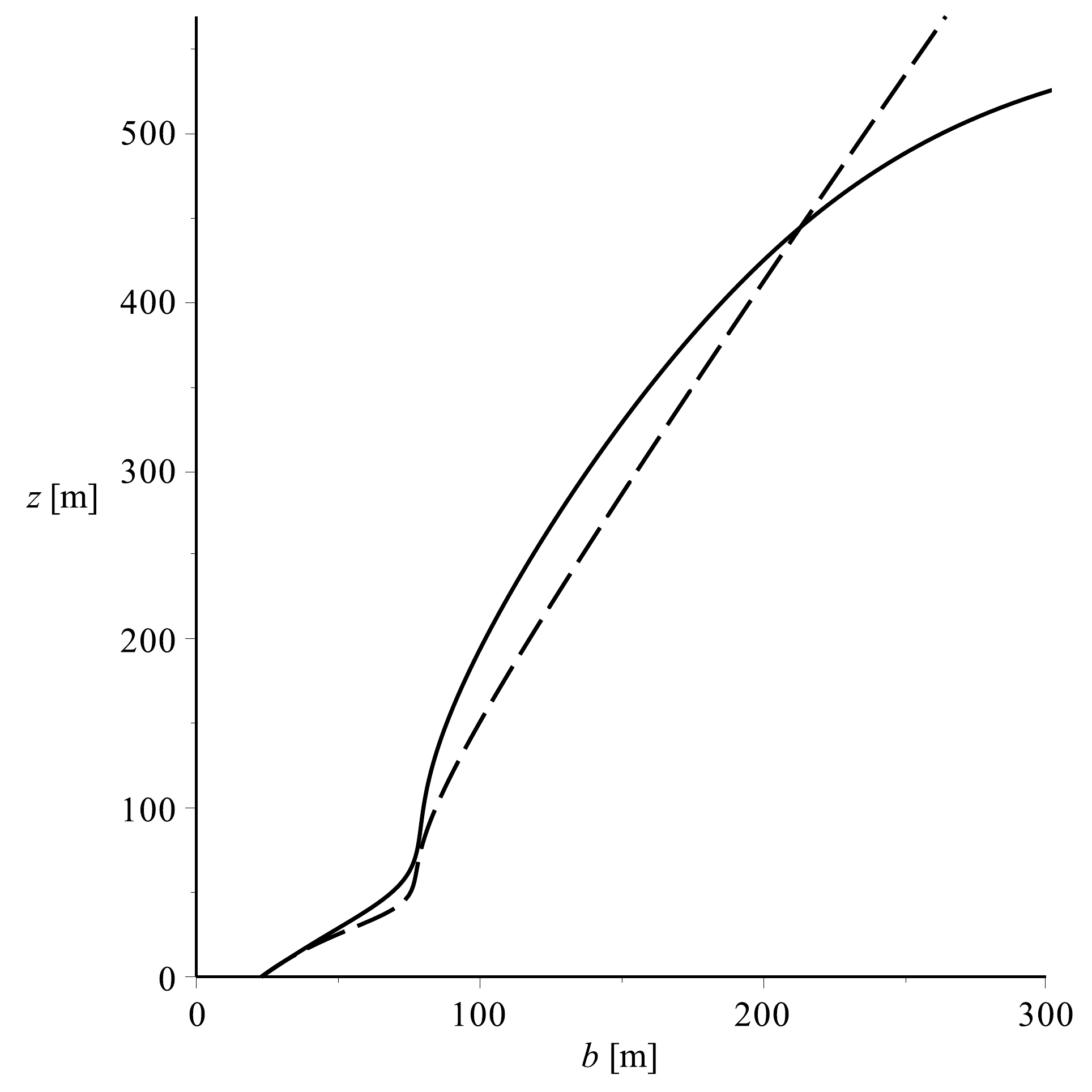}}\quad
\subfloat[][plume density]{\includegraphics[width=0.45\columnwidth]{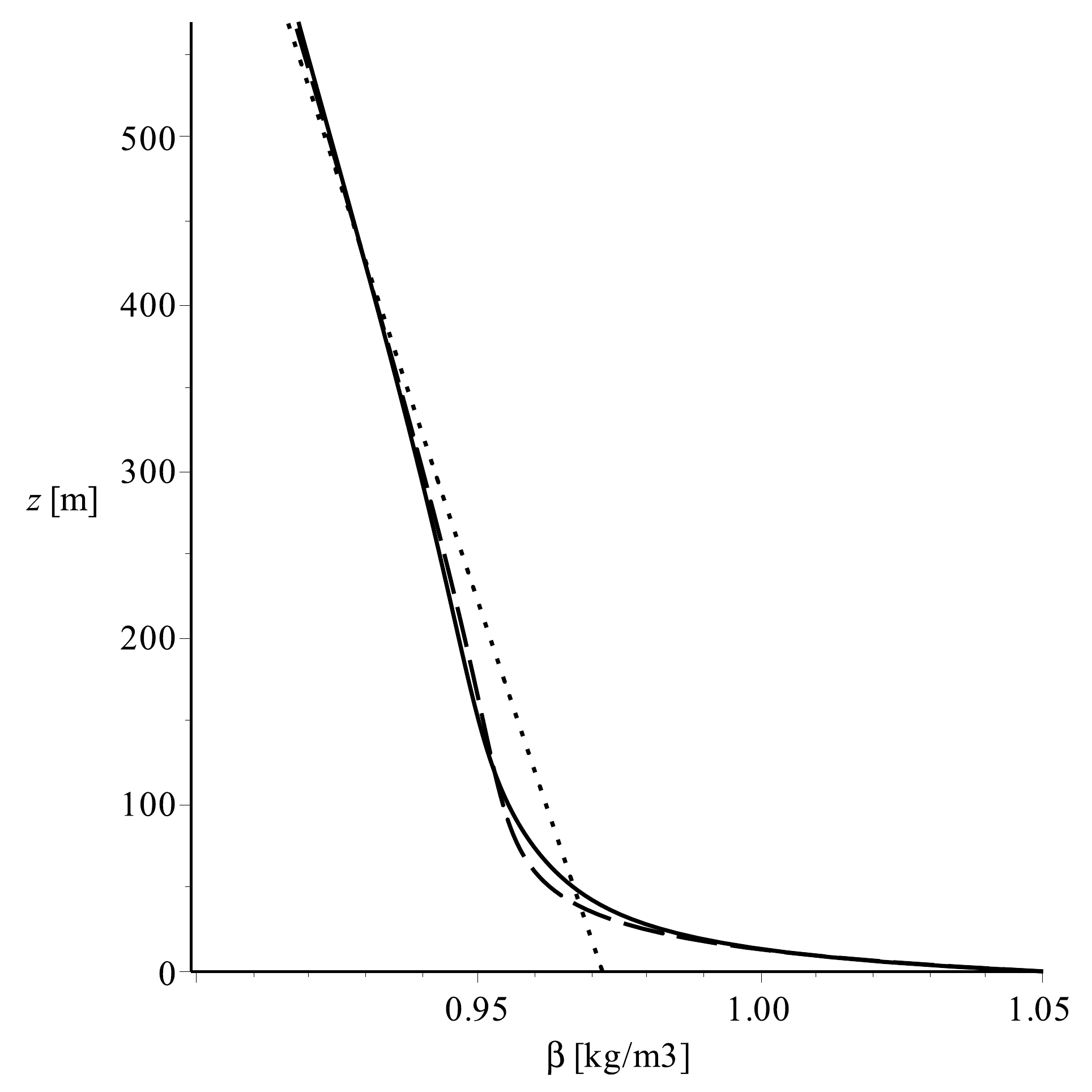}}\\
\subfloat[][plume temperature]{\includegraphics[width=0.45\columnwidth]{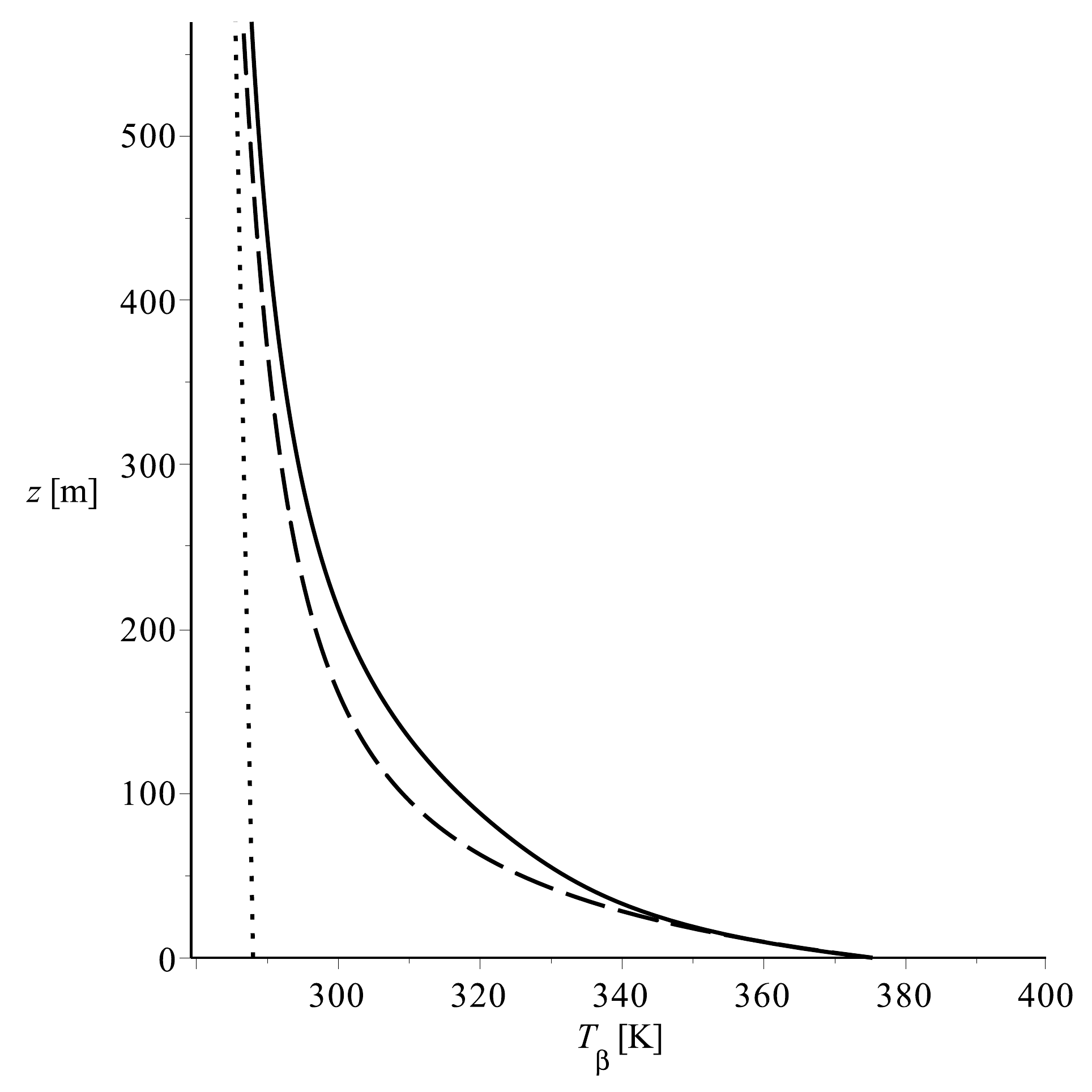}}\quad
\subfloat[][plume mass fractions]{\includegraphics[width=0.45\columnwidth]{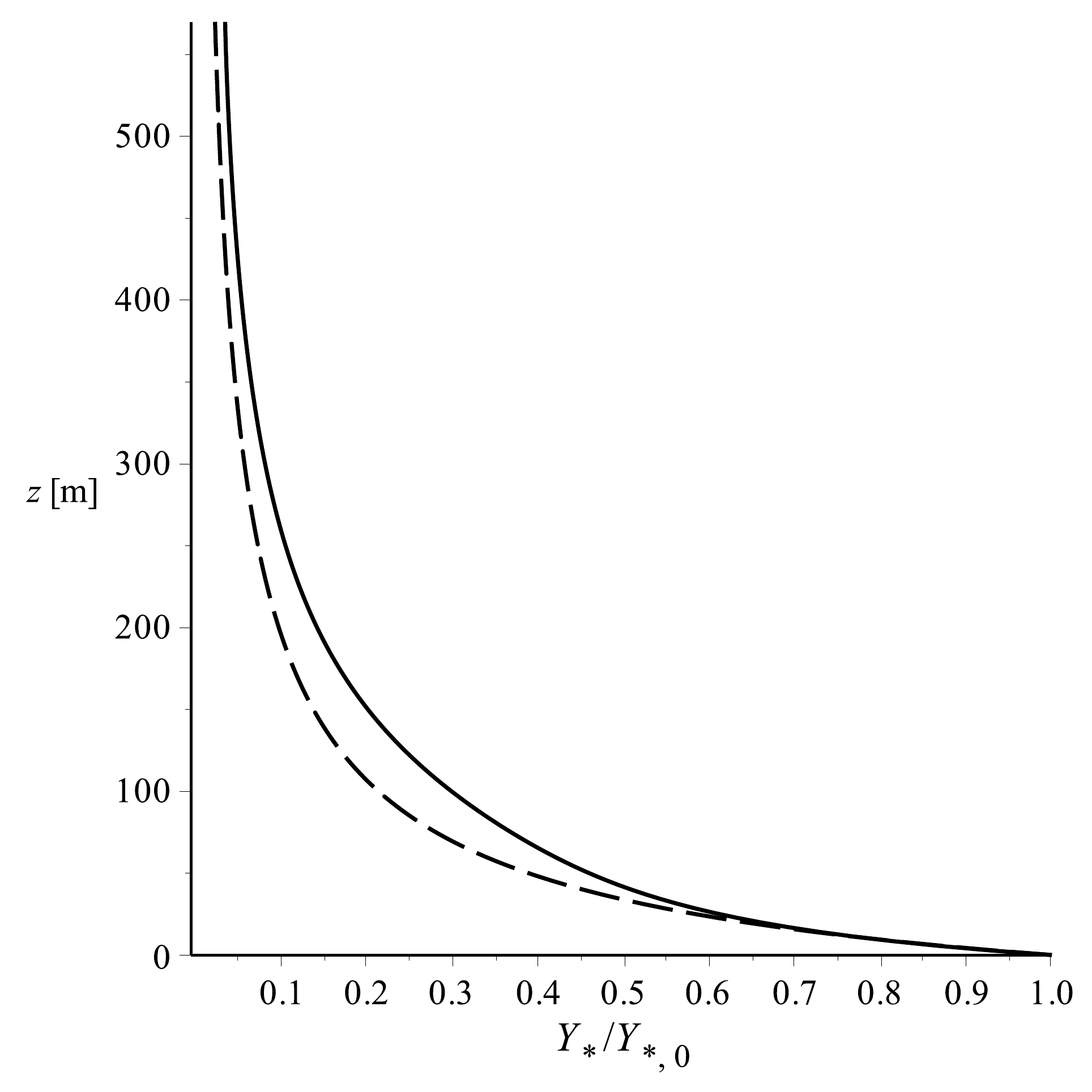}}
\caption{\textsf[{Santiaguito}]: Vertical evolution of the non-dimensional fluxes $q,\,m,\,f$ (log-linear scale) and of the dimensional physical parameters $U,\,b,\,\beta,\,T_\beta,\,Y_{\textup{e}\,(\textup{s})}$. Solid lines correspond to the numerical solution of model~\eqref{eq:fullModel}, while dashed lines are evaluated by using the analytic asymptotic solution Eqs.~\eqref{eq:asymptoticQ},~\eqref{eq:asymptoticM},~\eqref{eq:asymptoticF}.}
\label{fig:santiaguito_integral}
\end{figure}

\begin{figure}[p]
\centering
\subfloat[][mass, momentum and enthalpy fluxes]{\includegraphics[width=0.45\columnwidth]{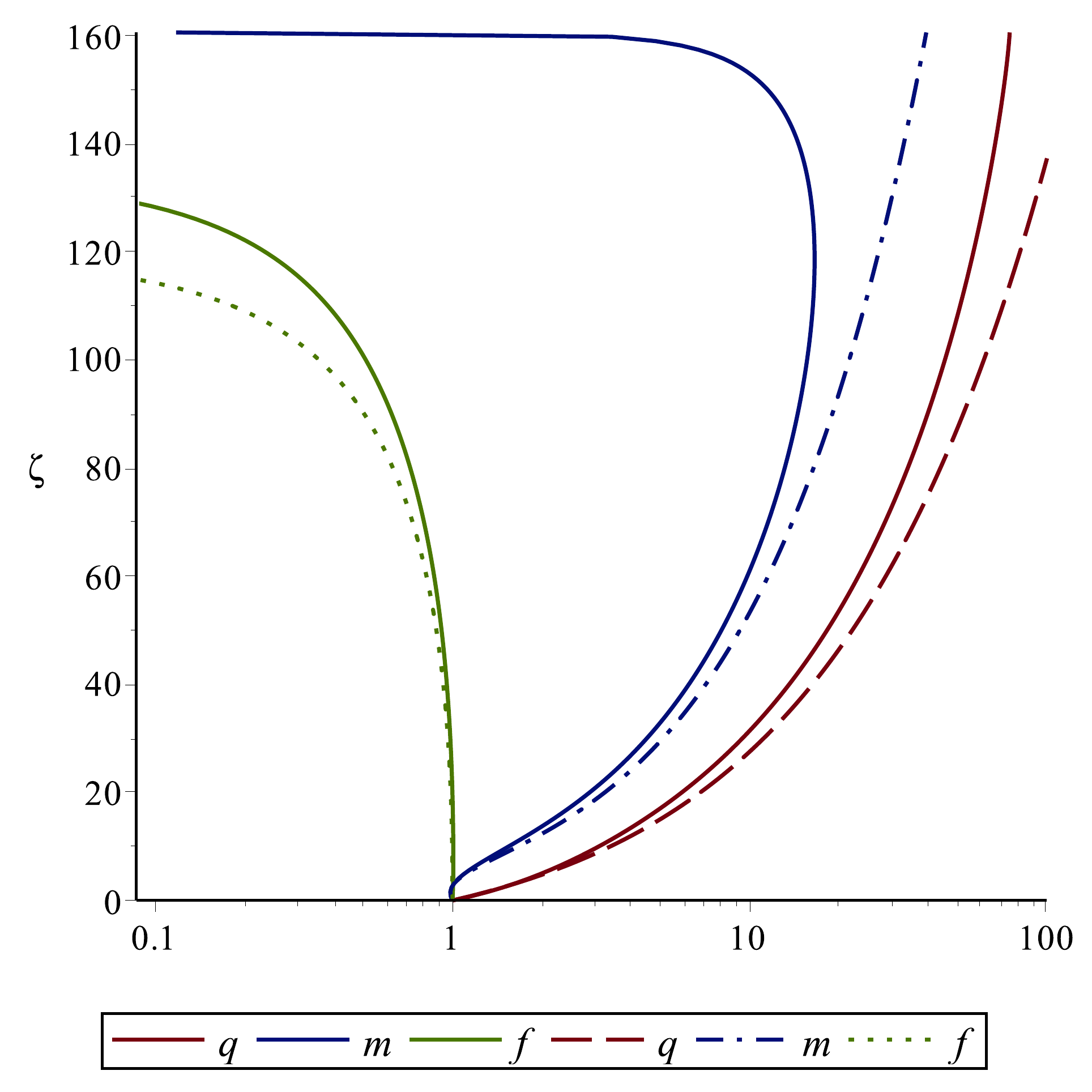}}\quad
\subfloat[][axial velocity]{\includegraphics[width=0.45\columnwidth]{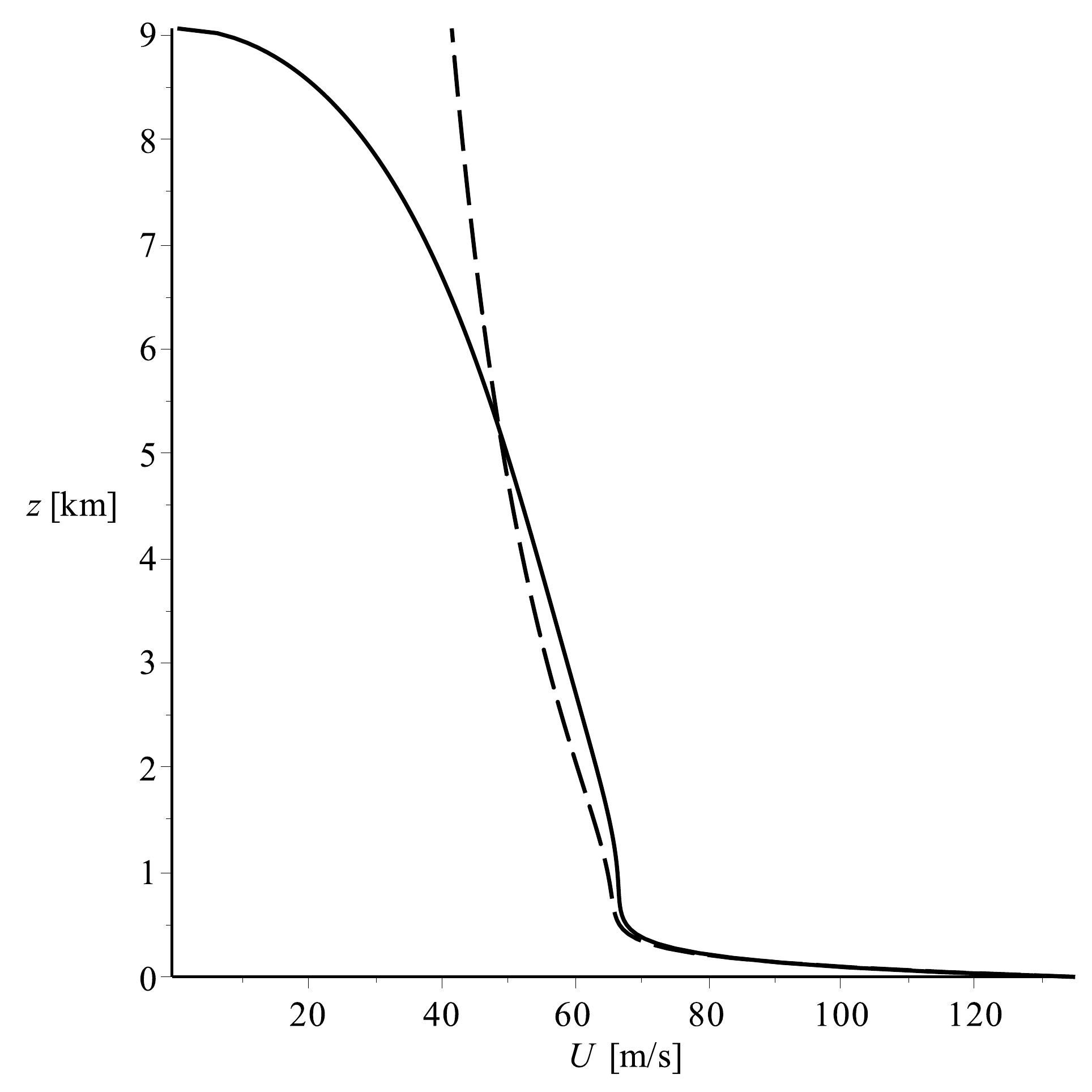}}\\
\subfloat[][plume radius]{\includegraphics[width=0.45\columnwidth]{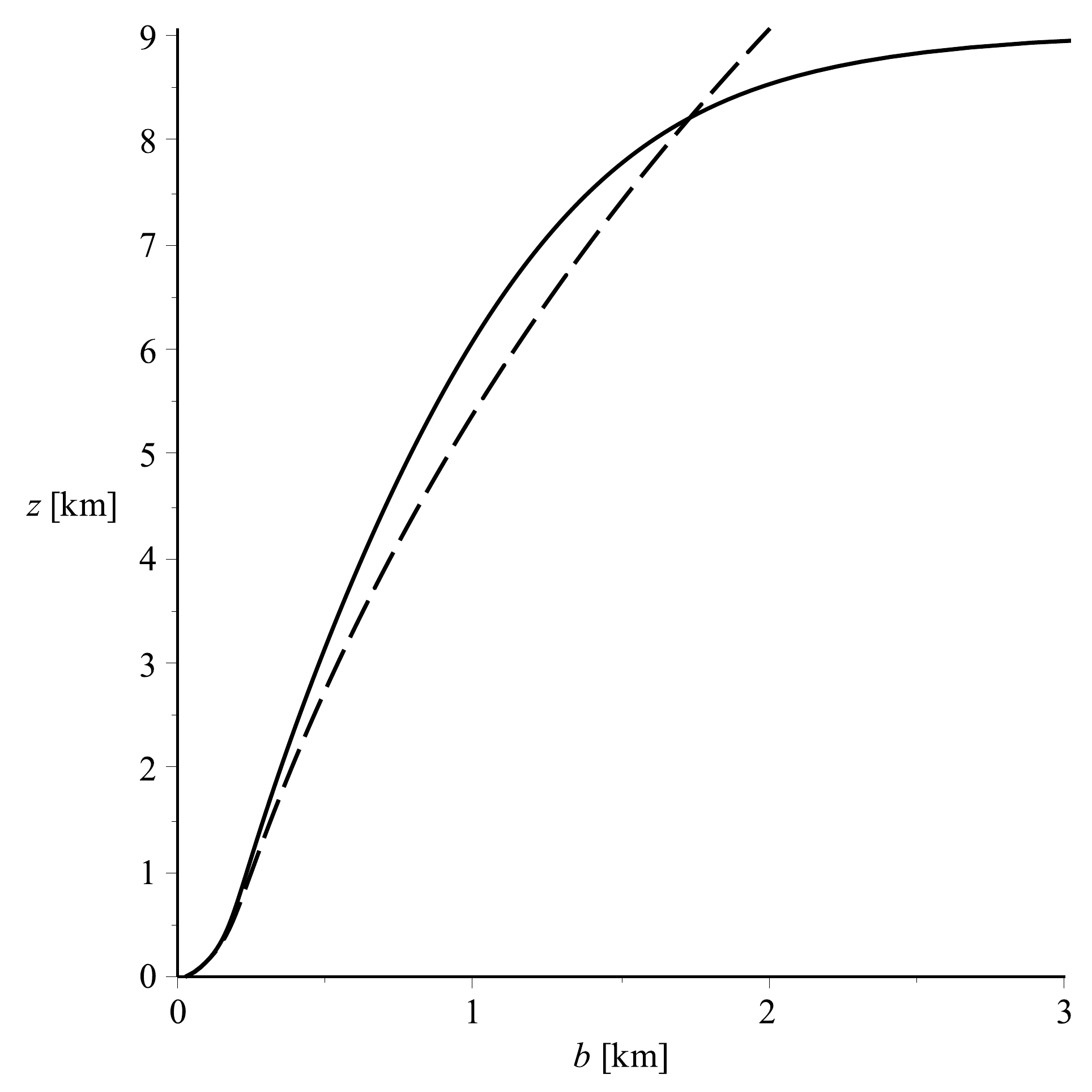}}\quad
\subfloat[][plume density]{\includegraphics[width=0.45\columnwidth]{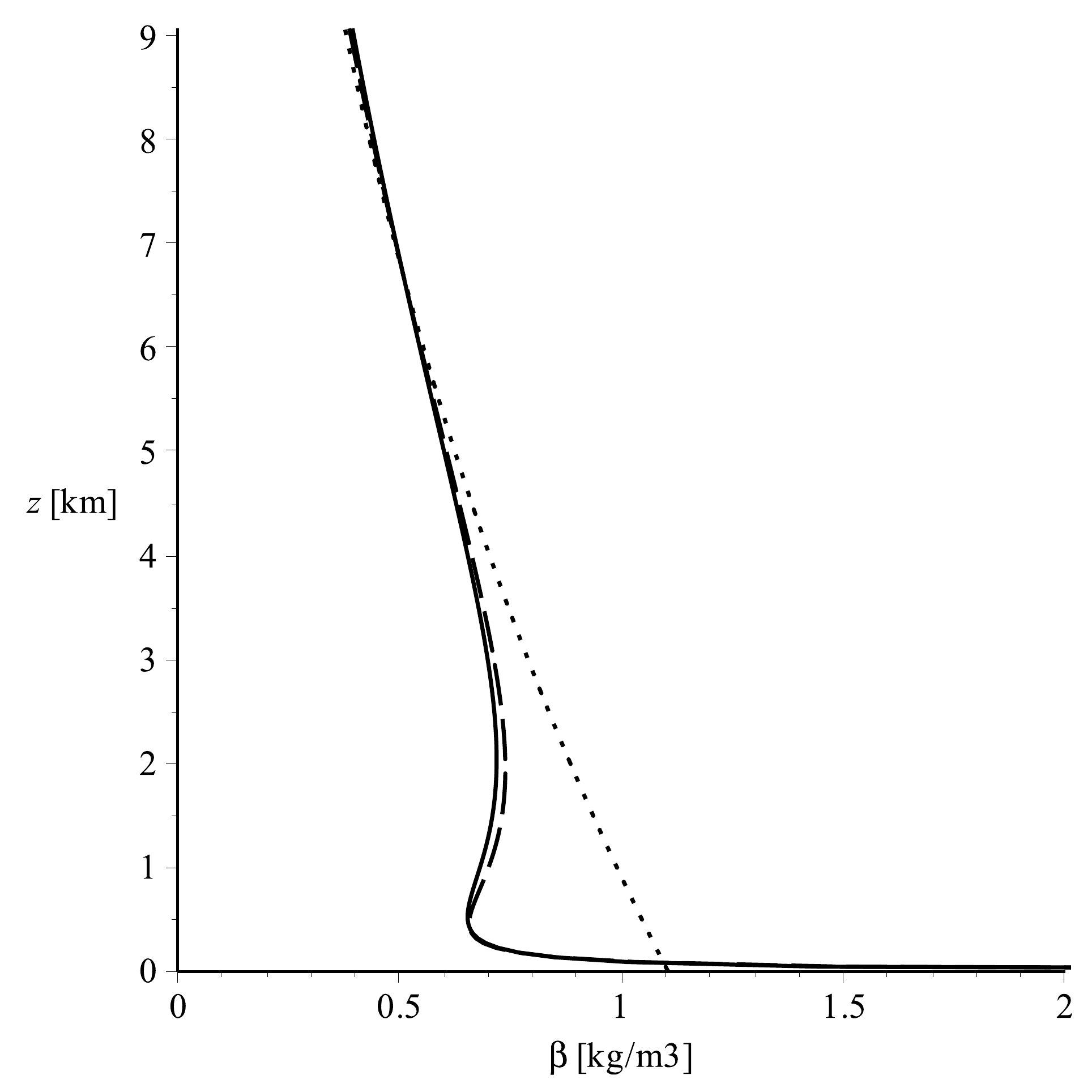}}\\
\subfloat[][plume temperature]{\includegraphics[width=0.45\columnwidth]{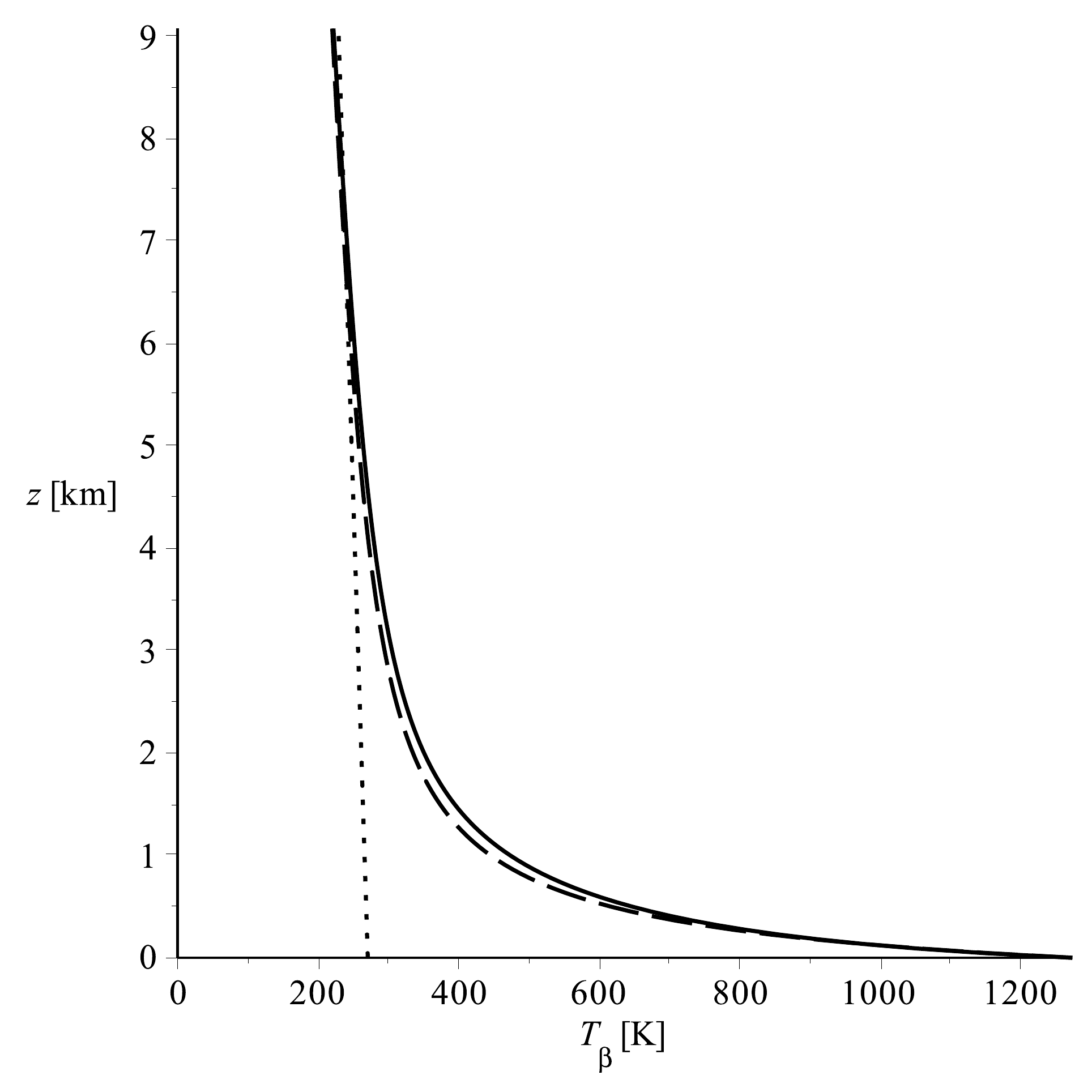}}\quad
\subfloat[][plume mass fractions]{\includegraphics[width=0.45\columnwidth]{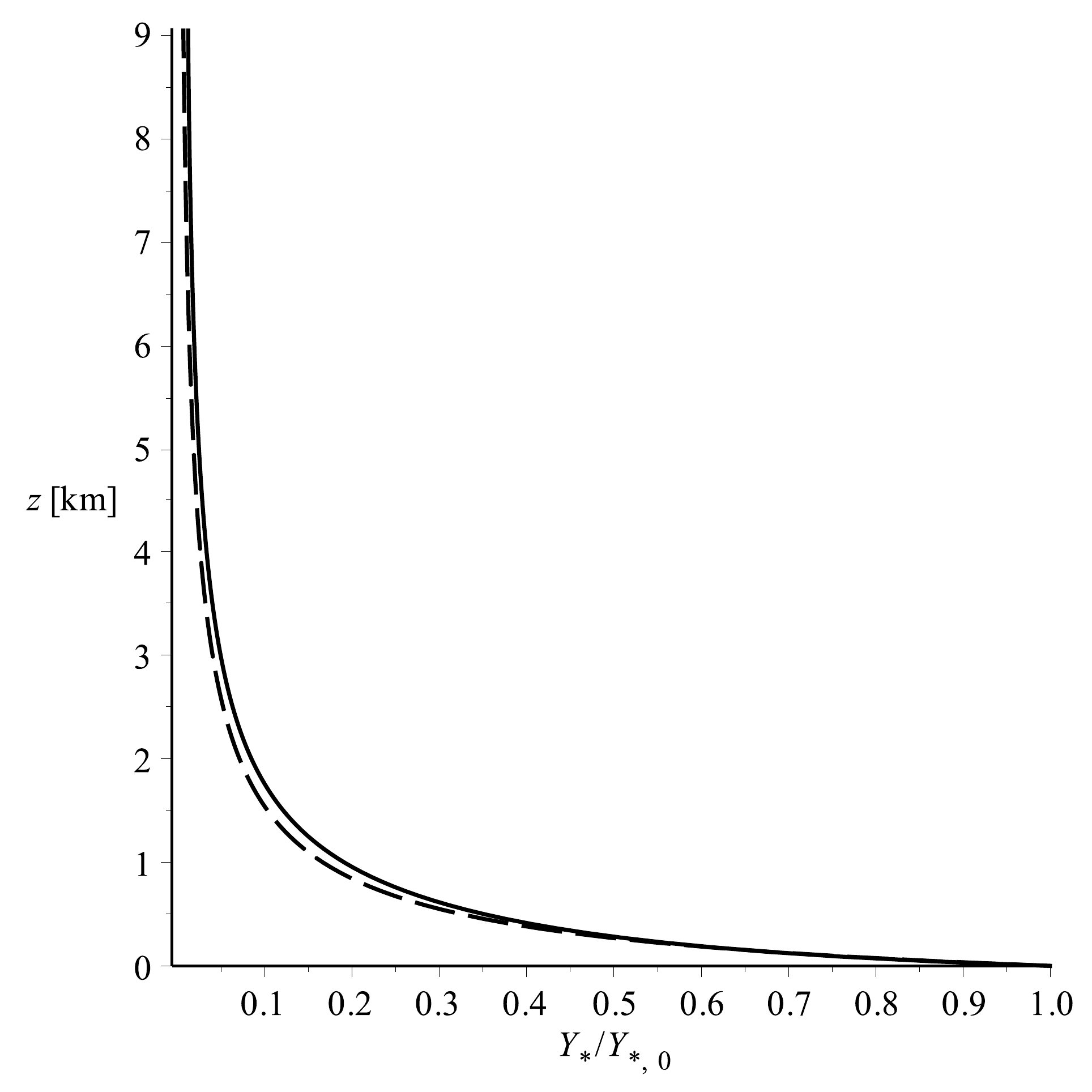}}
\caption{\textsf[{weakPlume}]: Vertical evolution of the non-dimensional fluxes $q,\,m,\,f$ (log-linear scale) and of the dimensional physical parameters $U,\,b,\,\beta,\,T_\beta,\,Y_{\textup{e}\,(\textup{s})}$. Solid lines correspond to the numerical solution of model~\eqref{eq:fullModel}, while dashed lines are evaluated by using the analytic asymptotic solution Eqs.~\eqref{eq:asymptoticQ},~\eqref{eq:asymptoticM},~\eqref{eq:asymptoticF}.}
\label{fig:weakplume_integral}
\end{figure}

\begin{figure}[p]
\centering
\subfloat[][mass, momentum and enthalpy fluxes]{\includegraphics[width=0.45\columnwidth]{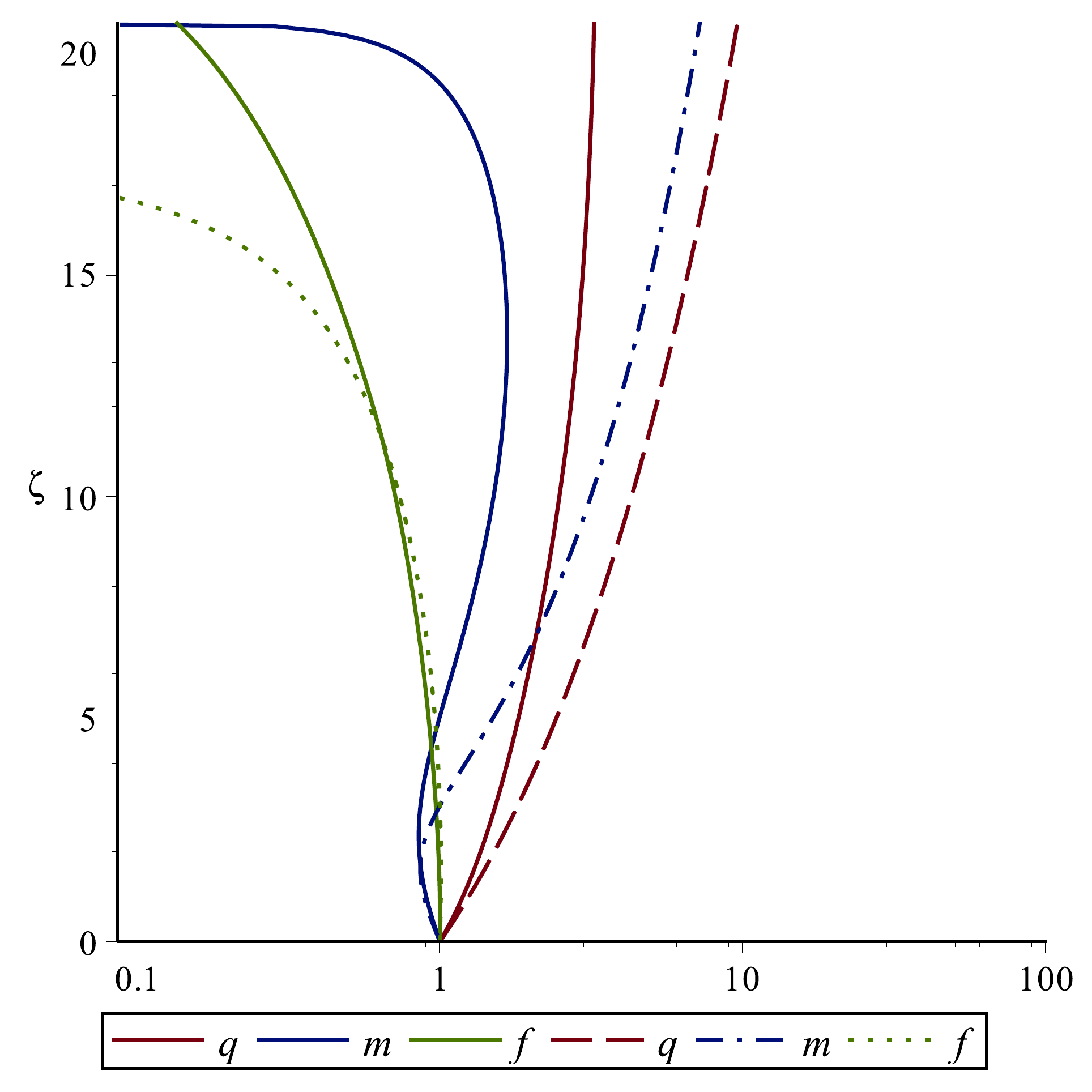}}\quad
\subfloat[][axial velocity]{\includegraphics[width=0.45\columnwidth]{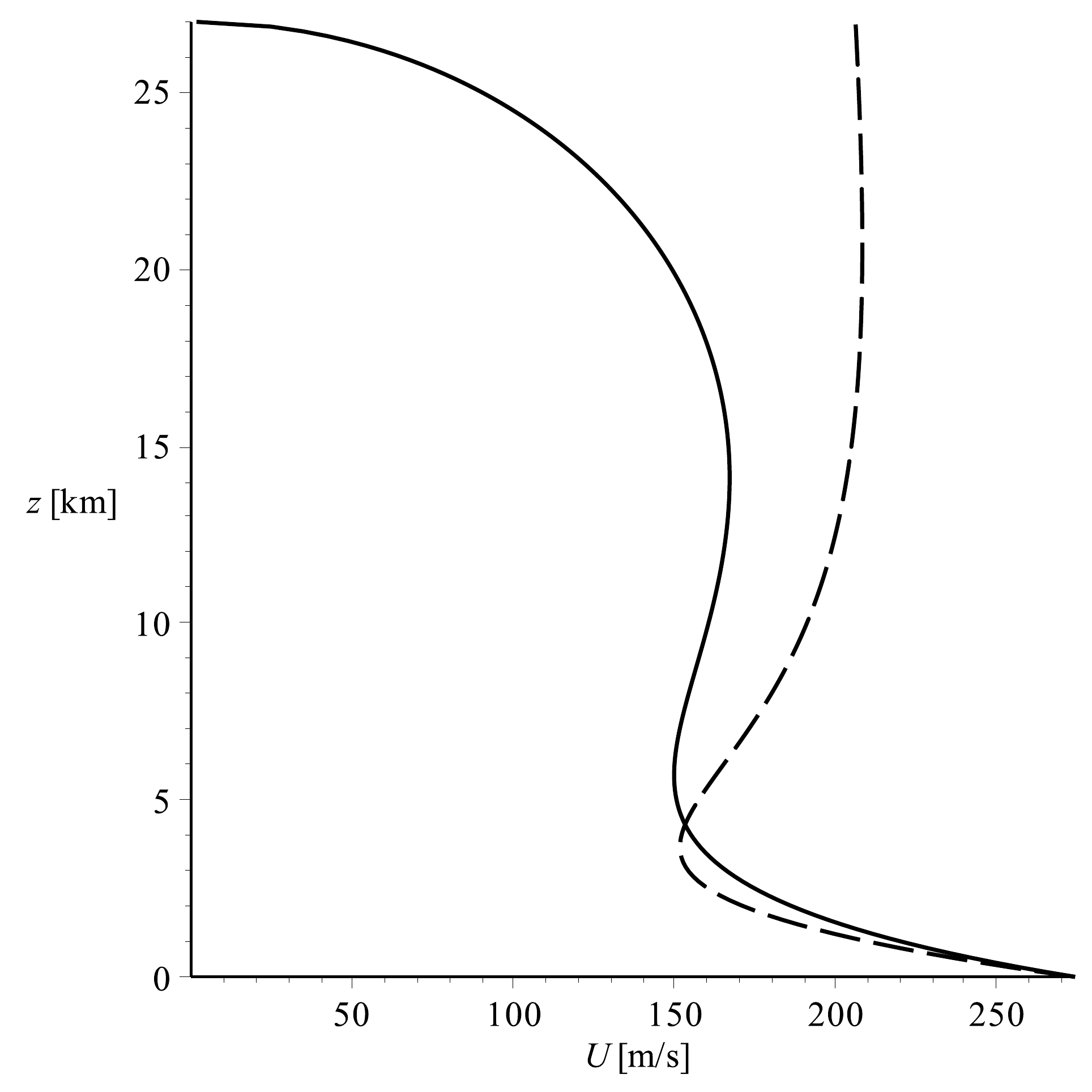}}\\
\subfloat[][plume radius]{\includegraphics[width=0.45\columnwidth]{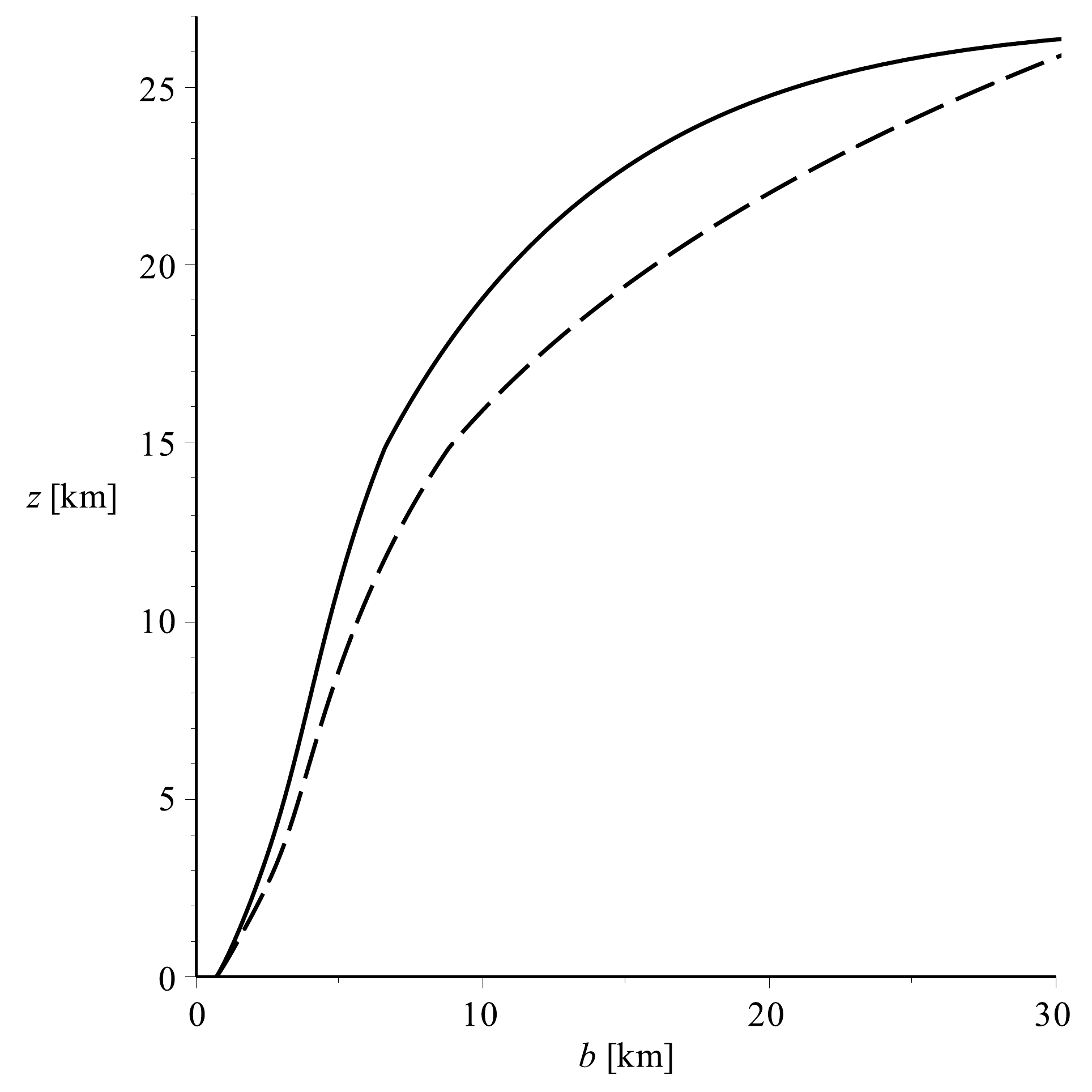}}\quad
\subfloat[][plume density]{\includegraphics[width=0.45\columnwidth]{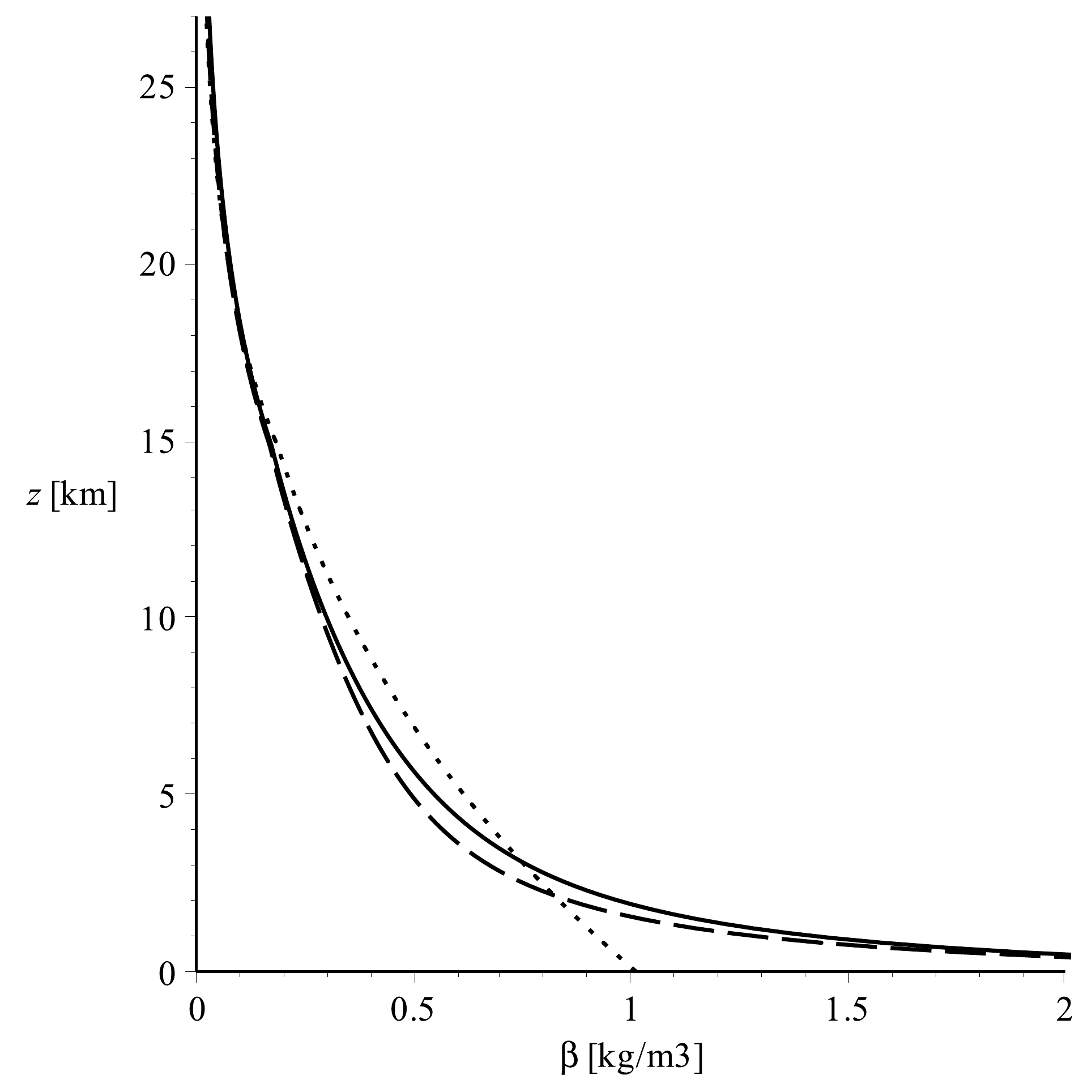}}\\
\subfloat[][plume temperature]{\includegraphics[width=0.45\columnwidth]{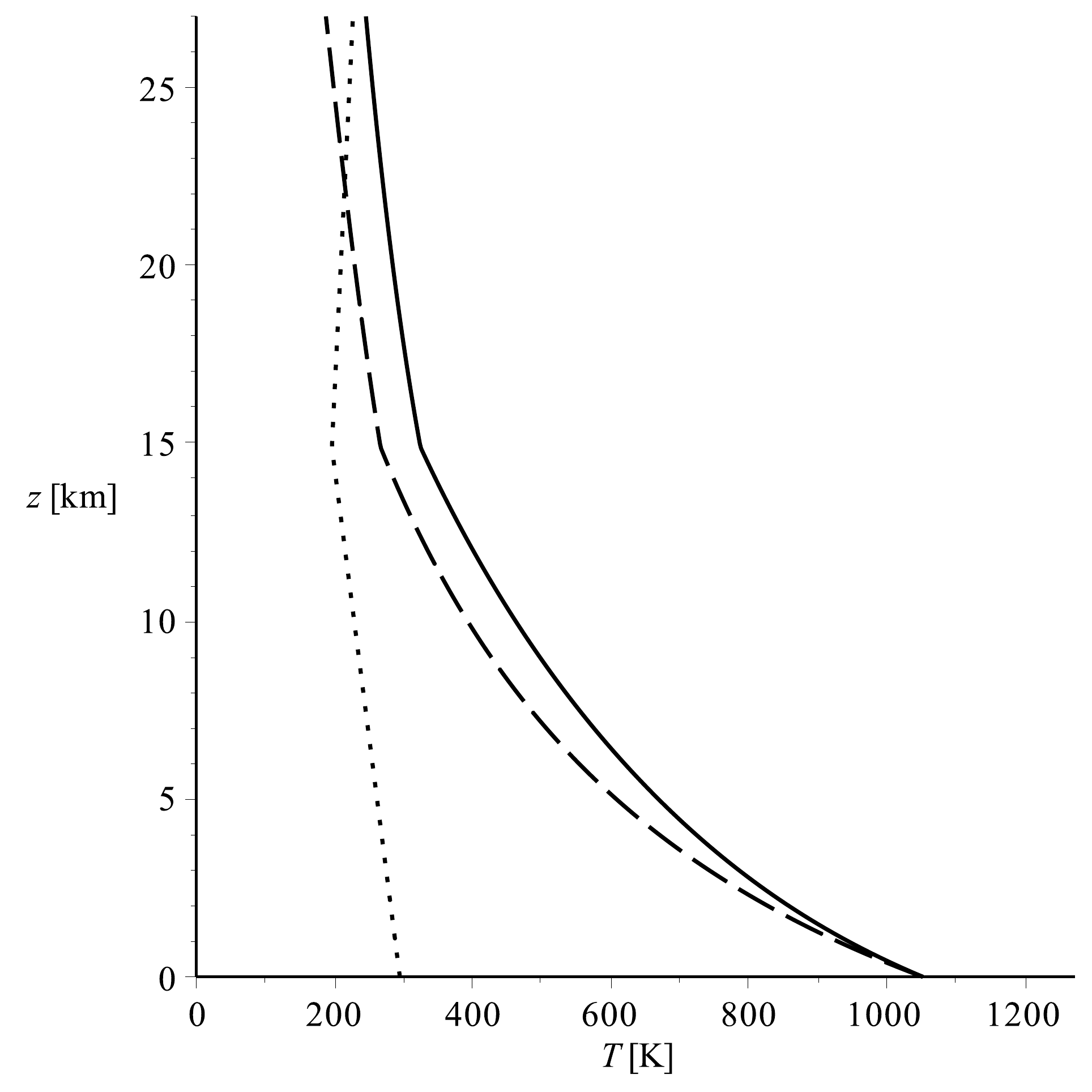}}\quad
\subfloat[][plume mass fractions]{\includegraphics[width=0.45\columnwidth]{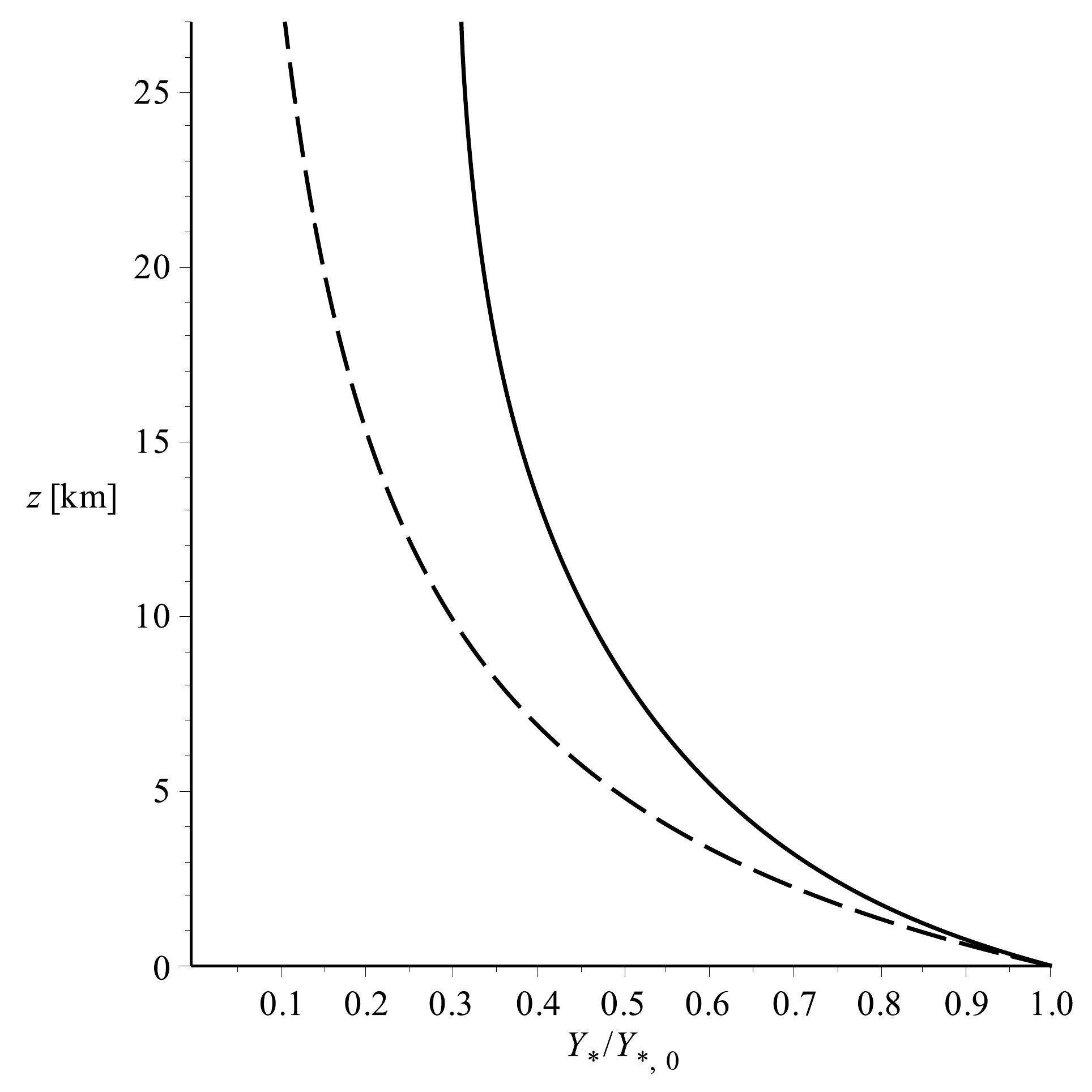}}
\caption{\textsf[{strongPlume}]: Vertical evolution of the non-dimensional fluxes $q,\,m,\,f$ (log-linear scale) and of the dimensional physical parameters $U,\,b,\,\beta,\,T_\beta,\,Y_{\textup{e}\,(\textup{s})}$. Solid lines correspond to the numerical solution of model~\eqref{eq:fullModel}, while dashed lines are evaluated by using the analytic asymptotic solution Eqs.~\eqref{eq:asymptoticQ},~\eqref{eq:asymptoticM},~\eqref{eq:asymptoticF}.}
\label{fig:strongplume_integral}
\end{figure}

Now we move to face the non-Boussinesq regime. The strategy we proposed in~\citet{Cerminara2015ir} is to use the asymptotic solution in the complete inversion formulas for $U$, $b$, $\beta$, $T_\beta$, $\Ye$ and $\Ys$ reported in Eq.~\eqref{eq:fullModel_physicalParameters}. The behavior of this approximation is showed in Figs.~\ref{fig:forcedplume_integral},~\ref{fig:santiaguito_integral},~\ref{fig:weakplume_integral},~\ref{fig:strongplume_integral}. There we notice that the solution works surprisingly well for all the presented plumes. In particular, the temperature and density profiles are well captured for all the cases. The best behavior is recorded in the non-Boussinesq monophase plume (recall $\phi = 0.893$). The asymptotic solution behaves worse for the plume radius and the plume axial velocity in the upper part, where the stratification play the most important role. Anyway, the plume height is captured with less than 10 \% of error for all the plumes. Systematically, the asymptotic mass flux is overestimated with respect model~\eqref{eq:fullModel}. This error present with more evidence in \textsf{strongPlume}, and directly reflects in the underestimation of the mass fractions along the plume axis.


\section{Comparison between results of 3D and integral plume models}
Integral models for plumes describe the evolution with height (the axial unity vector being $\hat{z}$) of three main variables: the flux of mass, momentum and buoyancy. The purpose of these kind of models is to reproduce -- as accurately as possible -- the behavior  of these three parameters under the hypothesis that the plume is stationary. Moving to the 3D models, they give us the plume variables as a function of time and space. In order to compare results, we have first of all to average the 3D result over a time window where the solution can be considered stationary. The second step to do in order to coherently compare the two kind of models is to define the three fluxes also in the 3D case. We choose to define it as described below.

Given $\Omega\times \mathcal{T}$, the space-time domain, we first average over $\mathcal{T}$ a generic 3D variable $\psi(\vec{x},t)$:
\begin{equation}
\bar{\psi} = \langle\psi\rangle_\mathcal{T}(\vec{x}) = \int_\mathcal{T} \psi(\vec{x},t)\,\de t\,.
\end{equation}
For keeping the notation as simple as possible, in this section we use $\bar{(\cdot)}$ in place of $\langle \cdot \rangle_\mathcal{T}$. We define a plume subset $\Oplm(z) \subset \Omega_z$, where $\Omega_z$ is the plane orthogonal to $\hat{z}$ at height $z$. Subset $\Oplm$ is identified by two thresholds: the averaged mixture velocity has positive axial component and the mass fraction of a tracer $\bar{y}_\textup{tracer}$ is larger than a minimum threshold $y_\textup{min}$:
\begin{equation}
\Oplm = \left\lbrace (x_1,x_2)\in\Omega_z\, |\quad \ubb\cdot\hat{z} \geq 0\quad \mbox{and}\quad \bar{y}_\textup{tracer} \geq y_\textup{min} \right\rbrace\,.
\end{equation}
We refer to the integral over this domain as:
\begin{equation}
\psi(z) = \langle\psi(\vec{x})\rangle_{\Oplm} \equiv \int_{\Oplm}\de x_1\de x_2\,\psi(x_1,x_2,z)\,.
\end{equation}
In particular we define respectively the mass flux, the kth mass fraction, the momentum flux and the buoyancy flux as follows:
\begin{subequations}
\begin{align}
& \pi Q = \left\langle \rhombar\, \ubb\cdot \hat{z} \right\rangle_{\Oplm} \equiv \pi \beta U b^2\\
& \pi Q_k = \left\langle \rhombar \bar{y}_k\, \ubb\cdot \hat{z} \right\rangle_{\Oplm} \equiv \pi \beta Y_k U b^2\\
& \pi M = \left\langle \rhombar (\ubb\cdot \hat{z})^2 \right\rangle_{\Oplm} \equiv \pi \beta U^2 b^2\\
& \pi F = \left\langle \left(\frac{1 + \sum_{k}(\chi_k - 1)\bar{y}_k}{1+\sum_{k}(\psi_k - 1)\bar{y}_k}\,\rho_\alpha - \rhombar\right) (\ubb\cdot \hat{z}) \right\rangle_{\Oplm} \equiv \pi \left(\frac{1 + Y_\chi}{1+Y_\psi}\alpha - \beta\right) U b^2\,,
\end{align}
\end{subequations}
where $Y_\psi = \sum_k (\psi_k -1) Y_k\,$, $Y_\chi = \sum_k (\chi_k -1) Y_k$ and $k \in \mathcal{I}\cup\mathcal{J}$ (with nil gas constant of the solid phase $\psi_j = 0$). Moreover, $\alpha(z) = \langle \rho_\alpha(\vec{x}) \rangle_{\Oplm}$. We choose this method for obtaining the one-dimensional integral fluxes because of two reasons: 1) it is the three-dimensional counterpart of what we have defined in Secs.~\ref{sec:buoyantPlumeSolution} and~\ref{sec:GasParticlePlume}, thus it holds even in non-Boussinesq regime
\footnote{A similar approach for the Boussinesq regime has been developed in~\citet{kaminski2005}.}; 2) it is independent on the shape of the radial profile of the plume.

By defining $Q_\psi = Y_\psi Q$ and $Q_\chi = Y_\chi Q$, we can recover the plume variables by using the same inversion formulas given in~\ref{eq:fullModel_physicalParameters}. We recall them in their dimensional form:
\begin{itemize}
\item plume radius $b(z) = \sqrt{\frac{Q(F+Q)(Q+Q_\psi)}{\alpha M (Q + Q_\chi)}}$
\item plume density $\beta(z) = \alpha \, \frac{Q(Q+Q_\chi)}{(F+Q)(Q+Q_\psi)}$
\item kth averaged mass fractions $Y_k(z) = \frac{Q_k}{Q}$
\item plume temperature $T(z) = T_\alpha\, \frac{F+Q}{Q+Q_\chi}$
\item plume velocity $U(z) = \frac{M}{Q}$
\item entrainment coefficient $\varkappa(z) = \frac{Q'}{2 \alpha U b}$
\end{itemize}
where $(\cdot)'$ is the derivative along the plume axis and $T_\alpha = p/R_\alpha\alpha$ is the atmospheric temperature profile.

It is worth noting that the methodology described in this section allows plume modelers to coherently compare results obtained from one-dimensional integral models with data obtained from complex three-dimensional simulations. Moreover, the entrainment coefficient $\varkappa$ -- the key empirical parameter for one-dimensional models -- can be easily obtained for three-dimensional fields. In~\cite{Cerminara2015ashee} we give some example of the results we obtain by using this averaging procedure for the post-processing of three-dimensional plume simulations. We have used the same procedure also for the IAVCEI (International Association of Volcanology and Geochemistry of the Earth Interior) plume model intercomparison initiative \citep{costa_etal_2015}, consisting in performing a set of simulations using a standard set of input parameters so that independent results could be meaningfully compared and evaluated, discuss different approaches, and identify crucial issues of state of the art of models.

\newpage 
\begin{appendices}
\appendixpage
\noappendicestocpagenum
\addappheadtotoc

\twocolumn
\section{Notation}
\begin{supertabular}{r p{0.8\columnwidth}}
\toprule
$\vec{a}$ & acceleration\\
$b$ & plume radius\\
$c$ & speed of sound\\
$C$ & specific heat\\
$C_\textup{D}$ & drag coefficient\\
$\Cp$ & specific heat at constant pressure\\
$\Cv$ & specific heat at constant volume\\
$\mathcal{C}$ & compressibility of the velocity field: $\langle|\Div\vec{u}|^2\rangle_\Omega/\langle|\nabla\vec{u}|^2\rangle_\Omega$\\
$d$ & particle diameter\\
$\mathsf{d}$ & spatial dimension\\
$D$ & vent diameter\\
$\mathcal{D}$ & strain rate tensor\\
$e$ & internal energy per unity of mass\\
$E$ & total energy per unity of mass\\
$\mathcal{E}$ & kinetic energy per unity of mass spectrum\\
$\vec{f}_j$ & drag force per unity of volume acting on the jth particle class\\
$F$ & buoyancy flux\\
$_2F_1$, $\mathfrak{F}$ & Gauss hypergeometric functions\\
$g$ & gravitational acceleration norm\\
$g'$ & reduced gravity\\
$\vec{g}$ & gravitational acceleration vector\\
$\hat{\vec{g}}$ & gravitational acceleration versor\\
$\mathcal{H}$ & enstrophy per unity of mass\\
$h$ & enthalpy per unity of mass\\
$H_\textup{max}$ & volcanic plume maximum height\\
$H_\textup{nbl}$ & volcanic plume neutral buoyancy level\\
$i$ & index running over all the chemical components in the fluid phase\\
$I$ & number of chemical components in the fluid phase\\
$\mathcal{I}$ & set of all the indexes $i$\\
$\mathbb{I}$ & identity tensor\\
$j$ & index running over all the particle classes\\
$J$ & number of particle classes\\
$\mathcal{J}$ & set of all the indexes $j$\\
$k$ & wavenumber\\
$\kg$ & thermal conductivity\\
$K$ & kinetic energy per unity of mass\\
$K_\textup{t}$ & subgrid-scale kinetic energy per unity of mass\\
$L$ & length scale\\
$m$ & mass\\
$N$ & number of grid cells\\
$\BV$ & Brunt-V\"ais\"all\"a frequency\\
$p$ & pressure of the fluid phase\\
$\vec{q}$ & heat flux\\
$r$ & radial coordinate\\
$\hat{r}$ & radial unity vector\\
$R$ & gas constant\\
$Q$ & mass flow rate\\
$\mathrm{Q}_j$ & heat per unity of volume exchanged from the fluid phase to the jth particle class\\
$\dot{Q}_\textup{W}$ & release of thermal energy from the vent\\
$\mathcal{Q}$ & subgrid-scale diffusivity vector for the temperature\\
$S$ & source term\\
$\mathbb{S}$ & rate-of-shear tensor\\
$\mathcal{S}$ & vorticity tensor\\
$t$ & time\\
$T$ & temperature\\
$\mathbb{T}$ & stress tensor\\
$\mathcal{T}$ & temporal domain\\
$\vec{u}$ & velocity vector\\
$U$ & velocity scale or mean plume velocity\\
$U_\epsilon$ & entrainment velocity\\
$V$ & volume\\
$\vec{w}$ & particle settling terminal velocity\\
$\mathcal{W}$ & WALE subgrid model operator\\ 
$\vec{x}$ & position vector\\
$y$ & mass fraction\\
$\mathcal{Y}$ & subgrid-scale diffusivity vector for the mass fraction\\
$z$ & axial coordinate\\
$\hat{z}$ & axial unity vector\\
\midrule
$\alpha$ & density of the atmosphere\\
$\beta$ & gas-particle mixture density for the integral plume model\\
$\beta_\rho$ & density ratio parameter\\
$\gamma$ & adiabatic index of the gas mixture\\
$\gammac$ & stability of the plume column\\
$\delta$ & grid scale\\
$\Delta x$ & smallest space scale of the dynamical problem\\
$\epsilon$ & volumetric concentration\\
$\ept$ & subgrid-scale energy dissipation\\
$\zeta$ & non-dimensional axial coordinate\\
$\etaK$ & Kolmogorov length scale\\
$\etaX$ & entrainment function\\
$\theta$ & atmospheric thermal gradient\\
$\vartheta$ & azimuth angle\\
$\kappa$ & dispersed on carrier mass ratio\\
$\varkappa$ & entrainment coefficient\\
$\lambda_\textup{T}$ & Taylor microscale\\
$\nu$ & fluid kinematic viscosity\\
$\xi$ & smallest resolved LES length scale\\
$\mu$ & fluid dynamic viscosity\\
$\mu_\textup{b}$ & fluid bulk viscosity\\
$\mut$ & subgrid-scale eddy viscosity\\
$\rho$ & bulk density\\
$\hat{\rho}$ & density\\
$\varrho$ & density scale\\
$\tau$ & typical time scale\\
$\tau_\textup{e}$ & eddy turnover time\\
$\tauK$ & Kolmogorov time scale\\
$\upsilon$ & molar fraction\\
$\phic$ & drag correction function\\
$\chi$ & ratio between specific heats\\
$\psi$ & ratio between the gas constants; generic function\\
$\Omega$ & spatial domain\\
\midrule
$\Co$ & Courant number\\
$\Ec$ & Eckert number\\
$\Eu$ & Euler number\\
$\Fr$ & Froude number\\
$\Ma$ & Mach number\\
$\Nu$ & Nusselt number\\
$\Pra$ & Prandtl number\\
$\Prat$ & subgrid-scale turbulent Prandtl number\\
$\Rey$ & Reynolds number\\
$\Ri$ & Richardson number\\
$\St$ & Stokes number\\
\midrule
$\langle\cdot\rangle$ & cell faces averaging\\
$\langle \cdot \rangle_\Omega$ & space domain averaging\\
$\langle \cdot \rangle_\mathcal{T}$ & temporal domain averaging\\
$\langle \cdot \rangle_j$ & jth mass fraction weight average over the domain\\
$\bar{(\cdot)}$ & filtered quantity\\
$\tilde{(\cdot)}$ & Favre-filtered quantity\\
$(\cdot)_\textup{dg}$ & dusty gas\\
$(\cdot)_\textup{e}$ & ejected gas phase\\
$(\cdot)_\textup{f}$ & fluid phase\\
$(\cdot)_\textup{g}$ & gas phase\\
$(\cdot)_i$ & ith chemical component of the fluid mixture\\
$(\cdot)_j$ & jth particle class\\
$(\cdot)_\textup{r}$ & correction due to particle decoupling\\
$(\cdot)_\textup{rms}$ & root mean square\\
$(\cdot)_\textup{s}$ & solid phase\\
$(\cdot)_\textup{Sth}$ & Sutherland law\\
$(\cdot)_\alpha$ & atmospheric\\
$(\cdot)_\textup{m}$ & gas - particle mixture\\
$(\cdot)_{\beta}$ & gas - particle mixture (integral model)\\
\bottomrule
\end{supertabular}

\onecolumn
\section{Gauss hypergeometric functions}\label{app:hypergeometric}
Gauss hypergeometric functions $_2F_1(\left[\cdot,\cdot\right];[\cdot];x)$ are useful in order to perform integrals of the form:
\begin{equation}
\int \left(x^c - a\right)^b \de x\,.
\end{equation}
$_2F_1(\left[\cdot,\cdot\right];[\cdot];x)$ is the hypergeometric function defined when $x \leq 1$ as:
\begin{align}
& _2F_1(a,b;c;x) = \sum_{n=0}^\infty \frac{(a)_n (b)_n}{(c)_n}\frac{x^n}{n!}\,,
\label{eq:hypergeometric_1}\\
& (a)_n =
\begin{cases}
1 & n=0\\
a(a+1)\dots(a+n+1) & n>0\,.
\end{cases}
\label{eq:hypergeometric_2}
\end{align}

In thesis we have to deal with integrals in which $c=2$, thus we define
\begin{align}
& \mathfrak{F}_b(x) \equiv \,_2F_1\left(\left[-b,\frac{1}{2}\right];\left[\frac{3}{2}\right];x\right)\\
& \mathfrak{G}_b(x) \equiv \,_2F_1\left(\left[-b,-b - \frac{1}{2}\right];\left[\frac{1}{2}-b\right];x\right)\,,
\end{align}
so that
\begin{align}
& \int (a-x^2)^b\de x =  a^b x\,\mathfrak{F}_b\left(\frac{x^2}{a}\right) + C & \mbox{if} \quad x^2 & < a\\
&\int (x^2-a)^b\de x =  \frac{x^{1+2b}}{1+2b}\,\mathfrak{G}_b\left(\frac{a}{x^2}\right) + C & \mbox{if} \quad x^2 & > a\,.
\end{align}
It is worth noting that $\mathfrak{F}_b(1)$ and $\mathfrak{G}_b(1)$ are finite and them value is tied to the Gamma function $\Gamma(x)$ as:
\begin{align}
& \mathfrak{F}_b(1) = \frac{\sqrt{\pi}\,\Gamma(1-b)}{2\,\Gamma(3/2-b)}\\
& \mathfrak{G}_b(1) = \frac{2^{2b}\sqrt{\pi}\,\Gamma(1-2b)}{\Gamma(1/2-2b)}\,.
\end{align}

\end{appendices} 
 
\addcontentsline{toc}{chapter}{\refname}
\bibliographystyle{apalike}
\bibliography{integralModels}

\nociteweb{*}
\bibliographystyleweb{alpha}
\bibliographyweb{web}
 
\end{document}